\documentclass[a4paper,12pt]{book}
\usepackage[ansinew]{inputenc}
\usepackage[T1]{fontenc} 
\usepackage[dvips]{graphics}
\usepackage{epsfig}
\usepackage{graphicx}
\renewcommand{\vec}[1]{{\bf #1}}
\usepackage{german}
\usepackage{pifont}
\usepackage{amsmath}
\usepackage{amssymb}
\usepackage{framed}
\usepackage{color}
\usepackage{courier}
\usepackage{makeidx}

\begin{document}

\begin{titlepage}

\LARGE

\begin{center}
{\bf Betrachtungen jenseits des Standardmodells der Teilchenphysik}\\
\large
\vspace{0.5cm}
{\bf Gravitation unter Einbeziehung zusätzlicher Dimensionen und nichtkommutative Geometrie}\\
\vspace{6.1cm}
Diplomarbeit\\
von\\
Martin Kober\\
\vspace{6.1cm}
Institut f\"{u}r Theoretische Physik\\
Johann Wolfgang Goethe Universit\"{a}t\\
Frankfurt am Main\\
\end{center}

\end{titlepage}

\tableofcontents

\chapter*{Einleitung}
\addcontentsline{toc}{chapter}{Einleitung}
\pagestyle{plain}

Die vorliegende Diplomarbeit besteht aus zwei Teilen. Im ersten Teil der Arbeit geht es um die Erweiterung der Allgemeinen Relativit\"{a}tstheorie durch die Annahme zus\"{a}tzlicher Dimensionen, w\"{a}hrend im zweiten Teil die Konsequenzen einer ver\"{a}nderten Raumzeitgeometrie unter der Annahme einer Vertauschungsrelation zwischen den Ortskoordinaten beschrieben werden. Beide Teile umfassen jeweils vier Kapitel, wobei die ersten drei Kapitel der Beschreibung des grunds\"{a}tzlichen Rahmens dienen, w\"{a}hrend die jeweils letzten Kapitel die neuen Ergebnisse dieser Diplomarbeit enthalten.  

Der erste Teil "`Gravitation und zus\"{a}tzliche Dimensionen"' beginnt mit einer Beschreibung der Allgemeinen Relativit\"{a}tstheorie. Am Anfang steht eine historische Einleitung, da ein wahres Verst\"{a}ndnis der Theorie wohl am besten in Bezug auf den historischen Zusammenhang erreicht werden kann. Die Darstellung muss sich zwangsl\"{a}ufig auf die grundlegenden Aussagen der Theorie beschr\"{a}nken, wobei zun\"{a}chst eine kurze Einleitung in die bei der Formulierung der Theorie verwendete Mathematik gegeben wird. Im darauffolgenden Kapitel wird die Idee der Annahme zus\"{a}tzlicher kompaktifizierter Dimensionen erl\"{a}utert, welche auf Kaluza und Klein zur\"{u}ckgeht. Damit sind die Voraussetzungen geschaffen, um im vierten Kapitel eine effektive Quantenfeldtheorie der Gravitation unter Einbeziehung der zus\"{a}tzlichen kompaktifizierten Dimensionen zu entwickeln, welche sich der Methode der Pfadintegralquantisierung bedient.  
Im letzten Kapitel des ersten Teiles soll dann schlie\ss lich die Konsequenz f\"{u}r einen Wechselwirkungsprozess der Teilchenphysik untersucht werden, n\"{a}mlich die Produktion eines ZZ-Paares durch Vernichtung zweier Protonen. Ein solcher Vorgang vermittelt durch Gravitonen k\"{o}nnte am LHC relevant werden. Abh\"{a}ngig von der Zahl der zus\"{a}tzlichen Dimensionen und dem Kompaktifizierungsradius und damit der modifizierten Planckmasse ergibt sich ein Beitrag zu dem Wirkungsquerschnitt f\"{u}r den genannten Prozess.
Dieser wird mit dem Wert verglichen, der bei alleiniger Zugrundelegung des Standardmodells erwartet w\"{u}rde, was R\"{u}ckschl\"{u}sse darauf zul\"{a}sst, ob die Existenz zus\"{a}tzlicher Dimensionen sich in einer signifikant erh\"{o}hten ZZ-Produktionsrate \"{a}u\ss ern k\"{o}nnte. Das Ergebnis der Untersuchung der ZZ-Produktionsrate im ADD Modell stellt das eigentlich neue wissenschaftliche Ergebnis des ersten Teiles dar und ist in \cite{Kober:2007bc} ver\"{o}ffentlicht.
Nat\"{u}rlich handelt es sich bei der hier verwendeten effektive Beschreibungsweise der Gravitation um einen Versuch, die Gravitation in den Rahmen der bisherigen empirisch best\"{a}tigten Teilchenphysik zu integrieren, welche durch relativistische Quantenfeldtheorien beschrieben wird. 
Dies rechtfertigt den Titel der Arbeit "`Betrachtungen jenseits des Standardmodells der Teilchenphysik"' auch in Bezug auf den ersten Teil. Es kann sich bei der hier verwendeten linearen Entwicklung des metrischen Feldes um die Metrik der flachen Minkowskiraumzeit a priori nur um eine N\"{a}herung handeln. Wichtiger aber ist, dass man es mit einer nichthintergrundunabh\"{a}ngigen Theorie zu tun hat. Damit wird sie der eigentlich konzeptionell wichtigsten Aussage der Allgemeinen Relativit\"{a}tstheorie nicht ganz gerecht, die darin besteht, dass die Raumzeit und ihre Struktur an sich eine dynamische Entit\"{a}t darstellt, das Gravitationsfeld also identisch mit der Raumzeit selbst ist.
Dennoch ist die Theorie durchaus ad\"{a}quat in Bezug auf die Untersuchung der Konsequenzen der in der hier beschriebenen Weise zus\"{a}tzlich eingef\"{u}hrten Dimensionen f\"{u}r die Teilchenphysik.

Das erste Kapitel des zweiten Teiles "`Nichtkommutative Geometrie"' beinhaltet zun\"{a}chst eine Darstellung des theoretischen Geb\"{a}udes von Eichtheorien. Diese f\"{u}hren die Bedeutung von Symmetrieprinzipien in der Teilchenphysik vor Augen, denen wohl grunds\"{a}tzlich eine konstitutive Rolle bei der Beschreibung der Wirklichkeit zukommt. In Anschluss daran wird die elektroschwache Theorie und der damit verkn\"{u}pfte Higgsmechanismus vorgestellt. Das folgende Kapitel gibt eine Einf\"{u}hrung in die Grundidee der nichtkommutativen Geometrie, welche unter anderem die Definition eines Sternproduktes f\"{u}r die Multiplikation von Feldern zur Folge hat. 
Schlie\ss lich wird sich speziell der Formulierung von Eichfeldtheorien auf einer nichtkommutativen Raumzeit zugewendet, die das Konzept von sogenannten Seiberg-Witten-Abbildungen beinhaltet, welche die Theorie der nichtkommutativen Raumzeit auf eine Theorie mit kommutativer Raumzeit abbilden.  
Der letzte Teil ist einer neuen Untersuchung gewidmet, welche sich auf den Higgsmechanismus bezieht und rein theoretischer Natur ist.
Es wird gezeigt, dass die spontane Symmetriebrechung auf der nichtkommutativen Raumzeit nach Abbildung auf die gew\"{o}hnlichen Felder zum gleichen Ergebnis f\"{u}hrt wie die spontane Symmetriebrechung nach der Abbildung, wie sie gew\"{o}hnlich in Betracht gezogen wird. 

\part{Gravitation und zus\"{a}tzliche Dimensionen}

\chapter{Die Allgemeine Relativit\"{a}tsheorie}

\section{Historische Einleitung}

\pagestyle{headings}

\begin{quote}
{\footnotesize "`General Relativity is the discovery that spacetime and the gravitational field are the same entity. What we call >spacetime< is itself a physical object, in many respects similar to the electromagnetic field."' (Carlo Rovelli)}
\end{quote}

Gegen Ende des neunzehnten Jahrhunderts gab es zwei Bereiche in der theoretischen Physik, die jeweils in sich widerspruchsfrei waren und den ihnen entsprechenden Erfahrungsbereich angemessen beschreiben konnten. Es handelte sich um die klassische Mechanik und die Elektrodynamik. Die klassische Mechanik stellt den ersten Versuch einer allgemeinen und exakten Naturtheorie dar. Newton gelang es bekanntlich, alle mechanischen Erscheinungen, von der Planetenbewegung bis hin zum harmonischen Oszillator in eine allgemeine begriffliche Form zu bringen, die durch die bekannten Newtonschen Axiome ausgedr\"{u}ckt wird. Sp\"{a}ter wurde sie durch Lagrange und Hamilton in einer noch eleganteren Weise mathematisch formuliert.
Die klassische Mechanik kennt, um es in der Sprache der Ontologie auszudr\"{u}cken, vier basale Entit\"{a}ten. Es handelt sich um Raum, Zeit, K\"{o}rper und Kr\"{a}fte. Raum und Zeit stellen so etwas wie eine starre B\"{u}hne dar, auf der sich das physikalische Geschehen abspielt, welche selbst jedoch nicht in das dynamische Geschehen miteinbezogen ist. Es sind die K\"{o}rper, welche den Gegenstand einer dynamischen Beschreibung darstellen. Gem\"{a}\ss\ dem zweiten Newtonschen Axiom ist die Beschleunigung eines K\"{o}rpers proportional zu der auf ihn einwirkenden Kraft. Damit ein physikalischer Vorgang eindeutig bestimmt ist, bedarf es neben den Anfangsbedingungen f\"{u}r die Geschwindigkeit und die Position des entsprechenden K\"{o}rpers eines speziellen Kraftgesetzes, dass eine Abh\"{a}ngigkeit der Kraft vom Raumzeitpunkt festlegt. Im Falle der Astronomie liefert dies das Newtonsche Gravitationsgesetz demgem\"{a}\ss\ zwischen zwei K\"{o}rpern eine Kraft proportional zum reziproken des Abstandsquadrates und dem Produkt der beiden Massen wirkt.  

Es stellte sich jedoch heraus, dass die klassische Mechanik nicht in der Lage war, auch die im achtzehnten und neunzehnten Jahrhundert untersuchten elektromagnetischen Ph\"{a}nomene zu erkl\"{a}ren. Den Schl\"{u}ssel zu einer allgemeinen Theorie, die genauso wie die Newtonsche Theorie im Bereich der Mechanik alle Ph\"{a}nomene im Bereich der Elektrodynamik umfasste, lieferte die Einf\"{u}hrung des v\"{o}llig neuartigen Begriffs des Kraftfeldes durch Faraday. Faraday hatte die Idee, die in der klassischen Mechanik als starr vorausgesetzten Kr\"{a}fte selbst einer inneren Dynamik zu unterwerfen.
Schlie\ss lich gelang Maxwell aufbauend auf der konzeptionellen Revolution durch Faraday die exakte mathematische Formulierung der Elektrodynamik. 
Er entwickelte ein System von vier Gleichungen, welche die Dynamik der elektromagnetischen Kraftfelder vollst\"{a}ndig beschreiben.  

Nun entdeckte jedoch Lorentz, dass die Maxwellschen Gleichungen nicht invariant unter der Transformationsgruppe der klassischen Mechanik, den Galileitransformationen, sondern invariant unter einer neuen Transformationsgruppe sind, den nach ihm benannten Lorentztransformationen.
Diese haben die Eigenschaft, dass sie die Lichtgeschwindigkeit invariant lassen. Klassische Mechanik und Elektrodynamik sind also invariant unter unterschiedlichen Transformationsgruppen und damit unvereinbar. Daneben stand das Experiment von Michelson und Morley, das ebenfalls die Unabh\"{a}ngigkeit der Lichtgeschwindigkeit vom Bezugssystem des Beobachters nahelegte. Es war nun eine der gro\ss en Leistungen Einsteins, dass er erkannte, dass sich der Widerspruch zwischen klassischer Mechanik und Elektrodynamik durch Aufgabe der bisherigen Bedeutung des Begriffs der Gleichzeitigkeit aufheben lie\ss. Einstein relativierte die Begriffe des absoluten Raumes und der absoluten Zeit, was zur speziellen Relativit\"{a}tstheorie und damit zur Vereinheitlichung von klassischer Mechanik und Elektrodynamik f\"{u}hrte \cite{Einstein:1905ve}.

Die Revolution der speziellen Relativit\"{a}tstheorie, welche erstmals auf konkret wissenschaftlichem Wege vor Augen f\"{u}hrte, was in der Erkenntnistheorie schon lange vorher behauptet wurde, dass n\"{a}mlich unsere grundlegenden Denkstrukturen und damit die Art wie wir die Welt wahrnehmen und denken zu unterscheiden sind von ihrer realen Beschaffenheit, h\"{a}tte bereits ausgereicht, um Einstein eine entscheidende Rolle in der Geistegeschichte der Menschheit zuzuordnen. Dieses Urteil beh\"{a}lt seine G\"{u}ltigkeit auch dann, wenn man einr\"{a}umt, dass es andere Wissenschaftler wie Poincar\'{e} gab, die \"{a}hnliche Gedanken hatten.  
Tats\"{a}chlich war Einstein mit der neu erreichten begrifflichen Einheit jedoch noch nicht zufrieden. Er sah grunds\"{a}tzlich zwei Probleme. Erstens trug die spezielle Relativit\"{a}tstheorie noch nicht die Gleichberechtigung beliebiger Bezugssysteme sondern nur die aller Inertialsysteme in sich. Zweitens war die Theorie der Gravitation in ihrer alten Form nicht mit der speziellen Relativit\"{a}tstheorie vereinbar, da sie eine instantane Wirkung zwischen Massen voraussetzte, was nat\"{u}rlich dem Postulat widerspricht, dass die Lichtgeschwindigkeit eine obere Grenze f\"{u}r den Austausch von Information darstellt. Einstein versuchte also, die Gravitation in Analogie zur Elektrodynamik als Feldtheorie zu formulieren. Es war allerdings h\"{o}chst bemerkenswert, dass im Falle der Gravitation die Gr\"{o}\ss e, von welcher die Wechselwirkung des K\"{o}rpers mit dem Feld abh\"{a}ngt, n\"{a}mlich die schwere Masse, \"{a}quivalent ist zur tr\"{a}gen Masse, also einer Gr\"{o}\ss e, welche allgemeine mechanische Eigenschaften eines K\"{o}rpers bestimmt. Dies verh\"{a}lt sich also anders als im Fall der Elektrodynamik, wo der Kopplungsparameter, n\"{a}mlich die Ladung, v\"{o}llig unabh\"{a}ngig ist von der tr\"{a}gen Masse eines K\"{o}rpers. 
Die Gleichheit des Kopplungsparameters der Gravitation mit der Gr\"{o}\ss e, welche bestimmt, wie sehr ein K\"{o}rper auf Kr\"{a}fte reagiert, brachte Einstein schlie\ss lich auf die geniale Idee, dass bei der Gravitation die Raumzeit selbst das Wechselwirkungsfeld sein k\"{o}nnte. Diese erhielte dann selbst eine innere Dynamik und tr\"{a}te mit den anderen Gegenst\"{a}nden physikalischer Beschreibung in Wechselwirkung.
Durch die Riemannsche Geometrie wurde die mathematische Sprache geliefert, um eine solche dynamische Raumzeit zu beschreiben. Der Gedanke, dass der physikalische Raum eine von der euklidischen Geometrie abweichende Struktur haben k\"{o}nnte, findet sich bereits bei Gauss, welcher Experimente anstellte, um genau dies  herauszufinden, aber keine Abweichung von der euklidischen Geometrie feststellen konnte. Freilich geht die Einsteinsche Beschreibungsweise der Raumzeit noch viel weiter \cite{Einstein:1916vd}.
Die Allgemeine Relativit\"{a}tstheorie stellt den Abschluss der klassischen Physik dar und ist wahrscheinlich der Gipfel begrifflich-konzeptionellen Denkens in der Geschichte der Naturwissenschaft \"{u}berhaupt. In der folgenden Darstellung soll zun\"{a}chst auf einige f\"{u}r die mathematische Formulierung der Allgemeinen Relativit\"{a}tstheorie bedeutsame differentialgeometrische Aspekte eingegangen werden, ehe mit ihrer Hilfe eine kurze Beschreibung der Theorie selbst folgt. Nat\"{u}rlich kann im Rahmen dieser Arbeit nur auf die basalen Grundprinzipien eingegangen werden. Ausf\"{u}hrlichere Darstellungen k\"{o}nnen beispielsweise in \cite{WeinbergGC},\cite{Wheeler} und \cite{DeFeliceClarke} gefunden werden.

\section{Mathematischer Formalismus}

F\"{u}r die folgende Betrachtung soll zun\"{a}chst einmal in Erinnerung gerufen werden, dass man es bei der mathematischen Beschreibung physikalischer Vorg\"{a}nge grunds\"{a}tzlich mit R\"{a}umen zu tun hat. Diese m\"{u}ssen nicht mit dem physikalischen Anschauungsraum zu tun haben. Es kann sich auch um rein abstrakte mathematische R\"{a}ume handeln, die bestimmte Zusammenh\"{a}nge in der Natur widerspiegeln sollen, wie etwa die Zustandsr\"{a}ume in der Quantenmechanik, welche eine Hilbertraumstruktur aufweisen. Hierbei ist es wichtig, dass die Struktur solcher R\"{a}ume nicht a priori gegeben ist, sondern dass man diese Struktur durch Einf\"{u}hrung bestimmter Konzepte zun\"{a}chst definieren muss. 

\subsection{Differenzierbare Mannigfaltigkeiten}

Am Anfang sei der Begriff der differenzierbaren Mannigfaltigkeit eingef\"{u}hrt, da er das mathematische Grundger\"{u}st darstellt, auf dem alle folgenden Konstruktionen der Allgemeinen Relativit\"{a}tstheorie aufbauen.

Es sei zun\"{a}chst eine abstrakte Punktmenge M gegeben, welche die Struktur eines topologischen Raumes aufweise, was bedeutet, dass ein System offener Mengen definiert ist, welches den Axiomen einer Topologie gen\"{u}gt. Als Karte bezeichnet man eine bijektive Abbildung $\phi$ einer offenen Teilmenge U aus M in eine offene Menge des $\mathcal{R}^d$

\begin{displaymath}
\phi : U \rightarrow \mathcal{R}^d.
\end{displaymath}
Durch eine solche Karte werden die einzelnen Punkte p aus U mit Koordinaten $x^\alpha$ bezeichnet, die den Punkten im $\mathcal{R}^d$ entsprechen. Die Karten fungieren also als Koordinatensysteme f\"{u}r die Mengen, auf denen sie definiert sind. Wenn man nun zwei verschiedene Karten $\phi$ und $\phi^{'}$ betrachtet, die auf zwei verschiedenen Mengen $U$ und $U^{'}$ definiert sind, die einander schneiden, so ist die Schnittmenge 
$U \bigcap U^{'}$ mit zwei Koordinatensystemen $x^\alpha$ und $x^{\alpha '}$ \"{u}berdeckt. Diese m\"{u}ssen also zwangsl\"{a}ufig direkt miteinander verkn\"{u}pft sein, was bedeutet, dass man die einen Koordinaten als Funktionen der anderen darstellen kann 

\begin{equation}
x^{\alpha '}=y^{\alpha '}(x^\beta) \quad,\quad 
x^{\alpha}=y^{\alpha}(x^{\beta '}).
\end{equation}
Als Atlas bezeichnet man eine Menge von Karten, welche den gesamten Raum M \"{u}berdecken. Eine differenzierbare $C^k$-Mannigfaltigkeit ist nun eine Menge, auf der ein solcher Atlas definiert ist, wobei f\"{u}r die Karten, die der  Atlas enth\"{a}lt, gelten muss, dass die Funktionen, welche auf der Schnittmenge zweier Karten die Koordinaten der einen Karte in die der anderen abbilden, k mal stetig differenzierbar sein m\"{u}ssen. Die Dimension der Mannigfaltigkeit ist durch die Anzahl der Koordinaten bestimmt, welche notwendig ist, um einen Punkt zu Kennzeichnen, also die Dimension d des $\mathcal{R}^d$, in welchen die Karten die Punkte der Mannigfaltigkeit abbilden.

\subsection{Tangentialr\"{a}ume und Tensorfelder}

Um den Begriff des Tangentialvektors bzw. des Tangentialraumes einzuf\"{u}hren, kann man zun\"{a}chst davon ausgehen, dass man es mit einer differenzierbaren d-dimensionalen Mannigfaltigkeit zu tun hat, auf der eine Funktion f definiert sei. In einer Umgebung eines Punktes p ist nun eine Karte definiert, welche es gestattet, den entsprechenden Punkt durch die d Koordinaten $(x^1(p),...,x^d(p))$ zu kennzeichnen. Desweiteren bezeichne k eine Kurve auf der Mannigfaltigkeit, welche durch s parametrisiert sei. In der Umgebung des Punktes p k\"{o}nnen die Koordinaten der Punkte, welche die Kurve durchl\"{a}uft, also in Abh\"{a}ngigkeit von s ausgedr\"{u}ckt werden. F\"{u}r die Werte der Funktion f entlang der Kurve gilt also

\begin{equation}
f(s)=f(x^1(s),...,x^d(s)).
\end{equation}
Es gelte $k(s_p)=p$. Man kann nun die \"{A}nderung der Funktion entlang der Kurve k im Punkt p betrachten, indem man sie dort nach dem Parameter s ableitet    

\begin{equation}
\frac{df}{ds}\mid_{s=s_p}=\frac{\partial f(x^1(s),...,x^d(s))}{\partial x^1}\frac{\partial x^1(s)}{\partial s}+...
+\frac{\partial f(x^1(s),...x^d(s))}{\partial x^d}\frac{\partial x^d(s)}{\partial s}\mid_{s=s_p}.
\end{equation}
Die Abbildung, welche der Funktion ihre Ableitung entlang einer Kurve zuordnet, bezeichnet man als Derivation. Man kann sie gewisserma\ss en durch den 
Operator 

\begin{equation}
\frac{d}{ds}=\frac{\partial}{\partial x^1}\frac{\partial x^1(s)}{\partial s}+...+\frac{\partial}{\partial x^d}\frac{\partial x^d(s)}{\partial s}
\end{equation}
darstellen. Da man solche Derivationen nun linear kombinieren kann und das Ergebnis wieder eine Derivation darstellt, bildet die Menge der Derivationen in einem Punkt p einen Vektorraum, den Tangentialvektorraum $T_p$ an p, wobei jede einzelne Derivation einem Tangentialvektor repr\"{a}sentiert, dessen Komponenten den Ableitungen der Koordinaten nach dem Parameter entsprechen $(\frac{\partial x^1}{\partial s},...,\frac{\partial x^d}{\partial s})$.
Die Menge der Linearformen $\lambda$ auf einem Vektorraum V, also Abbildungen der Form

\begin{displaymath}
\lambda: V \rightarrow \mathcal{R}\ \ \ \ \ mit\ \ \lambda(av_1+bv_2)=a\lambda(v_1)+b\lambda(v_2)\ \ v_1,v_2\ aus\ V,
\end{displaymath} 
bilden den Dualraum. Im Falle eines Tangentialvektorraumes bezeichnet man den Dualraum als Kotangentialraum. Die Menge aller Tangentialvektorr\"{a}ume bzw. Kotangentialvektorr\"{a}ume bezeichnet man als das Tangentialb\"{u}ndel bzw. Kotangentialb\"{u}ndel.
Ein Tensor W ist eine Multilinearform, also eine Abbildung der folgenden Form

\begin{displaymath}
W:\underbrace{V^* \times ... \times V^*}_{n-mal} \times \underbrace{V \times ... \times V}_{m-mal} \rightarrow \mathcal{R},
\end{displaymath}
wobei $V^*$ den Dualraum eines Vektorraums V von Verschiebungsvektoren, also im Falle einer Mannigfaltigkeit den Kotangentialvektorraum zum Vektorraum V darstellt.
Ein solcher Tensor wird als n-fach kontravariant und m-fach kovariant bezeichnet. Die Begriffe ko- und kontravariant r\"{u}hren daher, dass sich die Komponenten bei einer Basistransformation gem\"{a}\ss\ oder umgekehrt zu den Basisvektoren transformieren. Im folgenden wird ein Tensor durch die Indizes beschrieben werden, welche die einzelnen Komponenten bez\"{u}glich einer Basis kennzeichnen, wobei n Indizes oben und m Indizes unten stehen. Eine Abbildung, welche jedem Punkt einer Mannigfaltigkeit einen Tensor zuordnet, bezeichnet man als Tensorfeld.

\subsection{Affine Zusammenh\"{a}nge}

Als Parallelverschiebung von Vektoren wird eine Verschiebung bezeichnet, welche einen Vektor konstant l\"{a}sst.
In einem affinen Raum k\"{o}nnen Vektoren einfach von einem Punkt zu einem anderen verschoben werden, ohne dass eine besondere Vorschrift angegeben werden m\"{u}sste, wie zwei Vektoren an unterschiedlichen Punkten miteinander zu vergleichen sind. Daher ist dort die Verschiebung von Vektoren auch grunds\"{a}tzlich wegunabh\"{a}ngig. Dies ist jedoch anders im Falle gekr\"{u}mmter R\"{a}ume. Hier hat man es an jedem Raumpunkt mit einem lokalen Tangentialvektorraum zu tun und es muss eine Struktur definiert werden, die bestimmt, wie Vektoren an unterschiedlichen Raumpunkten miteinander verglichen werden m\"{u}ssen. Hierzu f\"{u}hrt man eine kovariante Ableitung ein. Diese ordnet einem kontravarianten Vektor einen einfach ko- und kontravarianten Tensor zu und ist durch ihre Wirkung auf einen Satz von Basisvektoren $e_\mu$ wie folgt bestimmt

\begin{equation}
\nabla_\mu e_\nu =\Gamma^\rho_{\mu\nu} e_\rho.
\end{equation}
Hierbei bezeichnet man die $\Gamma^\rho_{\mu\nu}$ als Zusammenhangkoeffizienten. Mit diesen kann man nun die Wirkung der kovarianten Ableitung auf einen kontravarianten Vektor mit den Komponenten $X^\nu$ bez\"{u}glich der Basis $e_\nu$ wie folgt ausdr\"{u}cken

\begin{equation}
\nabla_\mu X^\nu=\partial_\mu X^\nu+\Gamma^\nu_{\mu\rho} X^\rho.
\label{kovAbleitung1}
\end{equation}
F\"{u}r kovariante Vektoren mit den Komponenten $X_\nu$ ergibt sich 

\begin{equation}
\nabla_\mu X_\nu=\partial_\mu X_\nu-\Gamma^\rho_{\mu\nu} X^\nu.
\label{kovAbleitung2}
\end{equation}
Der Riemannsche Kr\"{u}mmungsstensor ist durch den Kommutator zweier kovarianter Ableitungen angewandt auf einen Verschiebungsvektor wie folgt definiert

\begin{equation}
(\nabla_\mu \nabla_\nu-\nabla_\nu \nabla_\mu)X^\sigma=R_{\mu\nu\rho}^{\ \ \ \sigma} X^\rho.
\label{Definition_Riemann-Tensor}
\end{equation}
Da die Nichtkommutativit\"{a}t der Parallelverschiebung von Vektoren die Kr\"{u}mmung eines Raumes beschreibt, ist der Riemanntensor ein Ma\ss\ f\"{u}r jene Kr\"{u}mmung. Man kann nat\"{u}rlich unmittelbar erkennen, dass er eine Antisymmetrie bez\"{u}glich $\mu$ und $\nu$ aufweist

\begin{equation}
R_{\mu\nu\rho}^{\ \ \ \ \sigma}=-R_{\nu\mu\rho}^{\ \ \ \ \sigma}.
\label{Riemann_Anti-Symmetrie}
\end{equation}
Desweiteren erf\"{u}llt er die zyklische Identit\"{a}t 

\begin{equation}
R_{[\mu\nu\rho]}^{\ \ \ \ \sigma}=0,
\label{Riemann_zyklische_Identitaet}
\end{equation}
und die sogenannte Bianchiidentit\"{a}t

\begin{equation}
\nabla_{[\epsilon} R_{\mu\nu]\rho}^{\ \ \ \ \sigma}=0.
\label{Bianchi-Identitaet}
\end{equation}
Diese spielt in Bezug auf die Einsteinschen Feldgleichungen eine entscheidende Rolle.
Wenn man ($\ref{kovAbleitung1}$) und ($\ref{kovAbleitung2}$) in ($\ref{Definition_Riemann-Tensor}$) verwendet, kann man den Riemanntensor in Abh\"{a}ngigkeit der Zusammenhangkoeffizienten ausdr\"{u}cken

\begin{equation}
R_{\mu\nu\rho}^{\ \ \ \ \sigma}=\partial_\mu \Gamma_{\nu\rho}^\sigma-\partial_\nu \Gamma^\sigma_{\mu\rho} 
+\Gamma^\sigma_{\mu\epsilon} \Gamma^\epsilon_{\nu\rho} -\Gamma^\sigma_{\nu\epsilon} \Gamma^\epsilon_{\mu\rho}.
\label{RiemannZusammenhangkoeffizienten}
\end{equation}
Wichtig ist schlie\ss lich noch die Definition der Torsion. Die Torsion entspricht der Nichtkommutativit\"{a}t der kovarianten Ableitungen bez\"{u}glich ihrer Anwendung auf skalarwertige Funktionen f. In der Allgemeinen Relativit\"{a}tstheorie werden ausschlie\ss lich torsionsfreie Zusammenh\"{a}nge betrachtet, f\"{u}r die demnach gilt

\begin{equation}
\nabla_\mu \nabla_\nu f=\nabla_\nu \nabla_\mu f.
\label{Torsionsfreiheit}
\end{equation}
Die Zusammenhangkoeffizienten $\Gamma^\rho_{\mu\nu}$ eines torsionsfreien Zusammenhangs sind symmetrisch bez\"{u}glich der unteren beiden Indizes.
Durch Verwendung von ($\ref{kovAbleitung1}$) und ($\ref{kovAbleitung2}$) in ($\ref{Torsionsfreiheit}$) ergibt sich n\"{a}mlich

\begin{equation}
\partial_\mu \partial_\nu f+\Gamma^\rho_{\mu\nu} \partial_\rho f=\partial_\nu \partial_\mu f+\Gamma^\rho_{\nu\mu}\partial_\rho f.
\end{equation}
Die gew\"{o}hnlichen Ableitungen sind vertauschbar, womit man direkt 

\begin{equation}
\Gamma^\rho_{\mu\nu}=\Gamma^\rho_{\nu\mu}
\label{Koeffizientensymmetrie}
\end{equation}
ablesen kann.

\subsection{Riemannsche Mannigfaltigkeiten}

Bisher wurden nur Mannigfaltigkeiten betrachtet, die mit einem affinen Zusammenhang ausgestattet waren. Auf solchen Mannigfaltigkeiten ist allerdings noch nicht zwangsl\"{a}ufig eine metrische Struktur definiert. Um Begriffen wie dem Abstand zweier Punkte oder der Orthogonalit\"{a}t zweier Tangentialvektoren einen Sinn zu verleihen, bedarf es der Einf\"{u}hrung einer Metrik. Eine Metrik g in einem Vektorraum ist ein symmetrischer zweifach kovarianter Tensor. Er ordnet also einem Paar zweier kontravarianter Vektoren eine reelle Zahl zu. Zwei Vektoren mit Komponenten $X^\mu$ und $X^\nu$ werden orthogonal genannt, wenn gilt 

\begin{equation}
g_{\mu\nu}X^\mu X^\nu=0.
\end{equation}
Ein metrisches Feld ist damit ein symmetrisches zweifach kovariantes Tensorfeld, das in jedem Punkt der Mannigfaltigkeit den entsprechenden Tangentialvektorraum mit einem inneren Produkt ausstattet. Mit der Auszeichnung eines solchen metrischen Feldes ist also die geometrische Struktur einer Mannigfaltigkeit vollkommen bestimmt. Die Metrik ordnet damit jedem Vektor $X^\mu$ in direkter Weise einen dualen Vektor $X_\nu$ zu

\begin{equation}
g_{\mu\nu}X^\mu X^\nu=X_\nu X^\nu. 
\end{equation}
Man bezeichnet eine solche Mannigfaltigkeit auch als Riemannsche Mannigfaltigkeit. 

Es ist nun m\"{o}glich, den affinen Zusammenhang, welcher verschiedene Tangentialvektorr\"{a}ume miteinander verbindet, 
mit Hilfe der Metrik zu definieren. Im Rahmen der Allgemeinen Relativit\"{a}tstheorie ist insbesondere der torsionsfreie Zusammenhang 
von Bedeutung, welcher durch die Bedingung definiert ist, dass innere Produkte zwischen Vektoren bei Verschiebung konstant gehalten werden sollen, was einem Verschwinden der kovarianten Ableitung bei Anwendung auf den metrischen Tensor entspricht. Dieser Zusammenhang ist eindeutig bestimmt und hei\ss t
Levy-Civita-Zusammenhang. Wenn man nun die Zusammenhangkoeffizienten des Levy-Civita-Zusammenhangs bestimmen m\"{o}chte, muss die Bedingung

\begin{equation}
\nabla_\mu g_{\rho\sigma}=0 
\end{equation}
entsprechend umformuliert werden. Durch Anwenden der kovarianten Ableitung auf jeden Index ergibt sich

\begin{equation}
\partial_\mu g_{\rho\sigma}-\Gamma_{\mu\rho}^{\nu} g_{\nu\sigma}-\Gamma_{\mu\sigma}^{\nu} g_{\rho\nu}=0.
\end{equation}
Zyklisches Vertauschen der Indizes f\"{u}hrt auf zwei weitere Gleichungen

\begin{eqnarray}
\partial_\mu g_{\rho\sigma}-\Gamma_{\mu\rho}^{\nu} g_{\nu\sigma}-\Gamma_{\mu\sigma}^{\nu} g_{\rho\nu}&=&0\nonumber\\
\partial_\rho g_{\sigma\mu}-\Gamma_{\rho\sigma}^{\nu} g_{\nu\mu}-\Gamma_{\rho\mu}^{\nu} g_{\sigma\nu}&=&0\nonumber\\
\partial_\sigma g_{\mu\rho}-\Gamma_{\sigma\mu}^{\nu} g_{\nu\rho}-\Gamma_{\sigma\rho}^{\nu} g_{\mu\nu}&=&0.
\end{eqnarray}
Wenn man nun die zweite Gleichung zur ersten addiert und die zweite subtrahiert, so ergibt sich unter Verwendung der Symmetrieeigenschaft des metrischen Tensors sowie ($\ref{Koeffizientensymmetrie}$) folgende Gleichung

\begin{equation}
-2\Gamma_{\mu\rho}^{\nu} g_{\nu\sigma}+\partial_\mu g_{\rho\sigma}+\partial_\rho g_{\sigma\mu}-\partial_\sigma g_{\mu\rho}=0.
\end{equation}
Durch umstellen erh\"{a}lt man

\begin{equation}
\Gamma_{\mu\rho}^{\nu}=\frac{1}{2} g^{\nu\sigma}(\partial_\mu g_{\rho\sigma}+\partial_\rho g_{\sigma\mu}-\partial_\sigma g_{\mu\rho}).
\label{Christoffelsymbole}
\end{equation}
In diesem speziellen Falle bezeichnet man die Zusammenhangkoeffizienten als Christoffelsymbole.
Wenn man den Kommutator der kovarianten Ableitung des Levy-Civita-Zusammenhangs auf die Metrik anwendet, so ergibt sich

\begin{equation}
(\nabla_\mu \nabla_\nu-\nabla_\nu \nabla_\mu)g^{\rho\sigma }=
R_{\mu\nu\lambda}^{\ \ \ \ \rho} g^{\lambda\sigma}+R_{\mu\nu\lambda}^{\ \ \ \ \sigma} g^{\rho\lambda}=R_{\mu\nu}^{\ \ \ \sigma\rho}+R_{\mu\nu}^{\ \ \  \rho\sigma}=0.
\end{equation}
Das bedeutet, dass der Riemanntensor in diesem Fall antisymmetrisch bez\"{u}glich der hinteren beiden Indizes ist

\begin{equation}
R_{\mu\nu}^{\ \ \ \sigma\rho}=-R_{\mu\nu}^{\ \ \ \rho\sigma}.
\label{Riemann_LCAntiSymmetrie}
\end{equation}

\section{Physikalische Prinzipien}

Das gesamte theoretische Geb\"{a}ude der Allgemeinen Rel\"{a}tivit\"{a}tstheorie geht aus dem \"{A}quivalenzprinzip hervor, dass ausgehend von der bereits in der Einleitung erw\"{a}hnten Gleichheit von schwerer und tr\"{a}ger Masse, die Unm\"{o}glichkeit der Unterscheidung eines beschleunigten Bezugssystems von der Wirkung eines Gravitationsfeldes beinhaltet. Dies bedeutet, dass man die Wirkung eines Gravitationsfeldes als Eigenschaft der Raumzeit selbst deuten kann, die eine in sich gekr\"{u}mmte Struktur aufweist, welche sich im Rahmen der Riemannschen Geometrie dann in einer von der euklidischen abweichenden Metrik \"{a}u\ss ert. Der oben entwickelte Formalismus wird also gewisserma\ss en auf die reale Raumzeit \"{u}bertragen.

\subsection{Die Postulate der Allgemeinen Relativit\"{a}tstheorie}

Die Allgemeine Relativit\"{a}tstheorie basiert auf zwei Grundpostulaten \cite{Einstein:1916vd}.\\

1) Die Raumzeit wird durch eine vierdimensionale Riemannsche Mannigfaltigkeit beschrieben, auf der demnach eine Metrik ausgezeichnet ist. Es handelt sich um eine Lorentzmetrik, die grunds\"{a}tzlich durch eine geeignete Koordinatentransformation lokal in die Gestalt $\eta_{\mu\nu}=diag(1,-1,-1,-1)$ gebracht werden kann.
Die genaue Geometrie der Raumzeit wird durch die Materieverteilung bestimmt, wobei dieser Zusammenhang in mathematisch exakter Weise durch die Einsteinschen Feldgleichungen ausgedr\"{u}ckt wird

\begin{equation}
G_{\mu\nu}=-8\pi G T_{\mu\nu}.
\end{equation}

2) Die Bewegung von kr\"{a}ftefreien K\"{o}rpern erfolgt auf der k\"{u}rzesten Linie durch die Raumzeit. Eine solche wird im verallgemeinerten Fall gekr\"{u}mmter R\"{a}ume als Geod\"{a}te bezeichnet. Weltlinien von K\"{o}rpern gehorchen also der Geod\"{a}tengleichung

\begin{equation}
\ddot X^\mu+\Gamma^\mu_{\rho\sigma}\dot X^\rho \dot X^\sigma=0,
\end{equation}
wobei die $\Gamma^\mu_{\rho\sigma}$ die Christoffelsymbole bezeichnen.

\subsection{Die Einsteinschen Feldgleichungen}

Bei der Suche nach den Feldgleichungen, welche die Raumzeitstruktur bestimmen, wurde Einstein durch das Analogon aus der Elektrodynamik geleitet. Hier koppelt das Elektromagnetische Feld an den elektrischen Viererstrom $J^\nu$, welcher die Ladungs- und Ladungsstromdichte enth\"{a}lt. Dieser Zusammenhang wird durch die inhomogenen Maxwellschen Gleichungen ausgedr\"{u}ckt, die in kovarianter Formulierung in folgender Gleichung enthalten sind

\begin{equation}
\partial_\mu F^{\mu\nu}=-J^\nu.
\end{equation}
Das Gravitationsfeld, also die Struktur der Raumzeit, kann aufgrund des klassischen Grenzfalles, der sich im Newtonschon Gravitationsgesetz ausdr\"{u}ckt, nur durch die Materieverteilung definiert sein, genauer gesagt die Energieverteilung. Diese aber ist im Rahmen der Relativit\"{a}tstheorie mit dem Impuls zum Viererimpuls verbunden. Es ist nun der Energie-Impuls-Tensor, bezeichnet als $T_{\mu\nu}$, welcher analog der Viererstromdichte, welche die Ladungsverteilung bestimmt, die Energie- und Impulsverteilung der Materie beschreibt.
Da die Energie-Impuls-Verteilung die Struktur der Raumzeit festlegen soll, muss es einen direkten Zusammenhang zu den Kr\"{u}mmungsgr\"{o}\ss en der Raumzeit geben. Hierbei ist nun folgendes wichtig.
Aufgrund des Energie- und Impulserhaltungssatzes muss der Energie-Impuls-Tensor divergenzfrei sein. Dies bedeutet nichts anderes als

\begin{equation}
\nabla^\mu T_{\mu\nu}=0.
\end{equation}
Daneben gibt es nur einen divergenzfreien Tensor zweiter Stufe, welcher sich aus den Kr\"{u}mmungsgr\"{o}\ss en konstruieren l\"{a}sst. Dieser kann unter Verwendung der Bianchiidentit\"{a}t ($\ref{Bianchi-Identitaet}$) gefunden werden. Aus dieser ergibt sich mit ($\ref{Riemann_Anti-Symmetrie}$)

\begin{equation}
\nabla_\epsilon R_{\mu\nu\rho}^{\ \ \ \sigma}+\nabla_\mu R_{\nu\epsilon\rho}^{\ \ \ \sigma}+\nabla_\nu R_{\epsilon\mu\rho}^{\ \ \ \sigma}=0.
\end{equation}
Durch erneute Anwendung von ($\ref{Riemann_Anti-Symmetrie}$) im zweiten Ausdruck und Kontraktion der Indizes $\nu$ und $\sigma$ erh\"{a}lt man

\begin{equation}
\nabla_\epsilon R_{\mu\rho}-\nabla_\mu R_{\epsilon\rho}+\nabla_\nu R_{\epsilon\mu\rho}^{\ \ \ \nu}=0.
\end{equation}
Wenn man nun im letzten Term ein weiteres Mal ($\ref{Riemann_Anti-Symmetrie}$) sowie ($\ref{Riemann_LCAntiSymmetrie}$) benutzt und anschlie\ss end $\rho$ mit $\epsilon$ kontrahiert, so ergibt sich 

\begin{equation}
\nabla^\rho R_{\mu\rho}-\nabla_\mu R+\nabla^\nu R_{\mu\nu}. 
\end{equation}
Weiteres Umformen f\"{u}hrt schlie\ss lich auf

\begin{equation}
\nabla^\nu(R_{\mu\nu}-\frac{1}{2}Rg_{\mu\nu})=0.
\label{Divergenzfreiheit}
\end{equation}
Der Tensor

\begin{equation}
G_{\mu\nu}=R_{\mu\nu}-\frac{1}{2}Rg_{\mu\nu}
\end{equation}
ist also divergenzfrei und wird als Einsteintensor bezeichnet. 
Der darin auftauchende Tensor $R_{\mu\nu}=R_{\mu\lambda\nu}^{\ \ \ \lambda}$ hei\ss t Riccitensor und die Gr\"{o}\ss e $R=g^{\mu\nu}R_{\mu\nu}$ wird als Ricciskalar bezeichnet.
Aufgrund der Eigenschaft ($\ref{Divergenzfreiheit}$) vermutete Einstein nun den Zusammenhang

\begin{equation} 
G_{\mu\nu}=\kappa T_{\mu\nu}.
\end{equation}
Dies ist die Einsteinsche Feldgleichung, wobei der Proportionalit\"{a}tsfaktor $\kappa$ durch Analogieschl\"{u}sse zum klassischen
Grenzfall, den die Einsteinschen Feldgleichungen nat\"{u}rlich imlizit enthalten m\"{u}ssen, bestimmt werden kann. Der Proportionalit\"{a}tsfaktor hat jedoch keine prinzipielle Bedeutung und deshalb soll er hier einfach nur angegeben werden, womit man schlie\ss lich

\begin{equation}
G_{\mu\nu}=-8\pi G T_{\mu\nu}
\end{equation}
erh\"{a}lt.

\subsection{Bewegung von K\"{o}rpern auf der Raumzeit}
 
Die k\"{u}rzeste Linie durch die Raumzeit zwischen zwei Punkten ist dadurch definiert, dass sich ein Weltlinientangentialvektor entlang der gesamten Kurve nicht \"{a}ndert, dass er also Parallelverschoben wird. Dies bedeutet, dass die kovariante Ableitung in Richtung der Kurve also des Tangentialvektors selbst angewandt auf diesen Vektor an jeder Stelle der Kurve gleich null ist. Ein Tangentialvektor ist durch die Ableitung der Koordinaten $X^\mu$, die zu einer Karte an einem bestimmten Punkt geh\"{o}ren, nach dem Kurvenparameter s charakterisiert, welche hier wie folgt notiert werden soll $\dot X^\mu=\frac{\partial X^\mu}{\partial s}$. Da die Ableitung in Richtung des Tangentialvektors von Interesse ist, werden die Ableitungen in Richtung der die Karte an einer bestimmten Stelle charakterisierenden Koordinaten mit den Komponenten des Tangentialvektors in eben diesen Koordinaten gewichtet. Dies bedeutet, dass sich die Forderung des Verschwindens der kovarianten Ableitung des Tangentialvektors entlang der Kurve

\begin{equation}
\nabla_s \dot X^\nu=0
\end{equation}
wie folgt ausdr\"{u}cken l\"{a}sst

\begin{equation}
\dot X^\nu \nabla_\nu \dot X^\mu=\dot X^\nu \partial_\nu \dot X^\mu+\dot X^\nu \Gamma^\mu_{\nu\rho} \dot X^\rho=\ddot X^\mu+\Gamma^\mu_{\nu\rho}\dot X^\nu \dot X^\rho=0.
\end{equation}
Dies ist aber die Geod\"{a}tengleichung, welche festlegt, wie sich der Tangentialvektor der Weltlinie eines K\"{o}rpers und damit die Bewegung des K\"{o}rpers bei vorgegebener Raumzeitstruktur verh\"{a}lt. Es handelt sich hierbei um eine Verallgemeinerung des Galileischen Tr\"{a}gheitsprinzips f\"{u}r R\"{a}ume mit beliebiger Geometrie. 

\chapter{Die Einf\"{u}hrung zus\"{a}tzlicher Dimensionen}

Die Allgemeine Relativit\"{a}tstheorie in der beschriebenen Form hatte zwar die Gravitation mit der speziellen Relativit\"{a}tstheorie in einer einheitlichen Theorie formuliert. Dennoch stellte auch sie, selbst bei Beschr\"{a}nkung auf den Rahmen der klassischen Physik, noch keine v\"{o}llig einheitliche Naturbeschreibung dar, denn die Elektromagnetischen Felder blieben zur geometrischen Beschreibungsweise des Gravitationsfeldes wesensfremd.
Dar\"{u}ber hinaus enth\"{a}lt die Einsteinsche Feldgleichung auf der rechten Seite den Energie-Impuls-Tensor, dessen spezielle Gestalt aus einer anderen
Theorie \"{u}bernommen werden muss, w\"{a}hrend die das Gravitationsfeld beschreibenden Gr\"{o}\ss en auf der linken Seite rein geometrischer Natur sind. Einstein strebte nun eine Beschreibungsweise an, bei der auch die Materie letztlich auf Geometrie zur\"{u}ckgef\"{u}hrt wird.

\section{Kaluza-Klein-Theorie}

Eine Erweiterung der Allgemeinen Relativit\"{a}tstheorie in dieser Richtung entwickelte der Mathematiker Theodor Kaluza \cite{Kaluza:1921tu}. Seine Theorie beinhaltete unter Einbeziehung einer f\"{u}nften Dimension auch den Elektromagnetismus, der damit also auch einer geometrischen Beschreibung zug\"{a}nglich gemacht wurde. Die Tatsache, dass die f\"{u}nfte r\"{a}umliche Dimension nicht direkt in Erscheinung tritt, kann durch den Ansatz Oskar Kleins auf elegante Art und Weise erkl\"{a}rt werden, welcher die Kaluzasche Theorie dahingehend modifizierte, dass er die zus\"{a}tzliche Dimension kompaktifizierte \cite{Klein:1926tv}. 

\subsection{Die Struktur der Raumzeit nach Kaluza und Klein}

Es wird also davon ausgegangen, dass der Raum neben den vier Dimensionen der gew\"{o}hnlichen Raumzeit, die mit griechischen Indizes bezeichnet seien,
welche von 0 bis 3 laufen, noch eine weitere Dimension $x^4$ enth\"{a}lt, die in der Weise kompaktifiziert ist, dass sie periodisch unter folgender Transformation ist

\begin{equation}
x^4 \rightarrow x^4+2\pi R,  
\end{equation}
was also der Kompaktifizierung zu einem Kreis entspricht. Die Metrik der vollst\"{a}ndigen Raumzeit $g_{MN}$ unter Einbeziehung der kompaktifizierten Dimension besteht also aus den Komponenten, die sich auf die gew\"{o}hnliche Raumzeit beziehen, den Komponenten, deren einer Index sich auf die f\"{u}nfte Raumzeitdimension bezieht und der Komponente, deren beide Indizes sich auf die kompaktifizierte Komponente beziehen. Es wird davon ausgegangen, dass die Metrik nur von den nichtkompaktifizierten Koordinaten abh\"{a}ngt. Der folgende Ansatz f\"{u}r den metrischen Tensor 

\begin{equation}
g_{MN}=\left(\begin{array}{cc}g_{\mu\nu}+g_{44}A_\mu A_\nu & 2g_{44}A_\mu\\ 2g_{44}A_\mu & g_{44}\end{array}\right)
\end{equation}
und damit f\"{u}r ein infinitesimales Linienelement

\begin{equation}
ds^2=g_{MN} dx^M dx^N=g_{\mu\nu}dx^\mu dx^\nu+g_{44}(dx^4+A_\mu dx^\mu)^2 
\end{equation}
ist invariant unter Transformationen der Form

\begin{equation}
x^d \rightarrow x^d+\lambda(x^\mu)\quad,\quad A_\mu \rightarrow A_\mu - \partial_\mu \lambda. 
\end{equation}
Dies entspricht also der Form nach einer Eichtransformation des Elektromagnetischen Feldes. 
Damit l\"{a}sst sich der Vierervektor $A_\mu$ mit dem Elektromagnetischen Potential identifizieren. Tats\"{a}chlich ergeben sich aus diesem Ansatz die Maxwellschen Gleichungen. Die Herleitung, aus der dies hervorgeht, sowie eine Darstellung des Kaluza-Klein-Formalismus im Allgemeinen findet sich in \cite{Polchinsky}.

\subsection{Generierung von Massen}

Es soll nun ein skalares Feld auf einer solchen f\"{u}nfdimensionalen Raumzeit aufgespalten werden in einen Anteil, welcher die Abh\"{a}ngigkeit von der vierdimensionalen Untermannigfaltigkeit der normalen Raumzeit ausdr\"{u}ckt, und eine Fourierentwicklung nach der kompaktifizierten Koordinate, 
deren Anregungen nat\"{u}rlich aufgrund der Periodizit\"{a}t der Koordinate gequantelt sind

\begin{equation}
\Phi_n(x^M)=\sum_{n=-\infty}^\infty \Phi (x^\mu) exp\left(\frac{inx^4}{R}\right).
\label{SkalarfeldKK}
\end{equation}
Bei einem masselosen Skalarfeld, dass der Gleichung $\partial_M \partial^M \phi_n(x^M)=0$ gen\"{u}gt, ergibt sich durch Aufspaltung des Operators  $\partial_M \partial^M$ in den d'Alembertoperator auf der vierdimensionalen Raumzeit und die zweite Ableitung nach der kompaktifizierten Koordinate
 
\begin{equation}
\partial_M \partial^M=\partial_\mu \partial^\mu+\partial_4 \partial^4
\end{equation}
und anschlie\ss ender Anwendung von $\partial_4 \partial^4$ die folgende Gleichung

\begin{equation}
\partial_\mu \partial^\mu \Phi_n (x^\mu)=\frac{n^2}{R^2} \Phi_n (x^\mu), 
\end{equation}
also der Form nach eine Klein-Gordon-Gleichung mit einem Massenterm $\frac{n^2}{R^2}$. Das Feld $\Phi_n$ erh\"{a}lt also eine Masse, die
dem Quadrat der Schwingungszahl der Anregung in der kompaktifizierten Dimension proportional ist. Auf die gleiche Weise erhalten im n\"{a}chsten Kapitel
Gravitonen eine Masse.

\section{Das ADD-Modell}

\subsection{Die Grundidee}

In einer anderen Weise werden im Rahmen des in j\"{u}ngerer Zeit entwickelten Randall-Sundrum-Modells \cite{Randall:1999ee},\cite{Randall:1999vf} und dem dazu verwandten ADD-Modell \cite{Antoniadis:1997zg},\cite{Arkani-Hamed:1998rs},\cite{Antoniadis:1998ig} zus\"{a}tzliche Dimensionen eingef\"{u}hrt. Hier geht man davon aus, dass die \"{u}blichen Materiefelder auf einer 3+1-dimensionalen Untermannigfaltigkeit eines h\"{o}herdimensionalen Raumes leben, der dem \"{u}blichen Raum entspricht und im Falle des Randall-Sundrum-Modells eine zus\"{a}tzliche und im Rahmen des ADD-Modells eine zun\"{a}chst nicht festgelegte Anzahl an zus\"{a}tzlichen Dimensionen enth\"{a}lt, die wie in der Theorie von Kaluza und Klein auf eine bestimmte Art und Weise kompaktifiziert sind. Im Gegensatz zu allen anderen Feldern ist das Gravitationsfeld jedoch nicht auf die Untermannigfaltigkeit des \"{u}blichen Raumes beschr\"{a}nkt. Dies k\"{o}nnte eine m\"{o}gliche L\"{o}sung des Hierarchieproblems liefern, welches in der Nichterkl\"{a}rbarkeit der unglaublichen Schw\"{a}che der Gravitation im Vergleich zu allen anderen Wechselwirkungen besteht. Aus dem Newtonschen Gravitationsgesetz ergibt sich eine neue Gr\"{o}\ss e, eine D-dimensionale Planckmasse $M_D$, die in folgender Relation zur gew\"{o}hnlichen Planckmasse $M_P$ steht

\begin{equation}
M_P^2=c M_D^{\delta+2} R^\delta,\quad c=const,
\end{equation} 
wobei $\delta$ die Zahl der zus\"{a}tzlichen Dimensionen und R den Kompaktifizierungsradius beschreibt.
Bei der Voraussetzung, dass die Gravitation im Falle der Existenz zus\"{a}tzlicher Dimensionen nicht auf die \"{u}bliche 3+1-dimensionale Raumzeit beschr\"{a}nkt ist, handelt es sich um eine vollkommen nat\"{u}rliche Annahme. Wenn man die Gravitation n\"{a}mlich, wie das in der Allgemeinen Relativit\"{a}tstheorie getan wird, als geometrische Eigenschaft der Raumzeit selbst deutet, so bedeutet die Annahme, dass die Gravitation auch auf die zus\"{a}tzlichen Dimensionen ausgedehnt ist, nichts anderes, als dass deren Geometrie ebenfalls einer gewissen Dynamik unterworfen ist.

Es wird also von einer D-dimensionalen Raumzeit ausgegangen. Ein beliebiger Punkt einer solchen Raumzeit wird durch ein D-Tupel an Koordinaten $z=(z_1,...,z_D)$ beschrieben. Alle zu den 3+1 Dimensionen der gew\"{o}hnlichen Raumzeit zus\"{a}tzlichen Dimensionen sollen nun torusf\"{o}rmig kompaktifiziert werden. Das D-Tupel an Koordinaten wird also aufgespalten in die Koordinaten, welche die Untermannigfaltigkeit der gew\"{o}hnlichen Raumzeit beschreibt, und jene der zus\"{a}tzlichen Dimensionen

\begin{equation}
z=(x_0,\vec x,y_1,...,y_\delta),\quad\delta=D-4, 
\end{equation}
wobei f\"{u}r die Koordinaten der zus\"{a}tzlichen Dimensionen die gleiche Periodizit\"{a}t wie im Falle der Theorie von Kaluza und Klein gilt

\begin{equation}
y_j\rightarrow y_j+2\pi R\quad,\quad j=1,...,\delta.
\end{equation}
Dies entspricht der Kompaktifizierung zu einem $\delta$-dimensionalen Torus, dessen Volumen durch $V_\delta=(2\pi R)^\delta$ gegeben ist. Damit gilt f\"{u}r die Relation zwischen $M_P$ und $M_D$

\begin{equation}
M_P^2=8\pi R^\delta M_D^{2+\delta}.
\end{equation}

\subsection{Metrische Struktur und Materieverteilung}

Die metrische Struktur des Raumes kann n\"{a}herungsweise durch die im letzten Kapitel beschriebenen Einsteinschen Feldgleichungen
erweitert auf D Dimensionen beschrieben werden

\begin{equation}
G_{MN}=-\frac{T_{MN}}{\bar M_D^{2+\delta}},
\end{equation}
wobei $\bar M_D=(2\pi)^{-\frac{\delta}{2+\delta}} M_D$. Damit ist der metrische Tensor $g_{MN}$ mit $M,N=0,...,D$ bestimmt.
Interessant ist jedoch nun der Zusammenhang der Metrik des h\"{o}herdimensionalen Raumes zur Metrik der 3+1-dimensionalen Untermannigfaltigkeit. F\"{u}r ein infinitesimales Linienelement gilt

\begin{eqnarray}
ds^2&=&G_{MN}(Y(x)) dY^M dY^N\nonumber\\
&=&G_{MN}(Y(x))\frac{\partial Y^M}{\partial x^\mu}dx^\mu \frac{\partial Y^N}{\partial x^\nu}dx^\nu.
\end{eqnarray}
Wenn man aber nun ber\"{u}cksichtigt, dass das Linienelement gleichzeitig durch $ds^2=g_{\mu\nu}dx^\mu dx^\nu$ beschrieben werden kann, bedeutet das f\"{u}r die Metrik $g_{\mu\nu}(x)$ der in den h\"{o}herdimensionalen Raum eingebetteten gew\"{o}hnlichen Raumzeit

\begin{equation}
g_{\mu\nu}=G_{MN}(Y(x))\partial_\mu Y^M \partial_\nu Y^N.
\end{equation}
Wie ist aber nun der die Materieverteilung bestimmende Energie-Impuls-Tensor zu beschreiben ? Wenn davon ausgegangen wird, dass die Materie- und Wechselwirkungsfelder auf der 3+1-dimensionalen Untermannigfaltigkeit leben und man das Gravitationsfeld und damit seine Selbstenergie als so schwach
annimmt, dass sie im Vergleich zu der Energie der \"{u}brigen Felder vernachl\"{a}ssigt werden kann, so muss auch der Energie-Impuls-Tensor auf die gew\"{o}hnliche Raumzeit beschr\"{a}nkt sein. Dies dr\"{u}ckt sich mathematisch in einer Deltafunktion bez\"{u}glich der zus\"{a}tzlichen Dimensionen aus. Der Energie-Impuls-Tensor der h\"{o}herdimensionalen Raumzeit besitzt also folgende Form

\begin{equation}
T_{MN}(z)=\eta^\mu_M \eta^\nu_N T_{\mu\nu} (x) \delta(y).
\end{equation}
Die Formulierung von Feldtheorien auf einer Raumzeit der beschriebenen Struktur wird in \cite{Sundrum:1998sj} gegeben.

\chapter{Eine Quantenfeldtheorie der Gravitation}

In diesem Kapitel soll es nun um die Formulierung der Gravitation als Quantenfeldtheorie unter Ber\"{u}cksichtigung der zus\"{a}tzlichen kompaktifizierten Dimensionen gehen. 
Grunds\"{a}tzlich besteht nat\"{u}rlich wie im klassischen Falle auch hier die Formulierung aus zwei Teilbereichen. Einerseits muss die Art und Weise festgelegt werden, wie die quantentheoretisch beschriebenen Materiefelder an das Gravitationsfeld koppeln und andererseits muss das Gravitationsfeld selbst quantentheoretisch beschrieben werden. Bez\"{u}glich des ersten Teilproblems gibt es eine unumstrittene Beschreibungsmethode, welche wie im Falle der anderen Wechselwirkungen auf ein Eichprinzip zur\"{u}ckgef\"{u}hrt werden kann. Im Gegensatz zu den anderen Wechselwirkungen, bei denen Invarianz unter lokalen Transformationen bez\"{u}glich innerer Symmetriegruppen gefordert wird, handelt es sich hierbei um die Forderung lokaler Invarianz unter Lorentztransformationen. Im Hinblick auf eine Vereinheitlichung mit den anderen Wechselwirkungen ist es sehr bemerkenswert, dass eine solche eichtheoretische Formulierung auch im Falle der Gravitation m\"{o}glich ist. Dies ist ein wichtiges Bindeglied zu den \"{u}brigen Wechselwirkungen. 

Die quantentheoretische Beschreibungsweise des Gravitationsfeldes selbst ist jedoch das eigentlich entscheidende Problem. Die hier verwendete Theorie geht von einer Entwicklung des metrischen Feldes um die Minkowskimetrik in linearer N\"{a}herung aus, welche dann in die den Einsteinschen Feldgleichungen korrespondierende Lagrangedichte eingesetzt wird. Ebenso wie bei der Quantisierung von \"{u}blichen Eichtheorien wird die Methode der Pfadintegralquantisierung verwendet. Diese von Feynman eingef\"{u}hrte quantenfeldtheoretische Beschreibungsweise f\"{u}hrt direkt von der Lagrangedichte auf die Propagatoren.
Es muss nat\"{u}rlich nicht erw\"{a}hnt werden, dass es sich hierbei nur um eine effektive Theorie handelt, die in keiner Weise den Anspruch hat, die Gravitation auf fundamentale Weise quantentheoretisch zu beschreiben. Dies ist wie bereits in der Einleitung erw\"{a}hnt schon aufgrund der fehlenden Hintergrundunabh\"{a}ngigkeit ausgeschlossen, welche durch die Entwicklung um die flache Minkowskimetrik von vorneherein nicht gegeben ist.   

\section{Eichtheorie der Gravitation}

Gem\"{a}\ss\ den anderen fundamentalen Wechselwirkungen kann wie bereits erw\"{a}hnt auch die Gravitation als Eichtheorie beschrieben werden. Man geht also ebenfalls zun\"{a}chst von einer freien Materiefeldgleichung aus, deren Kopplung an das Gravitationsfeld dann durch eine Symmetrieforderung bestimmt wird. Im Gegensatz zu der bei anderen Wechselwirkungen geforderten Invarianz unter lokalen Transformationen bez\"{u}glich innerer Symmetrien fordert man im Falle der Gravitation Invarianz unter lokalen Lorentztransformationen.

\subsection{Lokale Lorentzinvarianz und Vierbein}

Um das Verhalten von Fermionen in einem Gravitationsfeld eichtheoretisch zu beschreiben, muss zun\"{a}chst ein neuer mathematischer Begriff eingef\"{u}hrt werden. Es handelt sich um das Vierbein bzw. im Falle zus\"{a}tzlicher Dimensionen um das Vielbein.
Hierzu muss man sich das (schwache) \"{A}quivalenzprinzip in Erinnerung rufen, demgem\"{a}\ss\ ein beliebiges Gravitationsfeld durch Wahl geeigneter Koordinaten zum Verschwinden gebracht werden kann.
In dem ensprechenden Koordinatensystem gilt also

\begin{equation} 
g_{mn}=\eta_{mn}.
\end{equation}
Das Vierbein transformiert nun beliebige globale Koordinaten $x^\mu$ in diejenigen lokalen Koordinaten $y^m$, in welchen die Kr\"{u}mmung verschwindet. Das bedeutet

\begin{equation}
x^m=e^m_\mu y^\mu \quad,\quad \eta_{mn}=e^\mu_m e^\nu_n \eta_{\mu\nu}.
\end{equation}
Damit stellt das Vierbeinfeld eine vollst\"{a}ndige zum metrischen Feld \"{a}quivalente Beschreibungsweise des Gravitationsfeldes dar.
Man geht also von der freien Lagrangedichte des Diracfeldes aus 

\begin{equation} 
\mathcal{L}=\bar \Psi(i\gamma^\mu\partial_\mu-m)\Psi.
\end{equation}
Nun wird Invarianz unter lokalen Lorentztransformationen 

\begin{equation}
\psi\ \rightarrow\ U(x)\Psi\quad,\quad\partial_\mu\ \rightarrow\ \Lambda_\mu^\nu(x)\partial_\nu,
\end{equation}
gefordert. Da es sich bei $\Psi$ um ein Spinorfeld handelt, wird die Transformation durch die Generatoren der Lorentzgruppe in Diracspinordarstellung $\Sigma_{mn}=\frac{i}{4}[\gamma_m,\gamma_n]$ vermittelt.
Durch die Einf\"{u}hrung einer kovarianten Ableitung der folgenden Form

\begin{equation}
D_m=e_m^\mu(\partial_\mu+i\omega_\mu^{mn}\Sigma_{mn}),
\end{equation}
wobei die Zusammenhangkoeffizienten mit dem Vierbein in folgender Beziehung stehen 

\begin{equation}
\omega_\mu^{mn}=2e^{\nu [m}\partial_{[\mu}e_{\nu]}^{n]}+e_{\mu p}e^{\nu m}e^{\sigma n}\partial_{[\sigma}e_{\nu]}^p,
\end{equation}
kann man nun eine unter lokalen Lorentztransformationen invariante Lagrangedichte erhalten

\begin{equation}
\mathcal{L}=\bar \Psi(i\gamma^\mu D_\mu-m)\Psi.
\end{equation}
Eine ausf\"{u}hrlichere Beschreibung der Idee von Eichtheorien im Allgemeinen ist in Kapitel 5 zu finden.

\subsection{Wirkungen von Feldern im Gravitationsfeld}

Wenn man nun die zur Lagrangedichte geh\"{o}rige Wirkung formulieren m\"{o}chte, so muss man beachten, dass f\"{u}r ein infinitesimales Volumenelement $dV$ in einem Raum mit der Metrik $g_{\mu\nu}$ gilt

\begin{equation}
dV=\epsilon_{\mu\nu\rho\sigma} dx^\mu dx^\nu dx^\rho dx^\sigma,
\end{equation}
was zu folgendem Volumen eines Raumzeitbereiches $\omega$ f\"{u}hrt

\begin{equation}
V=\int_\omega \sqrt{-det(g_{\mu\nu})} dx^1 dx^2 dx^3 dx^4=\int_\omega \sqrt{-det(g_{\mu\nu})} d^4 x.
\end{equation}
Dies bedeutet, dass das Volumenelement in den lokalen Koordinaten, die zu einer flachen Metrik f\"{u}hren, mit den globalen Koordinaten in folgender Beziehung steht

\begin{equation}
d^4 y=\sqrt{-g} d^4 x,
\end{equation}
wodurch sich die folgende Wirkung ergibt

\begin{equation}
S_{Fermion}=\int d^4 x \sqrt{-g}\{\bar \Psi(\gamma^m e_m^\mu(\partial_\mu+i\omega_\mu^{mn}\Sigma_{mn})-m)\Psi\}.
\label{FermionGravWirkung}
\end{equation}
Hierbei steht $g$ f\"{u}r $det(g_{\mu\nu})$. F\"{u}r weitere Aspekte bez\"{u}glich der eichtheoretischen Formulierung der Gravitation und des Vierbeinformalismus sei auf \cite{Ramond},\cite{Nakahara} und \cite{Rovelli} verwiesen.
Die Formulierung der Wirkung von Bosonen und \"{u}blichen Eichfeldern ist bei weitem unproblematischer, da hier der metrische Tensor der Minkowskiraumzeit explizit auftaucht und durch den allgemeinen metrischen Tensor einer gekr\"{u}mmten Raumzeit ersetzt wird, was zu den folgenden Wirkungen f\"{u}hrt. F\"{u}r ein Boson ergibt sich 

\begin{equation}
S_{Boson}=\int d^4 x \sqrt{-g} \{g^{\mu\nu} D_\mu \Phi D_\nu \Phi-V(\Phi)\},
\label{BosonGravWirkung}
\end{equation}
wobei die kovarianten Ableitungen sich hier auf die \"{u}blichen Eichfelder beziehen. Die kovariante Ableitung in Bezug auf das Gravitationsfeld, also die Raumzeit, entspricht bei einem Skalarfeld nat\"{u}rlich der einfachen Ableitung. Die Wirkung der Eichfelder sieht wie folgt aus

\begin{equation} 
S_{Eichfeld}=\int d^4 x \{-\frac{g^{\mu\rho}g^{\nu\sigma}}{4}F_{\rho\sigma}F_{\mu\nu}\}.
\label{EichfeldGravWirkung}
\end{equation}

\section{Die Methode der Pfadintegralquantisierung}

Um das Gravitationsfeld selbst im Rahmen einer Quantenfeldtheorie zu beschreiben und einen Propagator herzuleiten, soll zun\"{a}chst eine kurze Beschreibung der Methode der Pfadintegralquantisierung gegeben werden, wie sie in \cite{WeinbergQTF1} gefunden werden kann. Feynman wurde bei der Suche nach einer quantenmechanischen Beschreibungsweise, welche vom Prinzip der kleinsten Wirkung ausgeht auf die M\"{o}glichkeit gef\"{u}hrt, Propagatoren direkt aus der Lagrangedichte herzuleiten \cite{Feynman:1948ur}. Ein Vorteil dieser Beschreibungsweise ist die Tatsache, dass sie explizit kovariant ist.

\subsection{Allgemeine Formulierung}

Gegeben sei ein quantenmechanisches System, dass durch einen Satz kommutierender hermitescher Operatoren $Q_a$ vollst\"{a}ndig beschrieben sei, wobei die kanonisch konjugierten Operatoren mit $P_a$ bezeichnet seien und durch die Vertauschungsrelationen der Heisenbergalgebra definiert sind

\begin{equation}
[Q_a,P_b]=i\delta_{ab} \quad,\quad [Q_a,Q_b]=[P_a,P_b]=0.
\end{equation}  
$| q \rangle$ sei Eigenzustand zu allen Operatoren $Q_a$ und die zeitliche Entwicklung werde durch den Hamiltonoperator $H$ beschrieben. Die Wahrscheinlichkeit ein Teilchen zur Zeit t' im Eigenzustand $| q' \rangle$ zu finden, wenn es sich zum Zeitpunkt $t$ im Eigenzustand $| q \rangle$ befunden hat, kann im Rahmen des Feynmanschen Pfadintegralformalismus wie folgt ausgedr\"{u}ckt werden 

\begin{eqnarray}
\langle q',t'|q,t \rangle=\lim_{d\tau \to 0} \int \left[ \prod_{k=1}^{N-1} \prod_a dq_{k,a}\right] \left[\prod_b \prod_{k=0}^{N-1} \frac{dp_{k,b}}{2\pi}\right]\nonumber\\
exp\left[i\sum_{k=1}^{N}\left(\sum_a(q_{k,a}-q_{k-1,a})p_{k-1,a}-H(q_k,p_{k-1})d \tau\right)\right],
\end{eqnarray}
wobei das Zeitintervall $t'-t$ in N Zeitintervalle $d\tau=\frac{t'-t}{N}$ unterteilt sei. Im Grenzfall $d\tau\rightarrow 0$ ergibt sich damit

\begin{eqnarray}
\langle q',t'|q,t \rangle=\int_{q_a(t)}^{q_a'} \prod_{\tau,a} dq_a(\tau) \prod_{\tau,b} \frac{dp_b(\tau)}{2\pi}\nonumber\\ 
exp\left[i\int_t^{t'} d\tau \left(\sum_a(\dot q_a (\tau) p_a (\tau)-H(q(\tau),p(\tau))d \tau\right)\right].  
\end{eqnarray}
Dieser Zusammenhang kann so interpretiert werden, dass ein quantenmechanisches System in gewissem Sinne alle m\"{o}glichen dynamischen Entwicklungen gleichzeitig durchl\"{a}uft. Die Wahrscheinlichkeit es zur Zeit t' in einem Zustand $| q' \rangle$ zu finden, wenn es sich zur Zeit t im Zustand $|q \rangle$ befunden hat entspricht der \"{U}berlagerung aller dynamischen Entwicklungen die zwischen diesen beiden Zeitpunkten formal m\"{o}glich sind. Im Exponenten steht ein Ausdruck, der in den physikalisch relevanten F\"{a}llen der Wirkung des Systems entspricht. Nun werden sich die Amplituden von m\"{o}glichen Entwicklungswegen, deren Wirkung sich stark unterscheidet, im Mittel gegenseitig aufheben, da sie aufgrund des Auftauchens der Wirkung in der Exponentialfunktion nicht koh\"{a}rent sind. Es tragen also im Wesentlichen nur die Entwicklungswege zum letztlich messbaren Betragsquadrat des inneren Produktes zwischen Anfangs- und Endzustand bei, denen ein Wert der Wirkung in der N\"{a}he des Maximums entspricht, da die Amplituden hier nahezu koh\"{a}rent sind und sich aufsummieren. Dies bedeutet im klassischen Grenzfall die Implikation des Hamiltonschen Prinzips der kleinsten Wirkung.
Im Rahmen einer quantenfeldtheoretischen Beschreibungsweise werden die Operatoren $Q_a$ und $P_a$ durch Feldoperatoren $\Phi_m (x)$ und ihre kanonisch konjugierten Feldoperatoren $\Pi_m (x)$ ersetzt, welche von den Raumzeitkoordinaten abh\"{a}ngen und einen Index m f\"{u}r den Spinfreiheitsgrad besitzen. Als Vakuum-Vakuum-Amplitude f\"{u}r ein zeitgeordnetes Produkt von Operatoren ergibt sich hier

\begin{eqnarray}
\langle Vac | T\{\mathcal{O}_A [\Phi_A(t_A), \Pi(t_A)], \mathcal{O}_B [\Phi_B(t_B), \Pi(t_B)], ...| Vac \rangle \nonumber\\
=|\mathcal{N}|^2 \int \left[\prod_{\tau,x,m} d\phi_m(x,\tau)\right] 
\left[\prod_{\tau,x,m} \frac{d\pi_m(x,\tau,m)}{2\pi}\right]\nonumber\\
\mathcal{O}_A [\pi(t_A),\phi(t_A)] \mathcal{O}_B [\pi(t_B),\phi(t_B)]\nonumber\\
\times exp\left[i\int_{-\infty}^{\infty}d\tau \left(\int d^3 x (\sum_m \dot \phi_m(x,\tau)\pi_m(x,\tau)-H[\phi(\tau),\pi(\tau)])+i\epsilon \right)\right].\nonumber\\
\end{eqnarray}
Der Integrand in der Exponentialfunktion hat die Gestalt einer Lagrangedichte. Da die Impulsoperatoren jedoch zun\"{a}chst unabh\"{a}ngige Gr\"{o}\ss en sind, darf dieser nicht einfach der Lagrangedichte gleichgesetzt werden. Dies ist nur dann der Fall, wenn der Hamiltonoperator eine quadratische Abh\"{a}ngigkeit von den kanonisch konjugierten Impulsoperatoren aufweist, denn man kann zeigen, dass dann gilt 

\begin{equation}
\dot \Psi(x,\tau)=\frac{\delta H[\Psi(\tau),\Pi(\tau)]}{\delta \pi(x,\tau)},
\end{equation}
was bedeutet

\begin{equation}
L[\Psi(\tau),\dot \Psi(\tau)]=\int d^3 x (\sum_n \dot \Psi(x,\tau) \Pi(x,\tau)-H[\Psi(\tau),\Pi(\tau)]),
\end{equation}
und damit

\begin{eqnarray}
\langle Vac | T\{\mathcal{O}_A [\Phi_A(t_A), \Pi(t_A)], \mathcal{O}_B [\Phi_B(t_B), \Pi(t_B)], ...| Vac \rangle\nonumber\\
=|\mathcal{N}|^2 \int \left[\prod_{\tau,x,m} d\phi_m(x,\tau)\right] \left[\prod_{\tau,x,m} \frac{d\pi_m(x,\tau,m)}{2\pi}\right]\nonumber\\
\mathcal{O}_A [\pi(t_A),\phi(t_A)] \mathcal{O}_B [\pi(t_B),\phi(t_B)] exp\left[i\int_{-\infty}^{\infty}d\tau L\left[\Psi(\tau),\dot \Psi(\tau)\right]+i\epsilon \right].
\end{eqnarray}

\subsection{Herleitung eines Propagators}

Um nun den Propagator des freien Feldes zu erhalten, zerlegt man die Lagrangedichte $\mathcal{L}$ in einen Term $\mathcal{L}_0$, welcher das freie Feld beschreibt, und einen Wechselwirkungsterm $\mathcal{L}_1$ 

\begin{equation}
L[\Psi(\tau),\dot \Psi(\tau)]=\int d^3 x [\mathcal{L}_0(\Psi(\vec x,\tau),\partial_\mu \Psi(\vec x,\tau))+
\mathcal{L}_1(\Psi(\vec x,\tau),\partial_\mu \Psi(\vec x,\tau))].
\end{equation}
Der Term $\mathcal{L}_0$, welcher das freie Feld beschreibt, ist eine quadratische Funktion des Feldes und daher kann der freie Teil der Wirkung wie folgt ausgedr\"{u}ckt werden

\begin{equation}
S_0=\int d^4 x \Psi_m (x) \mathcal{D}_{mm'} \Psi_{m'}.
\end{equation}  
Wenn man $\mathcal{D}$ nun im Impulsraum ausdr\"{u}ckt, so kann man durch eine recht lange Herleitung zeigen \cite{WeinbergQTF1}, dass f\"{u}r den Propagator im Ortsraum gilt

\begin{equation}
\Delta(y,x)=\frac{1}{(2\pi)^4}\int d^4 p e^{ip(y-x)}\mathcal{D}^{-1}(p).
\label{Propagator-Lagrangian}
\end{equation}
Im Falle einer Quantisierung von Eichtheorien ist noch die Einf\"{u}hrung eines Eichfixierungstermes von N\"{o}ten, der die
Gestalt einer Deltafunktion hat, sodass der Gesamtausdruck nur einen Beitrag liefert, wenn die Eichbedingung erf\"{u}llt ist.
Es ergibt sich in diesem Falle ein Ausdruck der folgenden Form f\"{u}r das Pfadintegral eines Feldes $\Psi$

\begin{equation}
\langle Vac | Vac \rangle = \int \left[\prod d\Psi(x)\right]\ \delta[(G(\Psi)]\ det|\frac{\delta G}{\delta \alpha}|\ exp(iS[\Psi(x)]),
\end{equation}
wobei $G$ die Eichfixierungsfunktion und $\alpha$ der Eichparameter ist. Wenn man die Eichfixierung und die Determinante in eine Exponentialfunktion umschreibt erh\"{a}lt man einen zus\"{a}tzliche Eichfixierungsterm im quadratischen Teil der Lagrangedichte sowie sogenannte Geisterfelder (siehe beispielsweise \cite{WeinbergQTF2} oder \cite{Rivers}). Aber diese Thematik spielt hier keine weitere Rolle, da es hier um Gravitonen gehen wird, deren Schwingungszust\"{a}nde in den zus\"{a}tzlichen Dimensionen ihnen eine Masse verleihen, wodurch die Eichinvarianz aufgehoben wird. 

\section{Quantisierung des Gravitationsfeldes}

\subsection{Die Lagrangedichte des Gravitationsfeldes}

Bez\"{u}glich der Herleitung eines Propagators f\"{u}r das Graviton, ist es zun\"{a}chst einmal wichtig, dass man die Einsteinsche Feldgleichung, wie alle anderen dynamischen Gleichungen durch eine Wirkung beschreiben kann, die durch das bekannte Hamiltonsche Variationsprinzip dann wieder auf die Einsteinschen Gleichungen f\"{u}hrt.
Es handelt sich um die Einstein-Hilbert-Wirkung, welche folgende Gestalt hat

\begin{equation}
S_{EH}=\frac{1}{16\pi G}\int d^4 x \sqrt{-g} R.
\end{equation}
Im allgemeineren Fall einer beliebigen Anzahl D an Raumzeitdimensionen hat man es mit der folgenden Wirkung zu tun

\begin{equation}
S=-\frac{1}{2}M^{D-2} \int d^D x \sqrt{g} R.
\end{equation}
Diese muss zun\"{a}chst in eine geeignete Gestalt gebracht werden, um anschlie\ss\ end den Propagator f\"{u}r das Graviton aus ihr ableiten zu k\"{o}nnen. 
Man verwendet nun eine lineare Entwicklung der Metrik $g_{MN}$ um die Minkowskimetrik der flachen Raumzeit 
\begin{equation}
g_{MN}=\eta_{MN}+2M^\frac{(D-2)}{2} h_{MN}.
\label{Entwicklung-Metrik}
\end{equation}
Die hier gegeben Darstellung der Quantisierung des Gravitationsfeldes folgt im Wesentlichen der in \cite{Callin:2004zm} gelieferten Herleitung. Eine Beschreibung der entsprechenden effektiven Quantenfeldtheorie im Falle einer gew\"{o}hnlichen Raumzeit wird hier \"{u}bergangen. Hierf\"{u}r sei auf \cite{Donoghue:1995cz} verwiesen.
In der Einstein-Hilbert-Wirkung taucht der Ricci-Skalar R auf, welchen man aus dem Riemanntensor durch Kontraktion erh\"{a}lt.
Da der Riemanntensor gem\"{a}\ss\ des Zusammenhangs aus dem ersten Kapitel ($\ref{RiemannZusammenhangkoeffizienten}$) \"{u}ber die Zusammenhangkoeffizienten bestimmt ist, welche wiederum im Falle der Allgemeinen Relativit\"{a}tstheorie als Christoffelsymbole \"{u}ber die Metrik definiert sind ($\ref{Christoffelsymbole}$), f\"{u}hrt ($\ref{Entwicklung-Metrik}$) auf einen Ausdruck f\"{u}r R in Abh\"{a}ngigkeit von h, womit sich folgender Ausdruck f\"{u}r die Lagrangedichte $\mathcal{L}_h=\sqrt{-g} R$ ergibt

\begin{equation}
\mathcal{L}_h=-\frac{1}{2}\partial_M \partial^M h+\frac{1}{2}\partial_R h_{MN} \partial^R h^{MN}
+\partial_M h^{MN}\partial_N h-\partial_M h^{MN} \partial_R h^R_N.
\end{equation}
Diese Lagrangedichte ist zun\"{a}chst invariant unter Eichtransformationen der folgenden Form

\begin{equation}
x^M \rightarrow x^M+2M^{2-D}{2} \alpha^M (x)\quad,\quad h_{MN} \rightarrow h_{MN}-(\partial_M \alpha_N+\partial_N \alpha_M).
\end{equation}
Dies w\"{u}rde das Hinzuf\"{u}gen eines Eichfixierungstermes der Gestalt $\mathcal{L}_{hg}=\frac{1}{\alpha}C_M C^M$ notwendig machen.
Im Rahmen dieser Arbeit sollen aber wie bereits erw\"{a}hnt massebehaftete Gravitonen betrachtet werden.  
Durch Einf\"{u}hrung eines sogenannten Fierz-Pauli-Termes $-\frac{1}{2}m^2(h^{MN} h_{MN}-h^2)$ erh\"{a}lt man folgende Lagrangedichte f\"{u}r
das Graviton, welches nun massebehaftet ist

\begin{eqnarray}
\mathcal{L}_h=-\frac{1}{2}\partial_M h \partial^M h+\frac{1}{2}\partial_R h_{MN} \partial^R h^{MN}
+\partial_\mu h^{MN}\partial_N h-\partial_M h^{MN} \partial_R h^R_N\nonumber\\
-\frac{1}{2}m^2 (h^{MN}h_{MN}-h^2).
\label{freieLagrangedichteMasse}
\end{eqnarray}
Diese besitzt nun keine Eichfreiheit mehr, womit keine zus\"{a}tzlichen Terme in das entsprechende Pfadintegral eingef\"{u}hrt werden m\"{u}ssen. 

\subsection{Der Gravitonpropagator}

Als Gravitonpropagator ergibt sich schlie\ss lich unter Verwendung der allgemeinen Relation zwischen dem Anteil des freien Feldes in der Lagrangedichte und dem entsprechenden Propagator ($\ref{Propagator-Lagrangian}$) der folgende Ausdruck f\"{u}r den Gravitonpropagator 

\begin{equation}
\Delta_{MNRS}(x,y)=\int \frac{d^D k}{(2\pi)^D}\frac{P_{MNRS}(k)}{k^2-m^2}e^{-ik(x-y)},
\label{GravitonpropagatorMasseD}
\end{equation}
wobei der Polarisationstensor $P_{\mu\nu\rho\sigma}$ wie folgt aussieht

\begin{eqnarray}
P_{MNRS}(k)=\frac{1}{2}(\eta_{MR}\eta_{NS}+\eta_{MS}\eta_{NR})
-\frac{1}{D-2}\eta_{MN}\eta_{RS}\\\nonumber
-\frac{1}{2m^2}(\eta_{MR}k_N k_\sigma+\eta_{NS}k_M k_R+\eta_{MR}k_N k_S+\eta_{NS}k_M k_R)\nonumber\\
+\frac{1}{(D-1)(D-2)}(\eta_{MN}+\frac{D-2}{m^2} k_M k_N)(\eta_{RS}+\frac{D-2}{m^2}k_R k_S).
\end{eqnarray}
Es wurde also ein Propagator f\"{u}r massebehaftete Gravitonen in einer D-dimensionalen Raumzeit hergeleitet. Innerhalb des ADD-Modells sind die zus\"{a}tzlichen Dimensionen aber kompaktifiziert. Deshalb w\"{a}hlt man f\"{u}r das Gravitationsfeld, das im Rahmen der obigen Entwicklung ($\ref{Entwicklung-Metrik}$) durch $h_{\mu\nu}$ beschrieben wird einen Ansatz, der dem bereits im letzten Kapitel beschriebenen f\"{u}r das Skalarfeld
entspricht ($\ref{SkalarfeldKK}$), mit dem Unterschied, dass von einer beliebigen Anzahl $\delta$ an zus\"{a}tzlichen Dimensionen ausgegangen wird. Das f\"{u}hrt auf folgenden Ausdruck f\"{u}r das Gravitationsfeld

\begin{equation}
h_{MN}(z)=\sum_{j=1}^\delta \sum_{n_\delta=-\infty}^{\infty} \frac{h_{MN}(x)}{\sqrt{V_\delta}}e^{i\frac{n^j y_j}{R}}.
\end{equation}
Gem\"{a}\ss\ der betrachteten Situation eines Skalarfeldes bei einer zus\"{a}tzlichen Dimension verleihen die Anregungen in den kompaktifizierten
Dimensionen dem Gravitationsfeld eine Masse, wobei die Masse nun der Summe der Quadrate des Verh\"{a}ltnisses der Schwingungszahlen zum Radius entspricht 

\begin{equation}
\sum_{j=1}^{\delta}|\hat n_j|^2 \quad,\quad |\hat n_j|^2=\frac{|n_j|^2}{R^2}. 
\end{equation}
Der in der obigen Betrachtung eingef\"{u}hrte Massenterm ergibt sich also unter der Annahme, dass das Gravitationsfeld in den zus\"{a}tzlichen kompaktifizierten Dimensionen Anregungszust\"{a}nde besitzt, die aufgrund der Periodizit\"{a}t gequantelt sein m\"{u}ssen \cite{Giudice:1998ck}, \cite{Han:1998sg}. Die Masse ist hier jedoch nicht wie gew\"{o}hnlich eine Konstante, sondern eben auch eine Eigenschaft eines Zustandes.
In dieser Beschreibungsweise wird die St\"{o}rung der Metrik, welche dem Gravitationsfeld entspricht, also aufgespalten in einen Anteil der die Geometrie der Untermannigfaltigkeit beschreibt, auf der die bekannten Materiefelder leben, und die zus\"{a}tzlichen Freiheitsgrade der weiteren Dimensionen. Bez\"{u}glich der Wechselwirkung mit den Materiefeldern ist also nur der erste Anteil von Bedeutung, wobei die Schwingungen in den zus\"{a}tzlichen Dimensionen nat\"{u}rlich das Verhalten der Gravitonen insofern beeinflussen als sie ihnen eben eine Masse verleihen. Dies bedeutet, dass sich letztlich f\"{u}r den Gravitonpropagator in dem hier vorausgesetzten Modell der oben hergeleitete Gravitonpropagator ($\ref{GravitonpropagatorMasseD}$) f\"{u}r den Fall von 4 Dimensionen ergibt

\begin{equation}
\Delta_{\mu\nu\rho\sigma}(x,y)=\int \frac{d^D k}{(2\pi)^D}\frac{P_{\mu\nu\rho\sigma}(k)}{k^2-m^2}e^{-ik(x-y)},
\label{GravitonpropagatorMasse}
\end{equation}
mit Polarisationstensor

\begin{eqnarray}
P_{\mu\nu\rho\sigma}(k)=\frac{1}{2}(\eta_{\mu\rho}\eta_{\nu\sigma}+\eta_{\mu\sigma}\eta_{\nu\rho})
-\frac{1}{2}\eta_{\mu\nu}\eta_{\rho\sigma}\\\nonumber
-\frac{1}{2m^2}(\eta_{\mu\rho}k_\nu k_\sigma+\eta_{\mu\sigma}k_\nu k_\rho+\eta_{\nu\rho}k_\mu k_\sigma+\eta_{\nu\sigma}k_\mu k_\rho)\\\nonumber
+\frac{1}{6}(\eta_{\mu\nu}+\frac{2}{m^2} k_\mu k_\nu)(\eta_{\rho\sigma}+\frac{2}{m^2}k_\rho k_\sigma),
\label{PolarisationstensorGraviton}
\end{eqnarray}
wobei f\"{u}r die Masse m gilt

\begin{equation}
m^2=\sum_{j=1}^{\delta}\left(\frac{n_j^2}{R^2}\right).
\end{equation}
Durch Fouriertransformation erh\"{a}lt man wie \"{u}blich den Propagator im Impulsraum 

\begin{equation}
\delta_{\mu\nu\rho\sigma}=\frac{P_{\mu\nu\rho\sigma}(k)}{k^2-m^2}.
\end{equation}
Der Polarisationstensor erf\"{u}llt die folgenden Relationen

\begin{equation}
\eta^{\mu\nu}P_{\mu\nu\rho\sigma}=0 \quad,\quad k^\mu P_{\mu\nu\rho\sigma}=0.
\end{equation}
Der Gravitonpropagator ($\ref{GravitonpropagatorMasse}$) wird in \cite{Giudice:1998ck} auf eine etwas andere Art und Weise hergeleitet.

\subsection{Kopplungen und Vertizes}

Wenn man nun die Wechselwirkung mit Materiefeldern konkret beschreiben will, so muss man den Energie-Impuls-Tensor in die Lagrangedichte integrieren. Die oben angegebene Einstein-Hilbert-Wirkung bezieht sich nur auf die Einsteingleichung ohne Materie, bei der also $T_{\mu\nu}=0$ angenommen wird. Es ist also notwendig, zur Lagrangedichte f\"{u}r das freie massebehaftete Graviton einen zus\"{a}tzlichen Term zu addieren, welcher auf den Energie-Impuls-Tensor in den Einsteinschen Feldgleichungen f\"{u}hrt. Insgesamt lautet damit die Lagrangedichte

\begin{eqnarray}
\mathcal{L}_h=-\frac{1}{2}\partial_\mu h \partial^\mu h+\frac{1}{2}\partial_R h_{\mu\nu} \partial^R h^{\mu\nu}
+\partial_\mu h^{\mu\nu}\partial_\nu h-\partial_\mu h^{\mu\nu} \partial_R h^R_\nu\nonumber\\
-\frac{1}{2}m^2 (h^{\mu\nu}h_{\mu\nu}-h^2)-\frac{1}{\bar M_P}h^{\mu\nu}T_{\mu\nu}.
\end{eqnarray}
Die spezifische Form des Energie-Impuls-Tensors ist nat\"{u}rlich durch die Wirkungen ($\ref{FermionGravWirkung}$),($\ref{BosonGravWirkung}$),($\ref{EichfeldGravWirkung}$) definiert, welche das Verhalten von Materiefeldern in gekr\"{u}mmten Raumzeiten beschreibt und um welche die Einstein-Hilbert-Wirkung des freien Gravitationsfeldes erweitert werden muss.
Aus den in diesen enthaltenen Kopplungstermen bestimmt man die Vertizes. F\"{u}r die Wechselwirkung eines Fermion bzw. Vektorbosons ergeben sich die
folgenden Vertizes \cite{Giudice:1998ck}.\\

\newpage

{\bf Fermion-Fermion-Graviton}

\begin{figure}[h]
\centering
\epsfig{figure=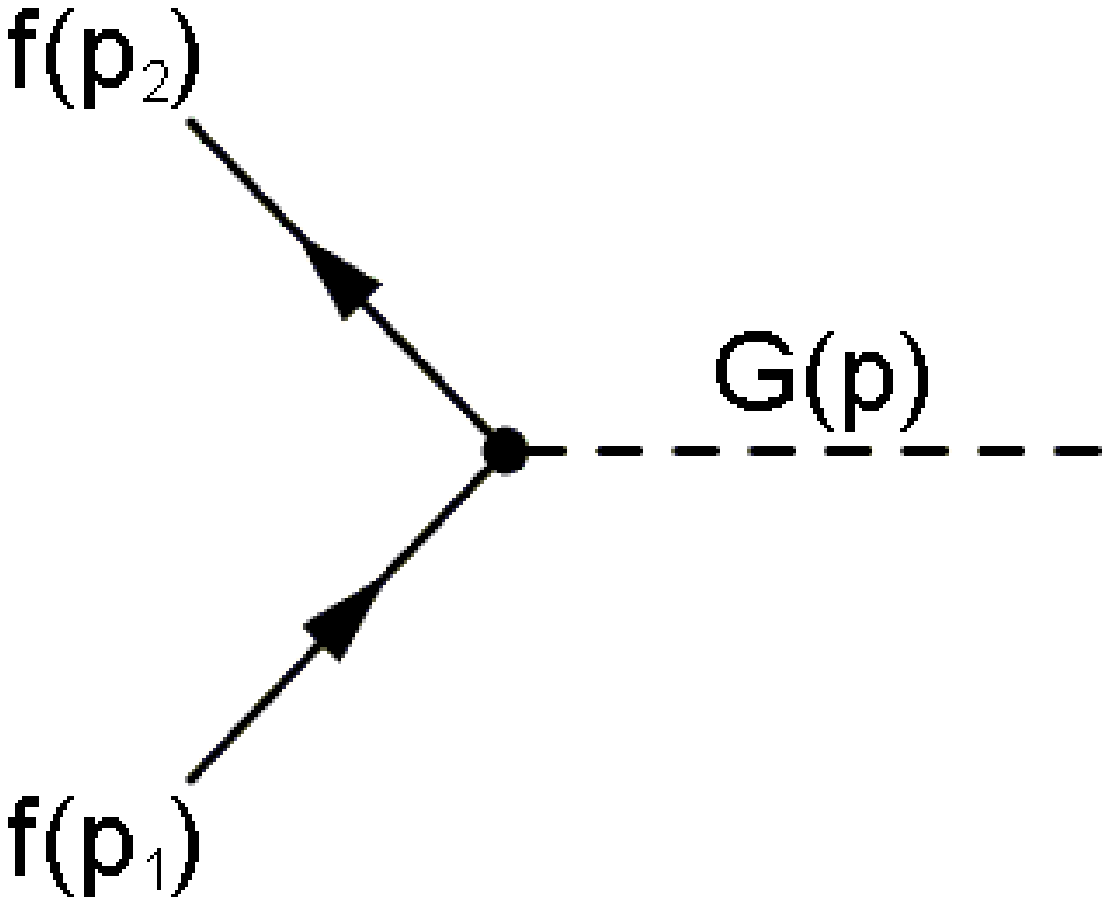,width=7cm}
\end{figure}

\begin{eqnarray}
-\frac{i}{4 \bar M_P}[W_{\mu\nu}+W_{\nu\mu}]\\\nonumber
\\\nonumber 
W_{\mu\nu}=(p_1+p_2)_\mu \gamma_\nu
\label{VertexFFG}
\end{eqnarray}
\\

{\bf Vektorboson-Vektorboson-Graviton}

\begin{figure}[h]
\centering
\epsfig{figure=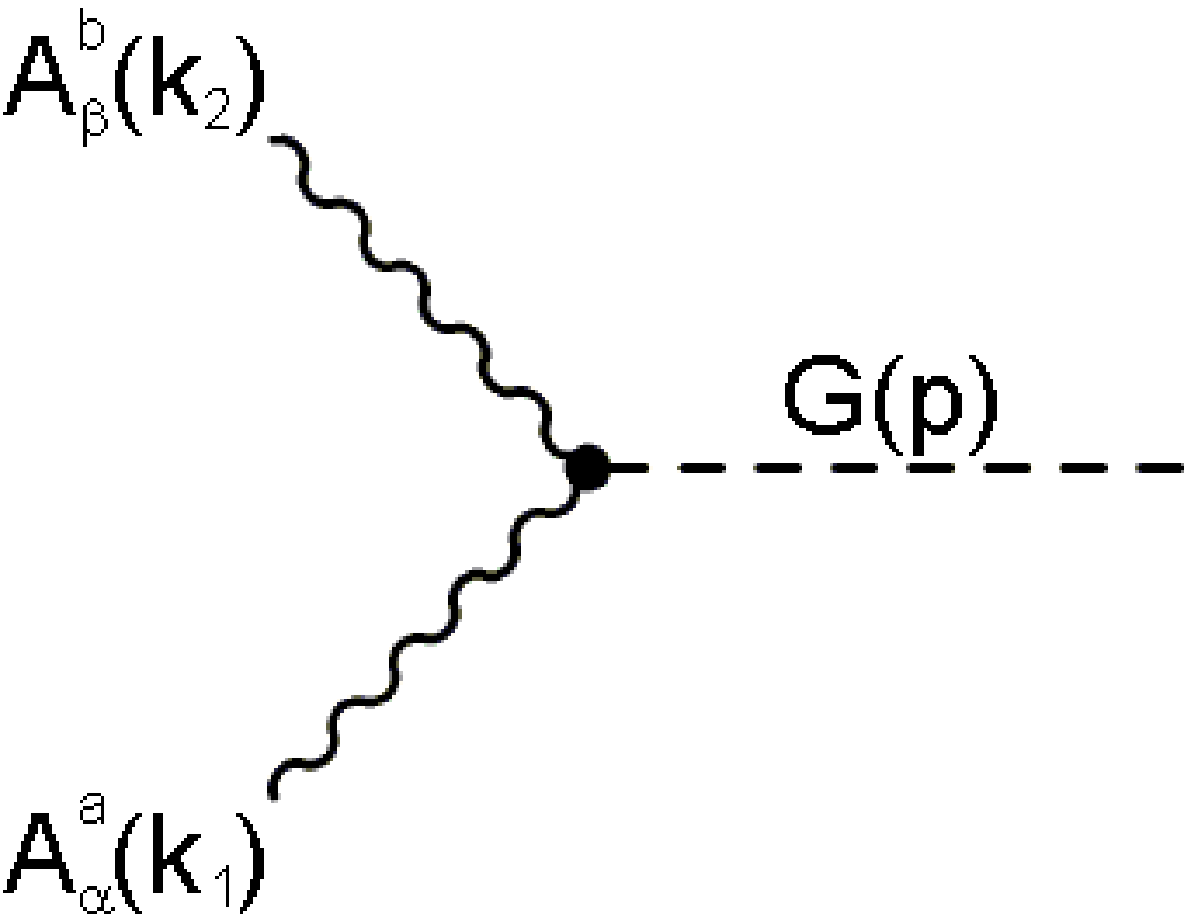,width=7cm}
\end{figure}

\begin{eqnarray}
&&-\frac{i}{4 \bar M_P}\delta^{ab}[W_{\mu\nu\rho\sigma}+W_{\nu\mu\rho\sigma}]\\\nonumber
\\\nonumber
W_{\mu\nu\rho\sigma}&=&\frac{1}{2} \eta_{\mu\nu}(k_{1\sigma}k_{2\rho}-k_1\cdot k_2 \eta_{\rho\sigma})+\eta_{\rho\sigma}k_{1\mu}k_{2\nu}\\\nonumber
&&+\eta_{\mu\rho}(k_1 \cdot k_2 \eta_{\nu\sigma}-k_{1\sigma}k_{2\nu})-\eta_{\mu\sigma}k_{1\nu}k_{2\rho}
\label{VertexVbVbG}
\end{eqnarray}
\\

{\bf Fermion-Fermion-Vektorboson-Graviton}

\begin{figure}[h]
\centering
\epsfig{figure=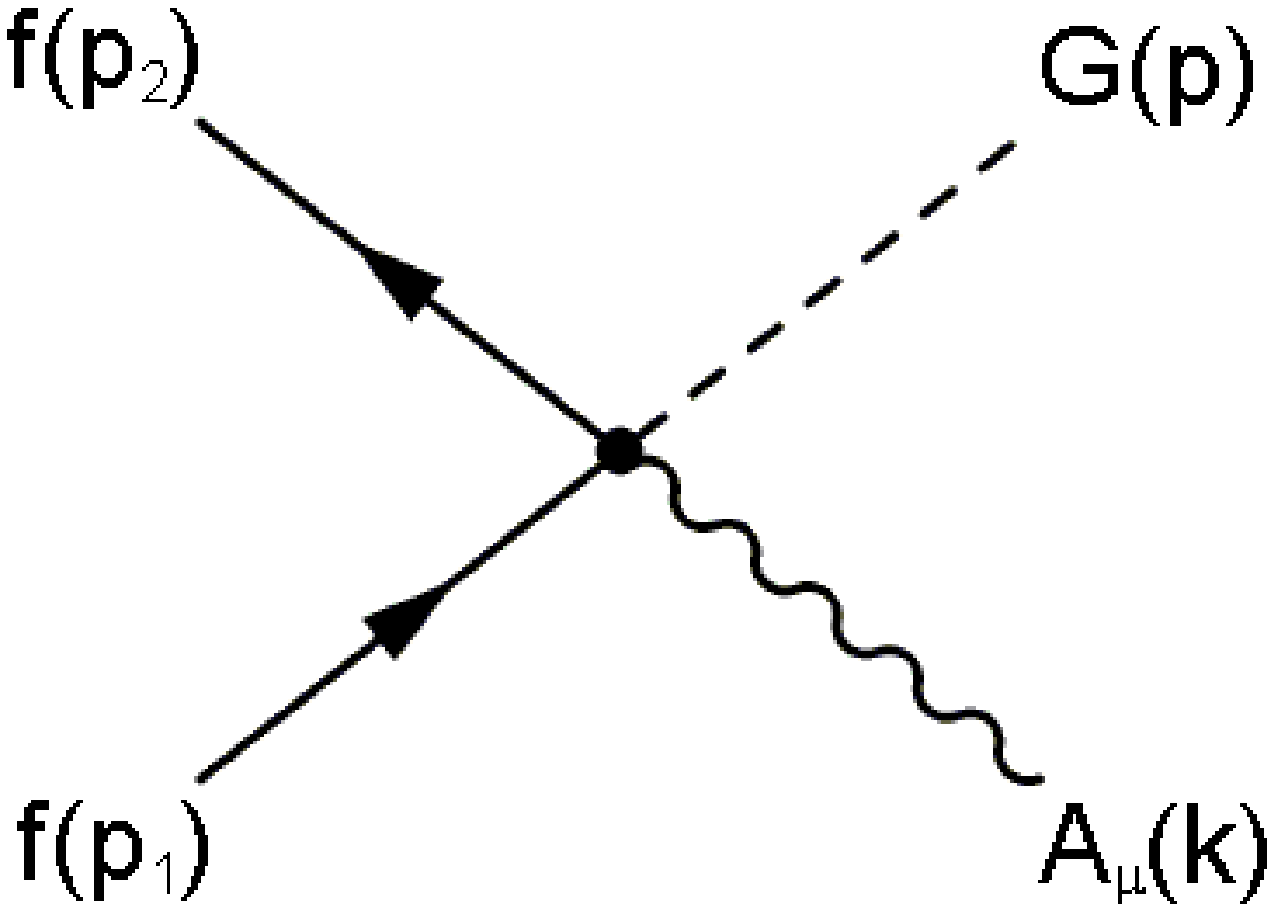,width=7cm}
\end{figure}

\begin{eqnarray}
-\frac{i}{2\bar M_P} gT^a (X_{\mu\nu\alpha}+X_{\nu\mu\alpha})\quad\quad\quad X_{\mu\nu\alpha}=\gamma_\mu \eta_{\nu\alpha}
\end{eqnarray}
\\

{\bf Vektorboson-Vektorboson-Vektorboson-Graviton}

\begin{figure}[h]
\centering
\epsfig{figure=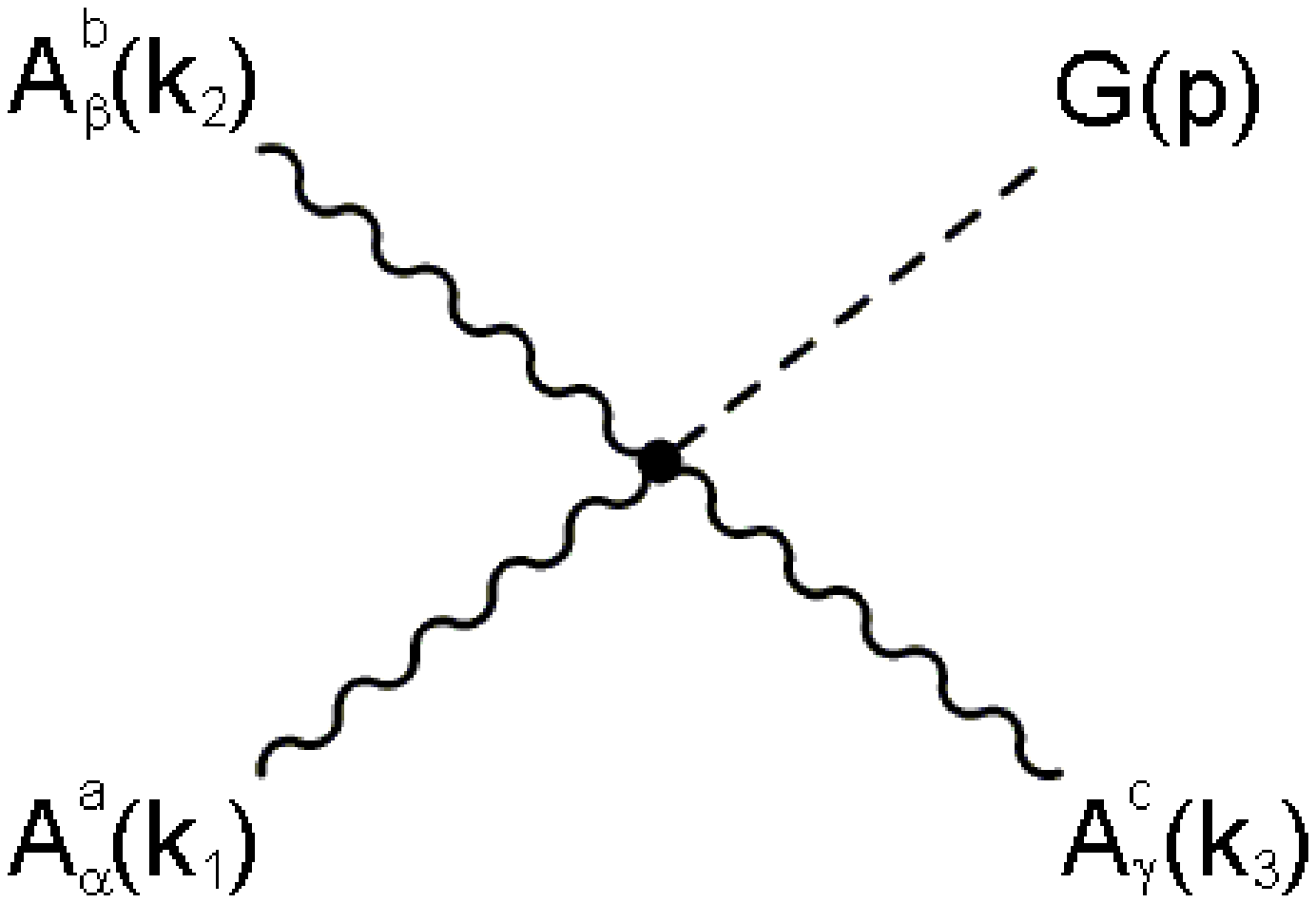,width=7cm}
\end{figure}

\begin{eqnarray}
&\frac{g}{\bar M_P}f^{abc}&[Y(k_1)_{\mu\nu\alpha\beta\gamma}+Y(k_2)_{\mu\nu\beta\gamma\alpha}+Y(k_3)_{\mu\nu\gamma\alpha\beta}\nonumber\\
&&+Y(k_1)_{\nu\mu\alpha\beta\gamma}+Y(k_2)_{\nu\mu\beta\gamma\alpha}+Y(k_3)_{\nu\mu\gamma\alpha\beta}]\nonumber\\
\nonumber\\
Y(k)&=&k_\mu(\eta_{\nu\beta}\eta_{\alpha\gamma}-\eta_{\nu\gamma}\eta_{\alpha\beta})\nonumber\\
&&+k_{\beta}\left(\eta_{\mu\alpha}\eta_{\nu\gamma}-\frac{1}{2}\eta_{\mu\nu}\eta_{\alpha\gamma}\right)-k_{\gamma}\left(\eta_{\mu\alpha}\eta_{\nu\beta}-\frac{1}{2}\eta_{\mu\nu}\eta_{\alpha\beta} \right)\nonumber\\
\end{eqnarray}

\chapter[ZZ-Produktion durch Gravitonen]{ZZ-Produktion durch Gravitonenvermittlung}

In diesem Kapitel soll die in den vorigen Kapiteln dargestellte Theorie nun auf einen konkreten physikalischen Vorgang angewandt werden. Es soll der Vorgang betrachtet werden, bei dem ein Teilchen und ein Antiteilchen sich gegenseitig vernichten und dabei ein virtuelles Graviton erzeugen, das sich anschlie\ss end in zwei Z-Teilchen umwandelt. Diese stellen neben den $W^+$ und den $W^-$-Teilchen die Austauschteilchen der schwachen Wechselwirkung dar (siehe Kapitel 5). Hierbei handelt es sich um einen Prozess, der im Standardmodell ziemlich unwahrscheinlich ist. Daher k\"{o}nnte der Gravitonenaustausch, welcher unter der Annahme zus\"{a}tzlicher Dimensionen und einer entsprechend modifizierten Planckmasse $M_D$ sehr viel wahrscheinlicher wird, zu einer deutlichen Abweichung gegen\"{u}ber dem gem\"{a}\ss\ dem Standardmodell erwarteten Wert f\"{u}r die Produktionsrate f\"{u}hren. 
Im Speziellen soll hier das Aufeinandertreffen zweier Protonen untersucht werden, da dies der Situation am LHC entspricht. Ein Proton setzt sich aus Quarks und Gluonen zusammen. Durch letztere wechselwirken die Quarks gem\"{a}\ss\ der Quantenchromodynamik miteinander und werden dadurch zusammengehalten. Zun\"{a}chst m\"{u}ssen die Wirkungsquerschnitte f\"{u}r die Einzelprozesse berechnet werden, ehe diese mit den Verteilungsfunktionen f\"{u}r Partonen innerhalb eines Protons gefaltet werden. Diese Verteilungsfunktionen sind aufgrund der unglaublichen Komplexit\"{a}t der physikalischen Verh\"{a}ltnisse innerhalb eines Protons nur durch Messungen bekannt, k\"{o}nnen also nicht selbst auf theoretischem Wege ermittelt werden.

\section[S-Matrix und Feynmanamplitude]{S-Matrix und Feynmanamplitude f\"{u}r die ZZ-Produktion durch Gluonen}

\begin{figure}[h]
\centering
\epsfig{figure=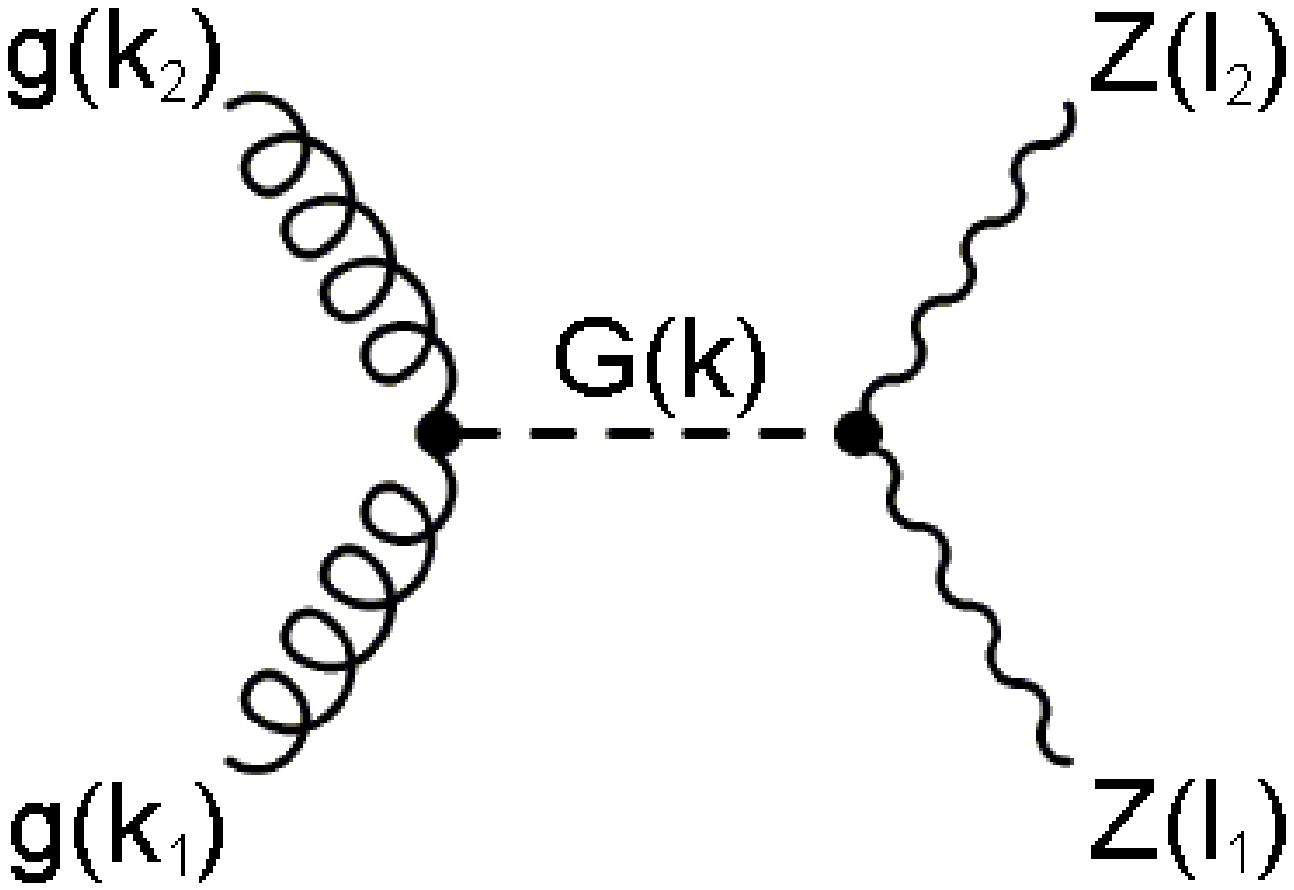,width=7cm}
\caption{\label{Graph1} Gluonen-Graviton-Z-Teilchen}
\end{figure}

\begin{figure}[h]
\centering
\epsfig{figure=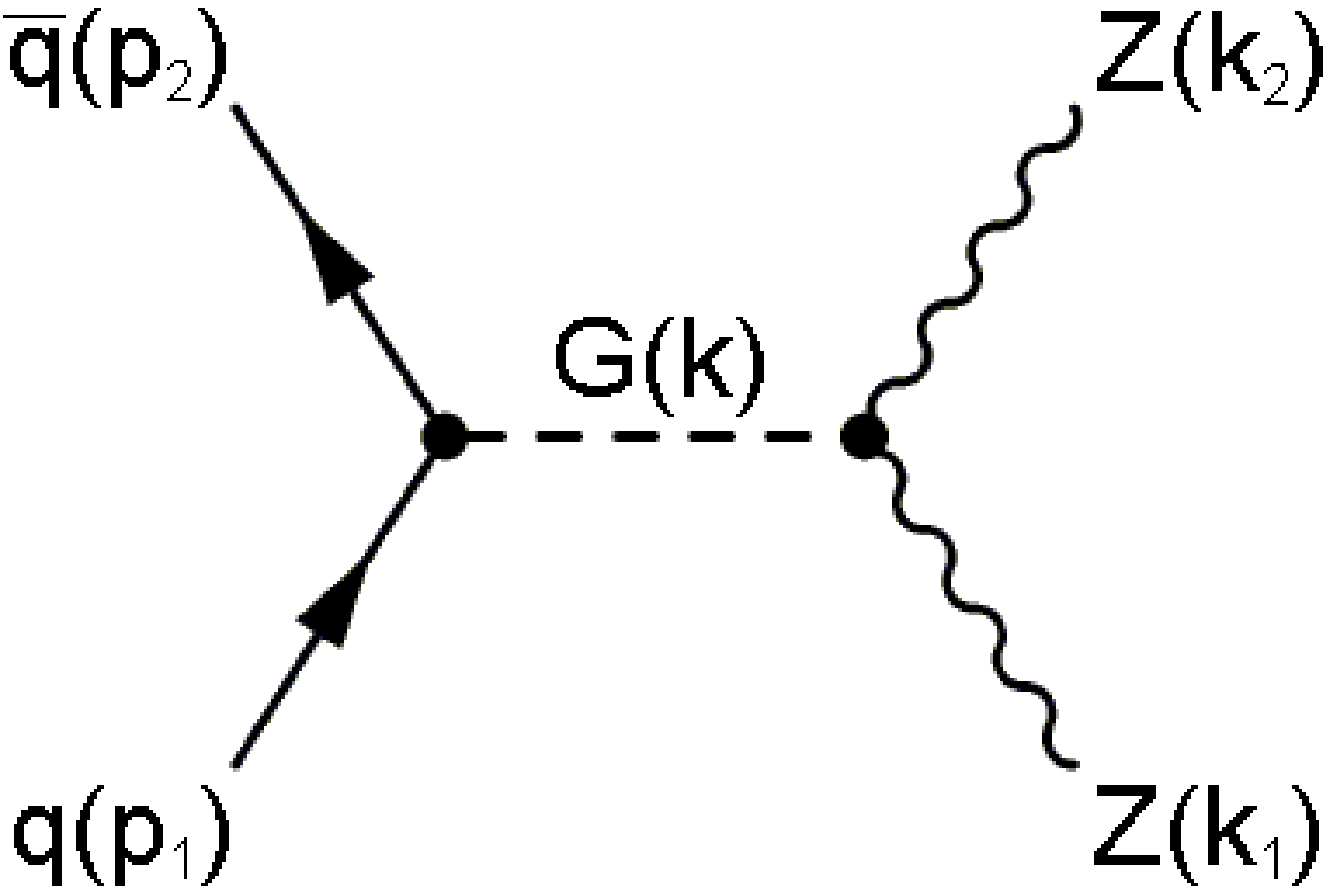,width=7cm}
\caption{\label{Graph2} Quarks-Graviton-Z-Teilchen}
\end{figure}

Der erste Schritt zur Berechnung der ZZ-Produktionsrate durch Vernichtung zweier Protonen  besteht in der Berechnung der S-Matrix f\"{u}r den ZZ-Produktionsprozess durch Vernichtung der im Proton enthaltenen Partonen. Hierbei wird die \"{u}bliche St\"{o}rungstheorie zu Grunde gelegt, die eigentlich in jedem Buch \"{u}ber Quantenfeldtheorie wie beispielsweise \cite{BjorkenDrellRQFT}, \cite{WeinbergQTF1} und \cite{PeskinSchroeder} zu finden ist und hier nicht weiter thematisiert werden soll. Es wird eine Rechnung in erster Ordnung durchgef\"{u}hrt. 
Zun\"{a}chst ist zu erw\"{a}hnen, dass alle hier auftauchenden Feynmangraphen, die einen Vierervertex der im letzten Kapitel angegebenen Art enthalten, entweder \"{u}berhaupt keinen Beitrag liefern oder im Rahmen einer St\"{o}rungsentwicklung in erster Ordnung nicht ber\"{u}cksichtigt werden m\"{u}ssen. Bei zwei einlaufenden Gluonen und zwei auslaufenden Z-Bosonen kann ohnehin nur ein Graviton ausgetauscht werden, da diese im Standardmodell nicht aneinander koppeln. Im Falle eines einlaufenden Quark-Antiquark-Paares gibt es zwei m\"{o}gliche Graphen mit Vierervertizes. Einerseits kann das Quark-Antiquark-Paar direkt in einen Vierervertex laufen und ein Graviton und ein Teilchen der schwachen Wechselwirkung erzeugen. In diesem Falle verschwindet der zweite Vertex mit den auslaufenden Z-Bosonen aufgrund der im Vertex auftauchenden Strukturkonstante $f^{abc}$, die bei zwei gleichen Indizes, die den gleichen zu den beiden Z-Teilchen geh\"{o}rigen Generatoren entsprechen, gleich null ist. Andererseits kann aber auch das eine Quark ein Z-Teilchen emittieren und gemeinsam mit einem im Vertex erzeugten Graviton weiterlaufen, um sich mit diesem und dem Antiquark dann im zweiten Vertex zu vernichten und dabei das andere Z-Boson zu erzeugen. In den beiden Vertizes steht jedoch neben der inversen Planckmasse auch noch die Kopplungskonstante der schwachen Wechselwirkung. Das bedeutet jedoch, dass dieser Beitrag von h\"{o}herer Ordnung ist und damit nicht ber\"{u}cksichtigt werden muss. Dies gilt nat\"{u}rlich nur dann, wenn die Gravitonenkopplung in der St\"{o}rungsentwicklung wie die Kopplung im Standardmodell, also in diesem Falle die der elektroschwachen Wechselwirkung, behandelt wird. Dies ist aber unumg\"{a}nglich, wenn eine Rechnung in erster Ordnung \"{u}berhaupt einen Sinn haben soll. Damit bleiben die Prozesse \"{u}brig, bei denen sich ein Quark-Antiquark- bzw. ein Gluon-Paar direkt vernichten und dabei ein einzelnes Graviton erzeugen, welches anschlie\ss end in zwei Z-Bosonen \"{u}bergeht (Graphen \ref{Graph1} und \ref{Graph2}).\newpage

Die S-Matrix des Feynmangraphen f\"{u}r den Quark-Antiquark-Prozess ergibt exakt 0 (siehe Anhang).
Da im Rahmen des Gravitonaustausches also einerseits nur die Gluonen entscheidend beitragen und im Standardmodell andererseits nur die Quarks \"{u}berhaupt einen Beitrag liefern, weil die Gluonen nur mit sich selbst und nicht elektroschwach wechselwirken (siehe Abschnitt 5.3), muss bez\"{u}glich des zus\"{a}tzlichen Beitrages nur ein Feynmandiagramm ausgewertet werden, um die entsprechende S-Matrix zu erhalten. Aufgrund der Unabh\"{a}ngigkeit der S-Matrix des Quarkprozesses von der des Gluonenprozesses, kann der zum Standardmodell zus\"{a}tzliche Wirkungsquerschnitt separat berechnet werden.
Zun\"{a}chst werden die Polarisationszust\"{a}nde f\"{u}r das einlaufende Gluon-Antigluon-Paar und das auslaufende ZZ-Paar, sowie die Ausdr\"{u}cke f\"{u}r die Vertizes und den Gravitonpropagator zusammengef\"{u}gt, die aus dem letzten Kapitel bekannt sind. Hierbei sollen die entsprechenden Ausdr\"{u}cke im Impulsraum verwendet werden. Entscheidend ist, dass im Gravitonpropagator \"{u}ber alle Anregungszust\"{a}nde in den zus\"{a}tzlichen kompaktifizierten Dimensionen summiert werden muss, die verschiedenen Massen des Gravitons entsprechen. Die Vertizes erhalten aufgrund der Impulserhaltung wie \"{u}blich eine $\delta$-Funktion. \"{U}ber den Impuls des virtuellen Gravitons muss integriert werden, da er keine direkte Messgr\"{o}\ss e darstellt. Damit erh\"{a}lt man folgenden Ausdruck f\"{u}r die S-Matrix

\begin{eqnarray}
S(g(k_1,g_1)+g(k_2,g_2) \rightarrow Z(l_1,Z_1)+Z(l_2,Z_2))\nonumber\\
=\frac{1}{(2\pi)^4}\int d^{4}k \frac{g^1_{\alpha a}}{(2\pi)^\frac{3}{2}\sqrt{k_{10}}} \frac{g_{2\beta b}}{(2\pi)^\frac{3}{2}\sqrt{k_{20}}}
\left(-\frac{i}{\bar M_P}\delta^{ab}\left[W^{\mu\nu\alpha\beta}+W^{\nu\mu\alpha\beta}\right] \right)\nonumber\\
\cdot (2\pi)^4\delta^4(k_1+k_2-k)
\sum_n \frac{iP_{\mu\nu\rho\sigma}}{k^2-m^2}
(2\pi)^4\delta^4(k-l_1-l_2)\nonumber\\
\cdot \left(-\frac{i}{\bar M_P}\delta^{cd}\left[W^{\rho\sigma\gamma\delta}+W^{\sigma\rho\gamma\delta}\right] \right)
\frac{Z_{1\gamma c}}{(2\pi)^\frac{3}{2}\sqrt{l_{10}}} \frac{Z_{2\delta d}}{(2\pi)^\frac{3}{2}\sqrt{l_{20}}}.
\label{S-Matrix}
\end{eqnarray}
Hierbei sind $W_{\mu\nu\alpha\beta}$ und $P_{\mu\nu\rho\sigma}$ gem\"{a}\ss\ ($\ref{VertexVbVbG}$) bzw. ($\ref{PolarisationstensorGraviton}$) wie folgt definiert

\begin{eqnarray}
W_{\mu\nu\alpha\beta}&=&\frac{1}{2} \eta_{\mu\nu}(k_{1\beta}k_{2\alpha}-k_1\cdot k_2 \eta_{\alpha\beta})+\eta_{\alpha\beta} k_{1\mu} k_{2\nu}\nonumber\\
&&+\eta_{\mu\alpha}(k_1 \cdot k_2 \eta_{\nu\beta}-k_{1 \beta} k_{2 \nu})-\eta_{\mu\beta} k_{1 \nu} k_{2 \alpha},
\label{W-Ausdruck}
\end{eqnarray}

\begin{eqnarray}
P_{\mu\nu\rho\sigma}&=&\frac{1}{2}(\eta_{\mu\alpha}\eta_{\nu\beta}+\eta_{\mu\beta}\eta_{\nu\alpha}-\eta_{\mu\nu}\eta_{\alpha\beta})\nonumber\\
&&-\frac{1}{2m^2}(\eta_{\mu\alpha}k_\nu k_\beta+\eta_{\nu\beta}k_\mu k_\alpha+\eta_{\mu\beta}k_\nu k_\alpha+\eta_{\nu\alpha}k_\mu k_\beta)\nonumber\\
&&+\frac{1}{6}\left(\eta_{\mu\nu}+\frac{2}{m^2}k_\mu k_\nu\right)\left(\eta_{\alpha\beta}+\frac{2}{m^2}k_\alpha k_\beta \right).
\label{P-Ausdruck}
\end{eqnarray}
Wichtig ist, dass der Tensor $P_{\mu\nu\rho\sigma}$ "`on mass-shell"' ist und damit $p^2=m^2$ gilt.
Der Ausdruck f\"{u}r die S-Matrix ($\ref{S-Matrix}$) kann umformuliert werden zu 

\begin{eqnarray}
S&=&\frac{-i}{(2\pi)^2 \sqrt{2k_{10}} \sqrt{2k_{20}} \sqrt{2l_{10}} \sqrt{2l_{20}}} \frac{1}{\bar M_P^2} \sum_n \frac{1}{p^2-m^2} \nonumber\\
&&\cdot g_{1\alpha} g_{2\beta} \left[W^{\mu\nu\alpha\beta}+W^{\nu\mu\alpha\beta}\right] P_{\mu\nu\rho\sigma} \left[W^{\rho\sigma\gamma\delta}+W^{\sigma\rho\gamma\delta}\right] Z_{1\gamma} Z_{2\delta} \nonumber\\
&&\cdot \delta^4(p-l_1-l_2).
\end{eqnarray}
Es wurde unter anderem \"{u}ber die erste Deltafunktion integriert und p als $p=k^1+k^2$ definiert.
Desweiteren soll nun der folgende Zusammenhang zwischen der S-Matrix und der Feynmanamplitude M verwendet werden

\begin{equation}
S=-2\pi iM\delta^4(p-l_1-l_2).
\label{SMatrixFeynman}
\end{equation}
Damit erh\"{a}lt man f\"{u}r die Feynmanamplitude M folgenden Ausdruck

\begin{eqnarray}
M&=&\frac{1}{(2\pi)^3 \sqrt{2k_{10}} \sqrt{2k_{20}} \sqrt{2l_{10}} \sqrt{2l_{20}}} \frac{1}{\bar M_P^2} \sum_n \frac{1}{p^2-m^2}\nonumber\\
&&\cdot g_{1\alpha} g_{2\beta} [W^{\mu\nu\alpha\beta}+W^{\nu\mu\alpha\beta}]
P_{\mu\nu\rho\sigma}
\left[W^{\rho\sigma\gamma\delta}+W^{\sigma\rho\gamma\delta}\right] Z_{1\gamma} Z_{2\delta}.
\end{eqnarray}
Wenn man A wie folgt definiert $A=\frac{1}{(2\pi)^3 \sqrt{2k_{10}} \sqrt{2k_{20}} \sqrt{2l_{10}} \sqrt{2l_{20}}}\sum_n \frac{1}{\bar M_P^2} \frac{1}{p^2-m^2}$, erh\"{a}lt man

\begin{equation}
M=A \cdot g_{1\alpha} g_{2\beta} \left[W^{\mu\nu\alpha\beta}+W^{\nu\mu\alpha\beta}\right]
P_{\mu\nu\rho\sigma}
\left[W^{\rho\sigma\gamma\delta}+W^{\sigma\rho\gamma\delta}\right] Z_{1\gamma} Z_{2\delta}.
\end{equation}
Da der Gravitonpropagator einerseits symmetrisch bez\"{u}glich der Indizes  $\mu$ and $\nu$ und andererseits symmetrisch bez\"{u}glich der Indizes $\rho$ and $\sigma$ ist, kann man obigen Ausdruck wie folgt schreiben   

\begin{equation}
M=4A \cdot g_{1\alpha} g_{2\beta} \cdot W^{\mu\nu\alpha\beta} \cdot P_{\mu\nu\rho\sigma} \cdot W^{\rho\sigma\gamma\delta} \cdot Z_{1\gamma} Z_{2\delta}.
\label{Feynman-Amplitude1}
\end{equation}
Indem man die Ausdr\"{u}cke f\"{u}r $W_{\mu\nu\alpha\beta}$ und $P_{\mu\nu\rho\sigma}$ aus ($\ref{W-Ausdruck}$) und ($\ref{P-Ausdruck}$)
in ($\ref{Feynman-Amplitude1}$) einsetzt, erh\"{a}lt man

\begin{eqnarray}
M&=&4A \cdot g_{1\alpha} g_{2\beta}
\cdot[\frac{1}{2}\eta^{\mu\nu}(k_1^\beta k_2\alpha-k_1 k_2\eta^{\alpha\beta})+\eta^{\alpha\beta}k_1^\mu k_2^\nu\nonumber\\
&&+\eta^{\mu\alpha}(k_1 k_2\eta^{\nu\beta}-k_1^\beta k_2^\nu-\eta^{\mu\beta}k_1^\nu k_2^\alpha]  \nonumber\\
&&\cdot[\frac{1}{2}(\eta_{\mu\rho}\eta_{\nu\sigma}+\eta_{\mu\sigma}\eta_{\nu\rho}+\eta_{\mu\nu}\eta_{\rho\sigma})\nonumber\\
&&-\frac{1}{2m^2}(\eta_{\mu\rho}p_{\nu}p_{\sigma}+\eta_{\nu\sigma}p_{\mu}p_{\rho}+\eta_{\mu\sigma}p_{\nu}p_{\rho}+\eta_{\nu\rho}p_{\mu}p_{\sigma})\nonumber\\
&&+\frac{1}{6}\left(\eta_{\mu\nu}+\frac{2}{m^2}p_{\mu}p_{\nu}\right)\left(\eta_{\rho\sigma}+\frac{2}{m^2}p_{\rho}p_{\sigma}\right)]\nonumber\\
&&\cdot[\frac{1}{2}\eta^{\rho\sigma}(k_1^\delta k_2\gamma-k_{1}k_{2}\eta^{\gamma\delta})+\eta^{\gamma\delta}k_1^\rho k_2^\sigma\nonumber\\
&&+\eta^{\rho\gamma}(k_1 k_2 \eta^{\sigma\delta}-k_1^\delta k_2^\sigma)-\eta^{\rho\delta}k_1^\sigma k_2^\gamma]\cdot Z_{1\gamma} Z_{2\delta}.
\end{eqnarray}
Die Polarisationsvektoren f\"{u}r die Gluonen und Z-Teilchen stehen immer senkrecht auf dem jeweiligen Impuls und erf\"{u}llen damit folgende Relationen 

\begin{equation}
k_1^\mu g_1^\mu=0\quad,\quad k_2^\mu g_2^\mu=0\quad,\quad l_1^\mu Z_{1\mu}=0\quad,\quad l_2^\mu Z_{2\mu}=0.
\end{equation}
Au\ss erdem soll die Rechnung von nun an im Schwerpunktsystem betrachtet werden, da sich hierdurch vieles vereinfacht. Dies bedeutet, dass gilt

\begin{equation}
\vec k_1=-\vec k_2\quad,\quad p=0\quad,\quad \vec l_1=-\vec l_2.
\end{equation}
(Die fettgedruckten Gr\"{o}\ss en sollen hier den r\"{a}umlichen Anteil der Vierervektoren bezeichnen.)
Damit erh\"{a}lt man f\"{u}r die Feynmanamplitude folgenden Ausdruck

\begin{eqnarray}
M&=&4A \cdot g_{1\alpha} g_{2\beta} \cdot
\left[-\frac{1}{2}\eta^{\mu\nu}(k_1 \cdot k_2)(g_1 \cdot g_2)+(g_1 \cdot g_2)(k_1\mu k_2\nu)+(k_1 \cdot k_2)g_1\mu g_2^\nu\right]\nonumber\\
&&\cdot[\frac{1}{2}(\eta_{\mu\rho}\eta_{\nu\sigma}+\eta_{\mu\sigma}\eta_{\nu\rho})-\frac{1}{3}\eta_{\mu\nu}\eta_{\rho\sigma}\nonumber\\
&&-\frac{1}{2m^2}(\eta_{\mu\rho}p_{\nu}p_{\sigma}+\eta_{\nu\sigma}p_{\mu}p_{\rho}+\eta_{\mu\sigma}p_{\nu}p_{\rho}+\eta_{\nu\rho}p_{\mu}p_{\sigma})\nonumber\\
&&+\frac{1}{3m^2}\eta_{\rho\sigma}p_\mu p_\nu+\frac{1}{3m^2}\eta_{\mu\nu}p_\rho p_\sigma+\frac{2}{3m^4} p_\mu p_\nu p_\rho p_\sigma]\nonumber\\
&&\cdot\left[-\frac{1}{2}\eta^{\rho\sigma}(l_1 \cdot l_2)(Z_1 \cdot Z_2)+(Z_1 \cdot Z_2)(l_1^\rho l_2^\sigma)+(l_1 \cdot l_2) Z_1^\rho Z_2^\sigma\right].
\label{Feynman-Amplitude2}
\end{eqnarray}

\section[Wirkungsquerschnitt f\"{u}r Gluonen]{Wirkungsquerschnitt f\"{u}r die ZZ-Produktion durch Gluonen}

\subsection{Von der Feynmanamplitude zum Wirkungsquerschnitt}

Der differentielle Wirkungsquerschnitt f\"{u}r den \"{U}bergang von einem Zustand $\alpha$ in einen Zustand $\beta$ hat allgemein folgende Gestalt

\begin{equation}
d\sigma(\alpha \rightarrow \beta)=(2 \pi)^4 u_\alpha^{-1} \sum_\sigma |M|^2 \delta^4(p_\beta-p_\alpha) d\beta,
\end{equation}
wobei gilt
\begin{equation}
u_\alpha=\frac{\sqrt{(p_1 \cdot p_2)^2-m_1^2 m_2^2}}{E_1 E_2}.
\end{equation}
(siehe \cite{WeinbergQTF1})\\
In dem Fall, der hier betrachtet wird, bedeutet dies

\begin{equation}
d \sigma=\frac{(2 \pi)^4}{2} \sum_\sigma |M|^2 \delta^4 (l_1+l_2-k_1-k_2)d^4 l_1 d^4 l_2.
\end{equation}
Durch Aufspaltung der $\delta$-Funktion und Ausnutzung der relativistischen Energie-Impuls-Beziehung erh\"{a}lt man

\begin{eqnarray}
d \sigma&=&\frac{(2 \pi)^4}{2} \sum_\sigma |M|^2 \delta^3 (\vec l_1+\vec l_2-(\vec k_1+\vec k_2))\nonumber\\
&&\cdot\delta(\sqrt{|\vec l_1|^2+m_Z^2}+\sqrt{|\vec l_2|^2 +m_Z^2}-2E) d^3 l_1 d^3 l_2.
\end{eqnarray}
Im Schwerpunktsystem gilt

\begin{equation}
\vec k_1+\vec k_2=0.
\end{equation}
Integrieren \"{u}ber $d^3 l_1$ liefert damit

\begin{equation}
d \sigma=\frac{(2 \pi)^4}{2} \sum_\sigma |M|^2 \delta(\sqrt{|\vec l_2|^2+m_Z^2}-E)d^3 l_2.
\end{equation}
Das ist gleichbedeutend mit

\begin{equation}
d \sigma=\frac{(2 \pi)^4}{2} \sum_\sigma |M|^2 \delta(\sqrt{|\vec l_2|^2+m_Z^2}-E)|\vec l_2|^2 sin(\theta) d|\vec l_2|d \Omega.
\end{equation}
Integrieren \"{u}ber $d|\vec l_2|$ liefert schlie\ss lich

\begin{equation}
d \sigma=\frac{(2 \pi)^4}{2} \sum_\sigma |M|^2 E \sqrt{E^2-m_Z^2} sin(\theta) d \Omega.
\end{equation}
Wenn man nun noch \"{u}ber den Azimutalwinkel integriert, bekommt man folgenden Ausdruck

\begin{equation}
d \sigma=\frac{(2 \pi)^5}{2} \sum_\sigma |M|^2 E \sqrt{E^2-m_Z^2} sin(\theta) d \theta.
\end{equation}
Der Wirkungsquerschnitt enth\"{a}lt nicht die Feynmanamplitude selbst, sondern das Betragsquadrat.
Nach Ausmultiplizieren des Ausdruckes f\"{u}r die Feynmanamplitude ($\ref{Feynman-Amplitude2}$) wird das im Wirkungsquerschnitt auftauchende Betragsquadrat der Feynmanamplitude gebildet. Da keine spezielle Polarisation der Gluonen angenommen wird und die Polarisation der Z-Teilchen keine Rolle spielt, muss \"{u}ber alle m\"{o}glichen Polarisationen summiert werden. Eine konkrete Berechnung von $\sum_\sigma |M|^2$ unter Ausnutzung der Relationen

\begin{equation}
\sum_\sigma g_\mu g_\nu^*=-\eta_{\mu\nu},
\end{equation} 
und

\begin{equation}
\sum_\sigma Z_\mu Z_\nu^*=\left(-\eta_{\mu\nu}+\frac{l_\mu l_\nu}{m_Z^2}\right)
\end{equation}
f\"{u}r die Polarisationsvektoren der masselosen Gluonen, sowie der massebehafteten Z-Teilchen, f\"{u}hrt nach Integration \"{u}ber den Streuwinkel $\theta$ schlie\ss lich auf folgende Ausdruck f\"{u}r den absoluten Wirkungsquerschnitt $\sigma$

\begin{equation}
\sigma=\frac{D^2}{\bar M_p^4}\frac{E \sqrt{E^2-m_Z^2}(3552 E^8-7400 E^6 m_Z^2+4977 E^4 m_Z^4-1257 E^2 m_Z^6+98 m_Z^8)}{30 \pi m_Z^4}.
\end{equation}
Hierbei ist D wie folgt definiert 

\begin{equation}
D=\sum_n \frac{1}{p^2-m^2},
\label{Kaluza-Klein-Summierung}
\end{equation}
wobei die Masse den Anregungen in den zus\"{a}tzlichen Dimensionen entspricht.
Die Energie bezieht sich nat\"{u}rlich auf das Schwerpunktsystem. Es ist daher sinnvoll, den Wirkungsquerschnitt mit Hilfe der lorentzinvarianten Mandelstamvariable s audszudr\"{u}cken. In Abh\"{a}ngigkeit von s ergibt sich folgender Ausdruck

\begin{eqnarray}
\sigma=\frac{D^2}{\bar M_p^4}\frac{s}{2}\sqrt{\frac{s}{4}-m_Z^2}
\ \ \ \ \ \ \ \ \ \ \ \ \ \ \ \ \ \ \ \ \ \ \ \ \ \ \ \ \nonumber\\
\cdot\frac{(13.875 s^4-115.625 s^3 m_Z^2+311.0625 s^2 m_Z^4-314.25 s m_Z^6+98 m_Z^8)}
{30 \pi m_Z^4}.
\label{WirkungsquerschnittsD}
\end{eqnarray}

\subsection{Summierung \"{u}ber die Kaluza-Klein-Anregungen}

Die Summation \"{u}ber die Kaluza-Klein-Zust\"{a}nde in (\ref{Kaluza-Klein-Summierung}) kann umformuliert werden, indem man die Tatsache ber\"{u}cksichtigt, dass die Masse $m$ des Gravitons durch die Kaluza-Klein-Anregungen \"{u}ber den Zusammenhang

\begin{equation}
m^2=\hat n^2=\sum_{j=1}^{\delta}|\hat n_j|^2
\label{Gravitonmasse_Anregungen}
\end{equation}
bestimmt ist. In \cite{Giudice:1998ck} ist die Summierung über die Kaluza-Klein Anregungen   ausgef\"{u}hrt. Dieser Darstellung soll hier gefolgt werden.
Die Summe \"{u}ber die einzelnen Anregungen kann n\"{a}herungsweise durch ein Integral \"{u}ber das Massenquadrat ersetzt werden, wobei eine Dichteverteilung bez\"{u}glich des Massenquadrats auftaucht, da unterschiedliche Massen unterschiedliche Gewichtungen haben. Wenn man Gleichung ($\ref{Gravitonmasse_Anregungen}$) betrachtet, sieht man, dass die Gewichtung des Massenquadrates der Zahl der Anregungskombinationen entspricht, die auf ein bestimmtes $\hat n^2$ f\"{u}hren. Diese entspricht im Limes eines gro\ss en $\hat n^2$ der Oberfl\"{a}che einer n-dimensionalen Sph\"{a}re. Man erh\"{a}lt den folgenden Ausdruck 

\begin{equation}
D=\sum_n \frac{1}{p^2-m^2}=\int_0^{\infty} dm^2 \frac{\rho(m)}{s-m^2} 
\end{equation}
mit

\begin{equation}
\rho(m)=\frac{R^n m^{(\delta-2)}}{(4\pi)^{\frac{n}{2}}\Gamma (\frac{\delta}{2})},
\end{equation}
wobei $\Gamma$ die \"{u}bliche Gammafunktion beschreibt.
Es wird nun das Verfahren dimensionaler Regularisierung angewandt. Unter Verwendung des
niedrigst dimensionalen Beitrages ($c_1=1$ and $c_i=0$ for $i \neq 1$), bei Wahl einer Regularisierungsskala in der Gr\"{o}\ss enordnung der neuen Massenskala $\Lambda=M_D$ und der
Relation zwischen der $(4+\delta)$-dimensionalen Planckmasse und der gew\"{o}hnlichen Planckmasse $M_P^2=8\pi M_D^{(2+\delta)}R^\delta$ wird man auf folgenden Ausdruck geführt
\begin{equation}
\sum_n \frac{1}{s-m_n^2}\approx\frac{\bar M_P^2\pi^{\frac{\delta}{2}}}{\Gamma (\frac{\delta}{2})M_D^4} \quad.
\label{KKSummierung}
\end{equation}
ein andere Näherungsverfahren für die Kaluza-Klein Summierung findet man in \cite{Han:1998sg}. Wir beschr\"{a}nken uns hier jedoch auf (\ref{KKSummierung}).\\
Das f\"{u}hrt schlie\ss lich auf den folgenden totalen Wirkungsquerschnitt 

\begin{equation}
\sigma(gg \rightarrow ZZ)=\frac{\pi^\delta \sqrt{\frac{\hat s}{4}}\sqrt{\frac{\hat s}{4}-m_Z^2}Z}{\Gamma^2 (\frac{\delta}{2}) M_D^8 30 \pi m_Z^4} 
\label{ADDWirkungsquerschnitt}
\end{equation}
mit

\begin{eqnarray}
Z&=&13.875 \hat s^4-115.625 \hat s^3 m_Z^2+311.0625 \hat s^2 m_Z^4\nonumber\\
&&-314.250 \hat s m_Z^6+98 m_Z^8\quad.
\end{eqnarray}
Das ist der totale Wirkungsquerschnitt als Funktion der Zahl der Extra Dimensionen $\delta$ und der Planckmasse $M_D$ in $4+\delta$ Dimensionen.

\section[Wirkungsquerschnitt f\"{u}r Protonen]{Wirkungsquerschnitt nach Faltung \"{u}ber die Verteilungsfunktionen im Proton und Vergleich mit dem Standardmodell} 

\subsection{Faltung \"{u}ber die Partonenverteilungsfunktionen}

Um nun den Wirkungsquerschnitt f\"{u}r den Proton-Proton-Prozess zu erhalten, m\"{u}ssen die Wirkungsquerschnitte f\"{u}r die einzelnen Partonenprozesse mit der entsprechenden Verteilungsfunktion f\"{u}r die Gluonen bzw. Quarks innerhalb des Protons gefaltet werden.
Die Formel f\"{u}r den Proton-Proton-Wirkungsquerschnitt bei gegebenen Partonenwirkungsquerschnitten lautet wie folgt

\begin{eqnarray}
\sigma(p^+(k_1) p^-(k_2)\rightarrow Z(l_1)Z(l_2))\ \ \ \ \ \ \ \ \ \ \ \ \ \ \ \ \ \ \ \ \ \ \ \ \ \ \ \ \ \ \ \ \ \ \ \ \ \ \ \ \ &&\nonumber\\
=\int_0^1 dx_2 \int_0^1 dx_1 \sum_f f_f(x_1) f_f(x_2)\sigma(q(x_1 k_1)\bar q(x_2 k_2)\rightarrow Z(l_1)Z(l_2)),&&
\end{eqnarray}
(siehe \cite{PeskinSchroeder})
wobei die $f_f$ die Partonenverteilungsfunktionen darstellen. Die Summierung bezieht sich auf die verschiedenen Sorten von Partonen, also die verschiedenen Quarksorten sowie die Gluonen. Da im obigen Ausdruck \"{u}ber die Beitr\"{a}ge der verschiedenen Partonen summiert wird, ist der durch die Gluonen bedingte Wirkungsquerschnitt und damit der Wirkungsquerschnitt der durch den Gravitonenaustausch hinzukommt weiterhin unbh\"{a}ngig von dem der Quarks. F\"{u}r die Differenz zum Standardmodellwirkungsquerschnitt gilt damit 

\begin{eqnarray}
\Delta\sigma(p^+(k_1) p^-(k_2)\rightarrow Z(l_1)Z(l_2))\ \ \ \ \ \ \ \ \ \ \ \ \ \ \ \ \ \ \ \ \ \ \ \ \ \ \ \ \ \ \ \ \ \ \ \ &&\nonumber\\
=\int_0^1 dx_2 \int_0^1 dx_1 f_g(x_1) f_g(x_2)\sigma(g(x_1 k_1) g(x_2 k_2)\rightarrow Z(l_1)Z(l_2)).&&
\end{eqnarray}

\subsection{ZZ-Produktion im Standardmodell}

\begin{figure}[h]
\centering
\epsfig{figure=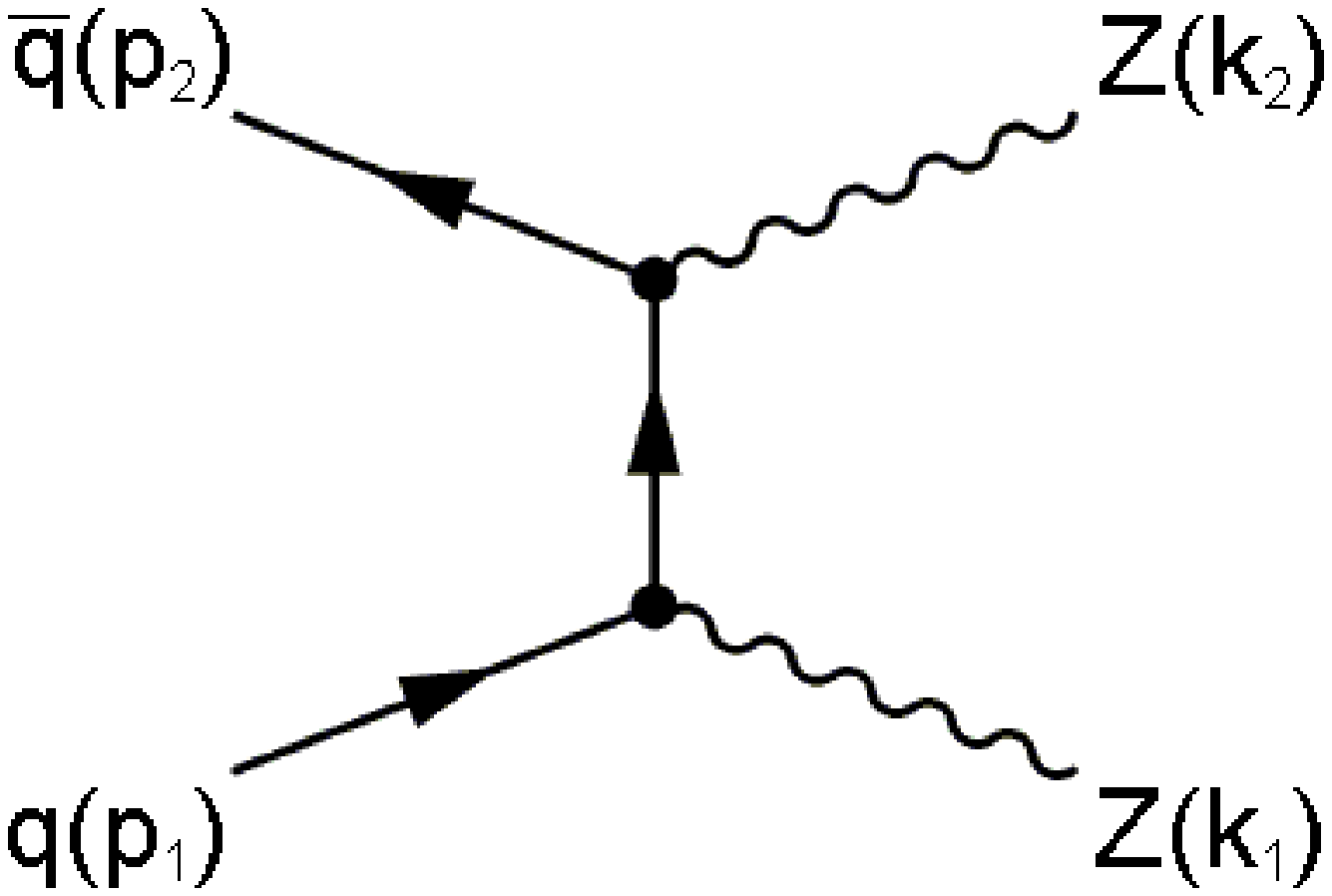,width=7cm}
\caption{\label{Graph3} Quarks - Z-Teilchen}
\end{figure}

\begin{figure}[h]
\centering
\epsfig{figure=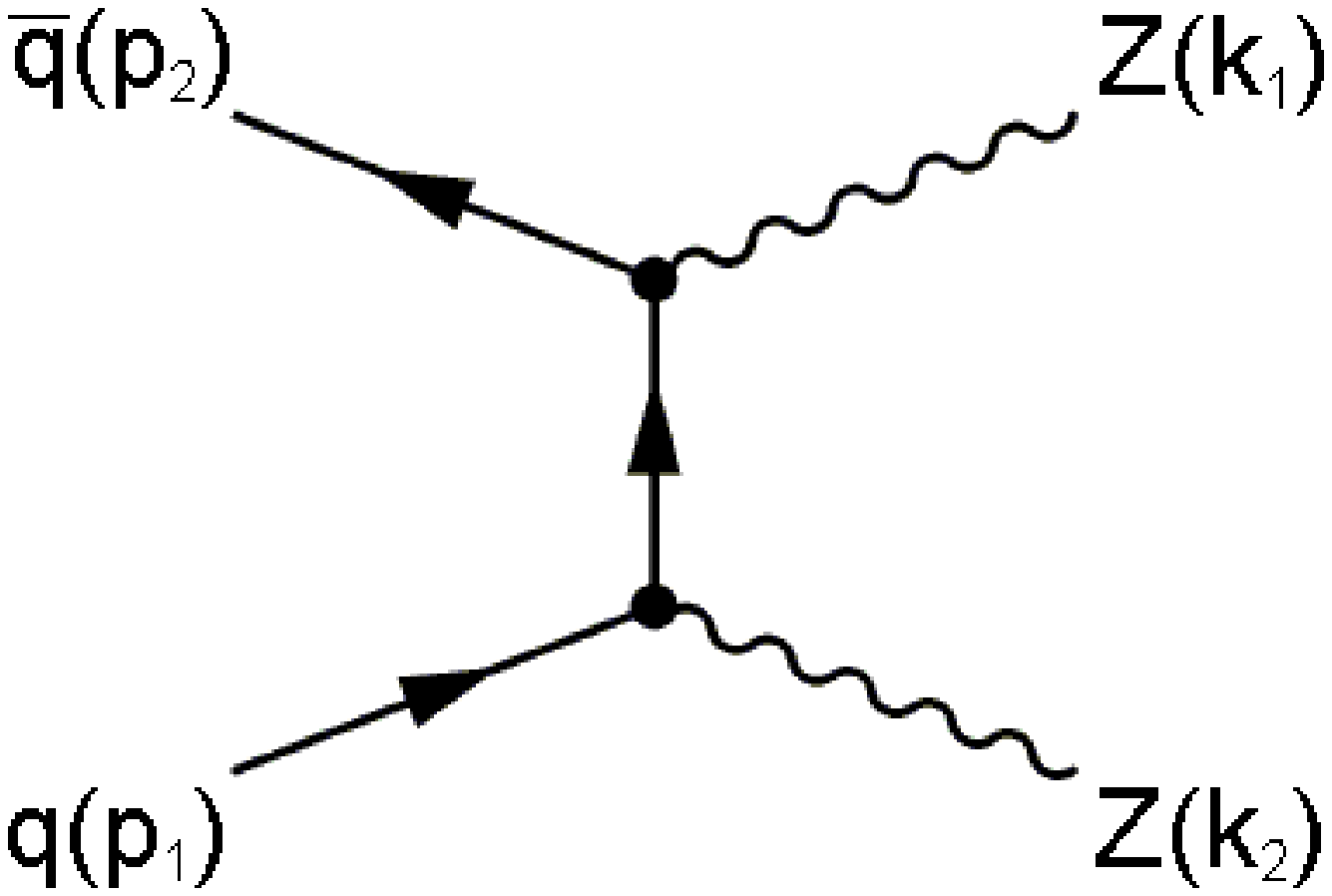,width=7cm}
\caption{\label{Graph4} Quarks - Z-Teilchen}
\end{figure}

\begin{figure}[h]
\centering
\epsfig{figure=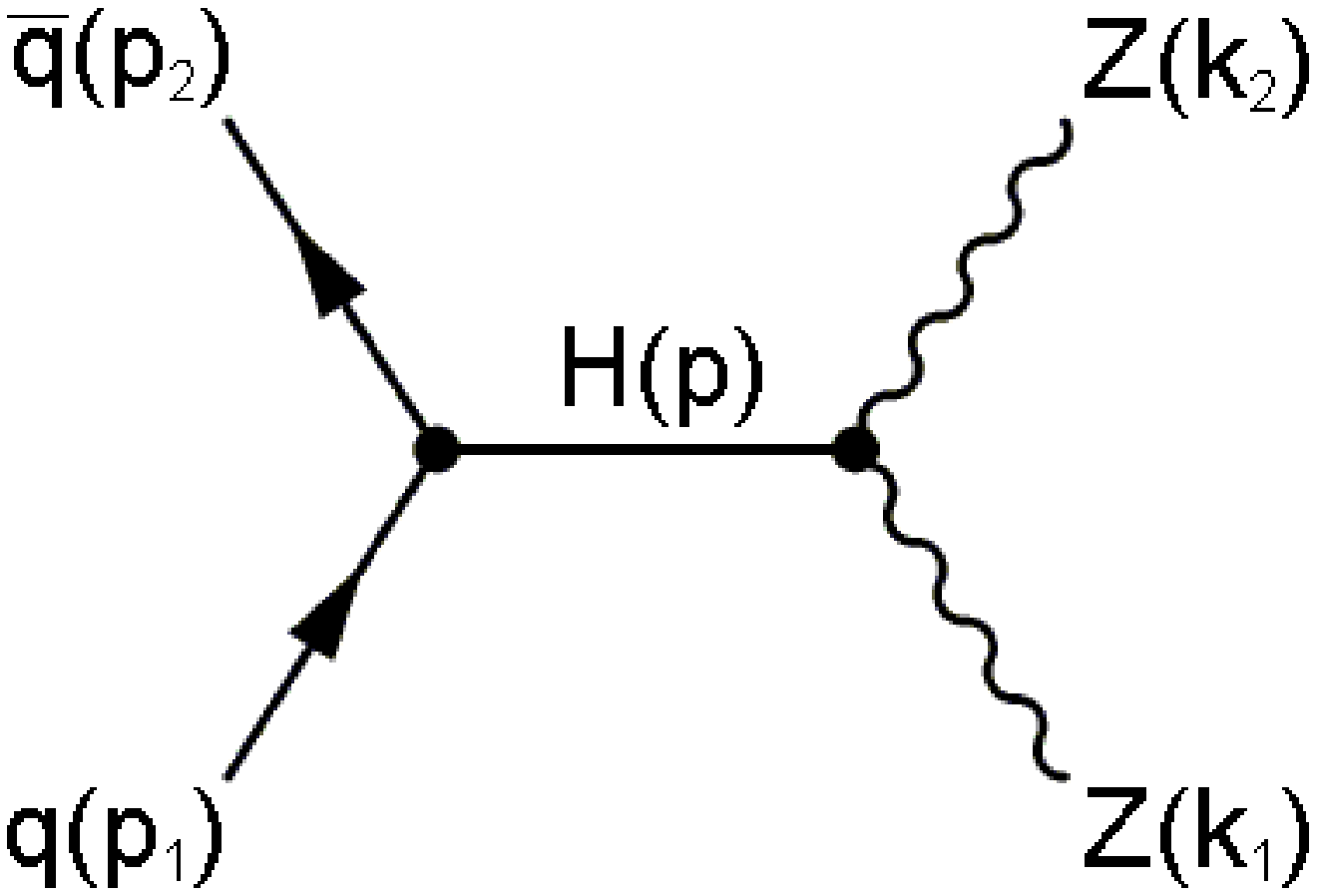,width=7cm}
\caption{\label{Graph5} Quarks - Higgs - Z-Teilchen}
\end{figure}

Um nun einen Vergleich des zus\"{a}tzlichen Wirkungsquerschnitts im ADD-Modell 
zu dem des Standardmodells herzustellen, muss analog der obigen Faltung \"{u}ber die Wirkungsquerschnitte f\"{u}r die Einzelprozesse innerhalb des Standardmodells gefaltet werden. Der \"{U}bergang von zwei Gluonen in zwei Z-Teilchen ist unm\"{o}glich, da Gluonen als Austauschteilchen der starken Wechselwirkung nur mit sich selbst also nicht schwach wechselwirken, worauf bereits hingewiesen wurde. Daher ergeben sich nur Wirkungsquerschnitte f\"{u}r den \"{U}bergang eines Quark-Antiquark-Paares in ein ZZ-Paar. Die folgenden Prozesse
(Graphen \ref{Graph3}, \ref{Graph4} und \ref{Graph5}) liefern im Standardmodell in erster Ordnung einen Beitrag zur ZZ-Produktion

Die Graphen (\ref{Graph3}),(\ref{Graph4}) und (\ref{Graph5}) beziehen sich jeweils auf ein Quark-Antiquark-Paar jeder Sorte.
Der Prozess, welcher dem Graphen (\ref{Graph3}) entspricht, ist aufgrund der gro\ss en Masse des Higgsteilchens zu vernachl\"{a}ssigen. Durch Auswertung dieser Feynmangraphen gelangt man zu den Wirkungsquerschnitten f\"{u}r das up-quark/charme-quark und das down-quark/strange-quark
Da aufgrund des vorausgesetzten hohen Impulses die Massen der Quarks in der relativistischen Energie-Impuls-Beziehung vernachl\"{a}ssigbar sind, kann von folgender Relation $E^2 \approx p^2$ ausgegangen werden. Dies bedeutet, dass der Wirkungsquerschnitt des charme-Quarks dem des up-quarks und der des strange-quarks dem des down-quarks entspricht. Die Wirkungsquerschnitte f\"{u}r das bottom- und das top-Quark k\"{o}nnen aufgrund ihrer riesigen Masse und damit viel geringeren Produktionswahrscheinlichkeit innerhalb des Protons vernachl\"{a}ssigt werden.
Da eine Raumzeit mit zwei oder weniger zus\"{a}tzlichen Dimensionen im Rahmen des ADD Modells empirisch ausgeschlossen ist, wird hier von drei oder mehr Dimensionen ausgegangen. Au\ss erdem wird im Allgemeinen angenommen, dass die modifizierte Planckmasse $M_D$ mindestens einen Wert von 1 TeV annimmt. 

\subsection{Ergebnisse und Diskussion}

Um den Wirkungsquerschnitt für den Proton-Proton Prozess zu erlangen, integriert man  \"{u}ber die Partonenverteilungsfunktionen gem\"{a}\ss dem vorletzten Unterkapitel. 
Die Partonenverteilungsfunktionen (gegeben in \cite{CTEQ6}) werden  
f\"{u}r den Prozess des Graphen (\ref{Graph1}) bei einer Skala von $Q=\sqrt{\hat s}$ 
und f\"{u}r die Prozesse der Graphen (\ref{Graph3}),(\ref{Graph4}) und (\ref{Graph5}) bei einer Skala von $Q=m_Z$ ausgewertet.  

\begin{figure}[ht]
\centering
\epsfig{figure=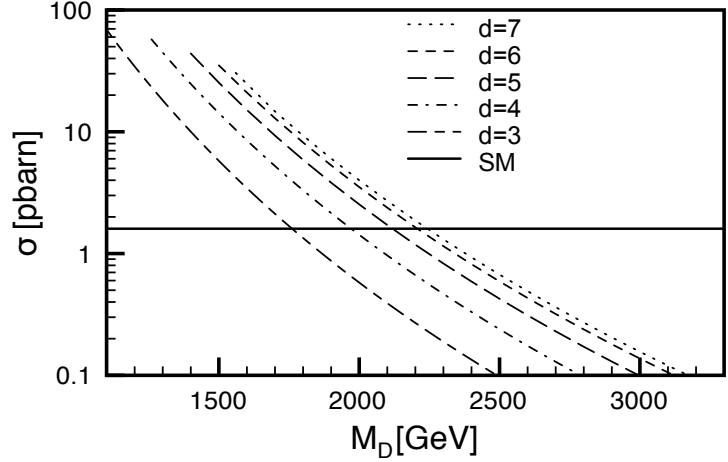,width=10cm}
\caption{\label{sig2}
Vergleich zwischen dem totalen Wirkungsquerschnitt der $p\bar{p}\rightarrow ZZ$-Produktion.  
f\"{u}r das Standardmodell \cite{Campbell:1999ah}, das Fermilab \cite{Acosta:2005pq} 
limit und das ADD model ($d=3,4,5,6,7$) f\"{u}r $\sqrt{s}=2000$~GeV
in Abh\"{a}ngigkeit von $M_D$.}
\end{figure}

In Graph (\ref{sig2}) ist der totale Wirkungsquerchnitt bei der Fermilabenergie von $2000$~GeV als eine Funktion der fundamentalen Massenskala $M_D$ aufgetragen
\cite{Giudice:1998ck}).
Man sieht hier, dass die Gravitonenvermittlung gem\"{a}\ss der in Kapitel 3 beschriebenen 
Theorie bei $M_D > 2500$~GeV  keinen beobachtbaren Einfluss auf die ZZ-Produktionsrate \cite{Acosta:2005pq} am Fermilab hat. 

\begin{figure}[ht]
\centering
\epsfig{figure=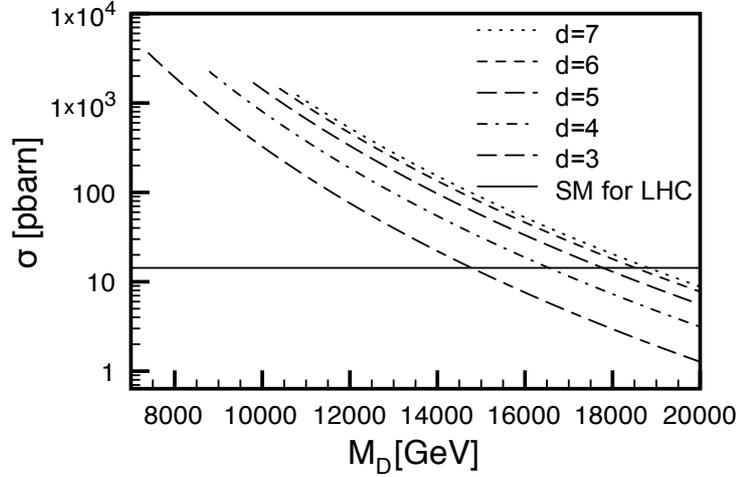,width=10cm}
\caption{\label{sig14}
Vergleich zweischen dem totalen $pp\rightarrow ZZ$ Produktions-Wirkungsquerschnitt
f\"{u}r das Standardmodell \cite{Ohnemus:1995gb} und das ADD model ($d=3,4,5,6,7$) f\"{u}r $\sqrt{s}=14000$~GeV in Abh\"{a}ngigkeit von $M_D$.}
\end{figure}

\begin{figure}[ht]
\centering
\epsfig{figure=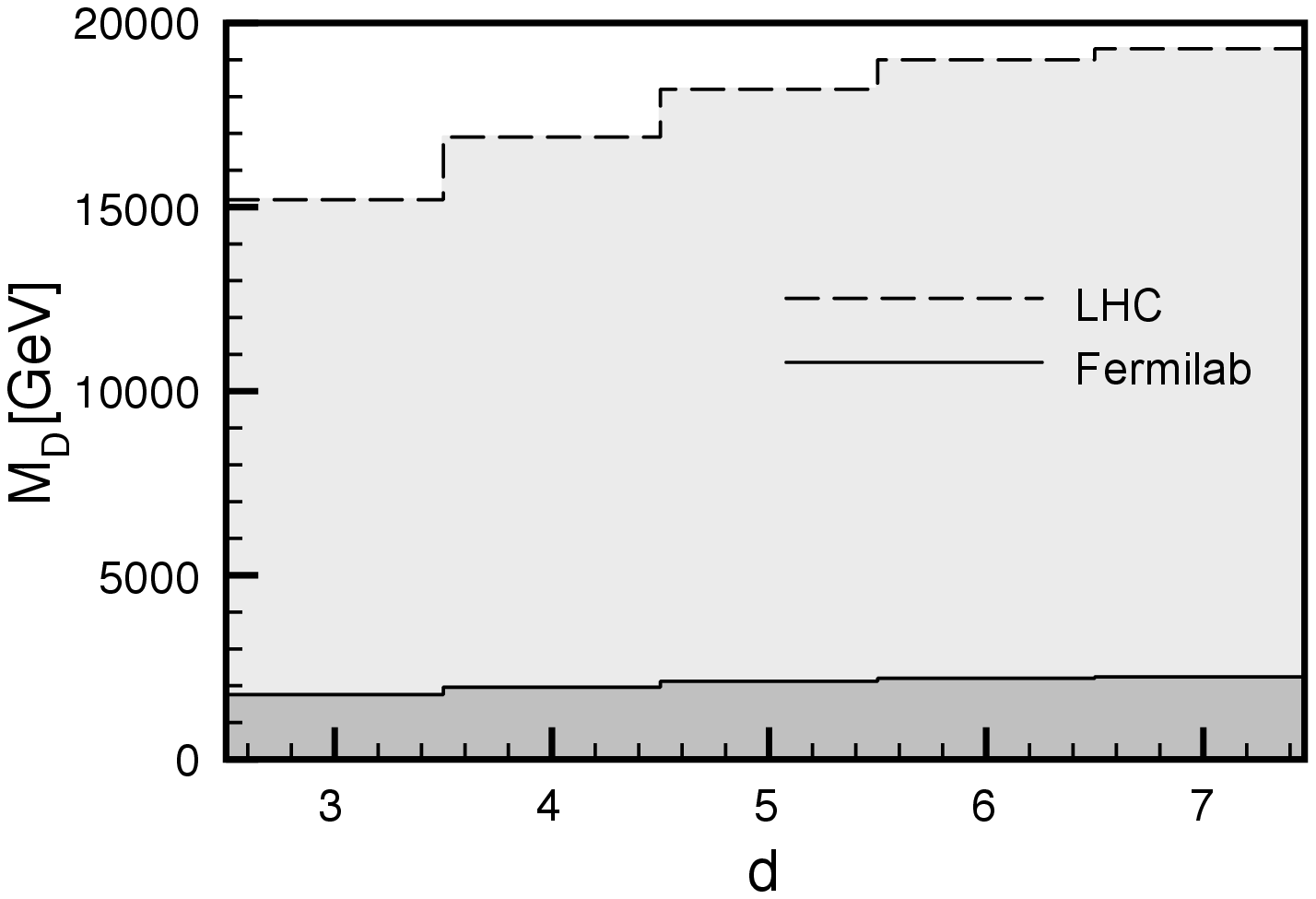,width=10cm}
\caption{\label{Mdschnitt}
Der zug\"{a}ngliche ADD model Parameter Raum (das bedeutet der Bereich, indem der zus\"{a}tzliche Wirkungsquerschnitt mindestens so gro\ss\ ist wie der Wirkungsquerschnitt im Standardmodell) im ZZ-Kanal f\"{u}r das Tevatron und den LHC.}
\end{figure}

Die gleiche Analyse wird in Graph (\ref{sig14}) für eine Proton-Proton-Reaktion bei einer Energie von $14000$~GeV gezeigt, wie sie am LHC verfügbar ist.
Hier k\"{o}nnte die drastische Differez zwischen der ZZ-Bosonen Rate des Standardmodells
und der gravitonenvermittelten ZZ-Bosonen Rate eine Beobachtung von Effekten gro\ss er zus\"{a}tzlicher Dimensionen sogar f\"{u}r eine fundamentale Skala von $M_D\sim 18000$~GeV erlauben.
Graph (\ref{sig14}) zeigt, dass das ADD-Resultat bei $\sqrt{s}$=14000 GeV
die Standardmodellvorhersage f\"{u}r kleines $M_D\ll \sqrt{s}$ \"{u}bertrifft. 
Dies k\"{o}nnte Ausdruck der Tatsache sein, dass die Regularisierungsmethode und der Zugang
der st\"{o}rungstheoretischen Quantenfeldtheorie in diesem Bereich ihre G\"{u}ltigkeit verlieren.
Deshalb ist es sinnvoll, die in den Graphen (\ref{sig2}) und (\ref{sig14}) aufgezeichneten Resultate nur nahe im Bereich ihrer G\"{u}ltigkeit zu verwenden $\sqrt{s}\sim M_D$.
Dies erlaubt eine Aussage dar\"{u}ber, ab welchem Wert f\"{u}r die modifizierte Planckmasse $M_D$ experimentelle Abweichungen von der ZZ-Produktionsrate des Standardmodells am LHC  \cite{Ohnemus:1995gb} erwartet werden sollten.
In Graph (\ref{Mdschnitt}) ist der \"{u}berpr\"{u}fbare Parameterraum sowohl f\"{u}r den LHC
alsauch das Tevatron aufgef\"{u}hrt.
F\"{u}r die experimentelle Beobachtung m\"{u}sste man nach zwei hochenergetischen und korrelierten Leptonenpaaren im Endzustand $Z\rightarrow l^+ l^-$ suchen.
Indem man den totalen Wirkungsquerschnitt mit dem Verh\"{a}ltnis $\eta$ multipliziert, kann der Wirkungsquerschnitt abgesch\"{a}tzt werden.
Dieses wiederum kann erhalten werden, indem man das Verh\"{a}ltnis   der Kopplungen in den Leptonen-Kan\"{a}len zu den Kopplungen in allen Fermionen-Kan\"{a}len betrachtet (es erscheint das Quadrat, weil beide Z-Bosonen zu einem di-leptonen Paar umgewandelt werden).
\begin{equation}
\eta=\left(\frac{(\frac{1}{2}-\sin^2(\theta_W))^2+\sin^4(\theta_W)}{2-4\sin^2(\theta_W)+\frac{16}{3}\sin^4(\theta_W)}\right)^2
\approx 0.01 \quad,
\end{equation}
wobei $\theta_W$ den Weinbergwinkel im Bereich der Z-Skala bezeichnet und desweiteren $\sin^2(\theta_W)\approx0.23$ gilt.

Es wurde also der zus\"{a}tzliche Beitrag zur ZZ-Produktion durch Gravitonenvermittlung innerhalb des ADD Modells in Proton-Proton Reaktionen bei hohen Energien berechnet. 
Die Rechnung wurde in niedrigster Ordnung (in $\sqrt{\alpha_{ew,strong}}$ und dem Verh\"{a}ltnis $m_{X}/M_D$) St\"{o}rungstheorie durchgef\"{u}hrt. St\"{o}rungstheorie h\"{o}herer Ordnung w\"{a}re aufgrund der Nichtrenormalisierbarkeit der effektiven Quantenfeldtheorie der Gravitation nicht durchf\"{u}hrbar gewesen. Aber auch der hier erreichte Genauigkeitsgrad l\"{a}sst sehr signifikante Aussagen zu.  
Es wurde gezeigt, dass der ZZ-Produktions-Wirkungsquerschnitt des Standardmodells im Vergleich zu diesem entscheidend erh\"{o}ht w\"{u}rde, wenn die fundamentale Massenskala des ADD-Modells kleiner als $15000$~GeV im Falle des LHC, beziehungsweise $1700$~GeV im Falle des Tevatron w\"{a}re. F\"{u}r den Fall von sieben Extra Dimensionen k\"{o}nnte sogar $M_D=18000$~GeV am LHC getested werden.

In Betracht dieser Ergebnisse ist es wichtig, den Leser daran zu erinnern, dass die Gr\"{o}\ss e des Gravitonenbeitrages des ADD Modells (\ref{ADDWirkungsquerschnitt}) direkt
von der gew\"{a}hlten Regulasrisierungsskala $\Lambda$ in Gleichung (\ref{KKSummierung}) abh\"{a}ngt.
Aber eine solche Wahl $\Lambda=M_D$ erscheint nat\"{u}rlich, da $M_D$ ja das analogon zur
Planckmasse darstellt, welche wiederum allgemein als eine absolute G\"{u}ltigkeitsgrenze
effektiver Quantenfeldtheorien angesehen wird. Wenn also eine Vergr\"{o}\ss erung des Wirkungsquerschnittes der ZZ-Rate im Standardmodell bei LHC-Energien beobachtet w\"{u}rde, k\"{o}nnte dies wichtige Einsichten in die m\"{o}glicherweise h\"{o}herdimensionale Struktur der Raumzeit liefern.

\part{Nichtkommutative Geometrie}

\chapter{Eichtheorien und Higgsmechanismus}

\section{Die Idee von Symmetrieprinzipien}

\begin{quote}
{\footnotesize "`>Am Anfang war die Symmetrie<, das ist sicher richtiger als die Demokritsche Behauptung >Am Anfang war das Teilchen<. Die Elementarteilchen verk\"{o}rpern die Symmetrien, sie sind ihre einfachsten Darstellungen, aber sie sind erst eine Folge der Symmetrien."' (Werner Heisenberg)}
\end{quote}

Grunds\"{a}tzlich kann man die fundamentalen Teilchen des Standardmodells in zwei Klassen einteilen, in die Klasse der 
eigentlichen Materieteilchen und die der Austauschteilchen der Wechselwirkungsfelder. Alle Teilchen der ersten Klasse 
sind Fermionen und alle der zweiten Klasse Bosonen. Fermionen unterscheiden sich dadurch von Bosonen, 
dass sie einen halbzahligen Spin besitzen, w\"{a}hrend der Spin letzterer ganzzahlig ist. Dies bedeutet, 
dass Fermionen im Gegensatz zu Bosonen dem Paulischen Ausschlie\ss ungsprinzip gehorchen, 
das besagt, dass sich zwei Elementarteilchen mit halbzahligem Spin niemals im gleichen Zustand befinden d\"{u}rfen. 
Gem\"{a}\ss\ dem von Pauli bewiesenen Spin-Statistik-Theorem gehorchen Fermionen daher einer anderen Statistik als Bosonen. 

Die unterschiedlichen Spins der einzelnen Elementarteilchen spiegeln sich mathematisch in einer verschiedenen Struktur 
der sie beschreibenden Felder wider. Nun m\"{u}ssen aber alle diese Felder, da sie nun einmal auf der Raumzeit leben,
invariant unter der Symmetriegruppe der speziellen Relativit\"{a}tstheorie sein, also der Poincar\'{e}gruppe. Die 
Transformationen der Poincar\'{e}gruppe m\"{u}ssen also auf dem entsprechenden mathematischen Raum dargestellt werden, 
der die Spinstruktur des Teilchens beschreibt. 

In Rahmen relativistischer Quantenfeldtheorien beschreibt man Elementarteilchen bzw. die ihnen entsprechenden Felder 
daher als irreduzible Darstellungen der Poincar\'{e}gruppe. Neben der Spinstruktur spielt hier nat\"{u}rlich 
auch noch die Frage eine Rolle, ob es sich um massebehaftete oder Teilchen ohne Ruhemasse handelt. Das Quadrat der 
Ruhemasse des Teilchens ist neben dem Spin Casimiroperator der Poincar\'{e}gruppe und stellt damit wie der Spin 
eine lorentzinvariante Eigenschaft des Teilchens selbst dar. Die Struktur der Poincar\'{e}gruppe wird durch die 
folgende Liealgebra beschrieben

\begin{eqnarray}
i[J^{\mu\nu},J^{\rho\sigma}]&=&\eta^{\nu\rho}J^{\mu\sigma}-\eta^{\mu\rho}J^{\nu\sigma}
-\eta^{\sigma\mu}J^{\rho\nu}+\eta^{\sigma\nu}J^{\rho\mu},\nonumber \\
i[P^{\mu}, J^{\rho\sigma}]&=&\eta^{\mu\rho} P^{\sigma}-\eta^{\rho\sigma} P^{\rho},\nonumber\\
i[P^{\mu},P^{\sigma}]&=& 0.
\end{eqnarray}
Die $J^{\mu\nu}$-Generatoren beschreiben die homogene Lorentzgruppe. Diejenigen Generatoren, deren Indizes Werte 
zwischen 1 und 3 annehmen, entsprechen den Drehimpulsoperatoren, also den Generatoren der Drehungen in den drei 
Raumrichtungen, wohingegen die Generatoren mit einer 0 als Index die Lorentzboosts in die drei Raumrichtungen 
darstellen. Aufgrund der Antisymmetrie verschwinden die Elemente mit zwei gleichen Indizes. Insgesamt gibt es also 
sechs unabh\"{a}ngige Symmetrietransformationen innerhalb der homogenen Lorentzgruppe. 
Hinzu kommen die vier Raumzeittranslationen, welche durch die Impulsoperatoren $P^{k}$ und den Hamiltonoperator $P^0$
beschrieben werden und untereinander kommutieren. Insgesamt ergibt sich also eine zehndimensionale Gruppe.
  
Jede Sorte Elementarteilchen gehorcht einer bestimmten Feldgleichung, die \"{u}ber das Hamiltonsche Wirkungsprinzip mit
einer Lagrangedichte verbunden werden kann. Die Dynamik jedes freien Feldes, das einem bestimmten Teilchen entspricht, 
kann also mit Hilfe der entsprechenden Feldgleichung beschrieben werden. Nun beruht aber das reale physikalische Geschehen
im Wesentlichen auf Wechselwirkung zwischen verschiedenen Objekten. Es dr\"{a}ngt sich also die Frage auf, wie die 
Kopplung zwischen den verschiedenen Materieteilchen bzw. Feldern und den Wechselwirkungsfeldern beschrieben werden kann. 
Eine der faszinierendsten Tatsachen der modernen Physik ist, dass man die Wechselwirkung der Felder 
untereinander auch auf Symmetrieprinzipien zur\"{u}ckf\"{u}hren kann. Hierzu muss zun\"{a}chst einmal erw\"{a}hnt werden, 
dass Felder neben der \"{a}u\ss eren raumzeitlichen Struktur und dem Spin als irreduzible Darstellungen der 
Poincar\'{e}gruppe auch noch zus\"{a}tzliche innere Strukturen aufweisen. Die Idee der zus\"{a}tzlichen inneren Symmetrien 
geht auf Heisenberg zur\"{u}ck, der im Jahr 1932 den Isospin einf\"{u}hrte. Der Unterschied zwischen dem Proton und dem 
Neutron stellt sich hier in den beiden unterschiedlichen Einstellungen des Isospins dar. In Bezug auf die starke Wechselwirkung verhalten sich die beiden Teilchen vollkommen gleich. Die Symmetrie der Naturgesetze bez\"{u}glich des Isospins wird also durch die unterschiedliche Ladung und die leicht unterschiedlichen Massen gebrochen. 

Eine \"{a}hnliche Idee liegt der Beschreibung der Wechselwirkung von Teilchen im Rahmen von Eichtheorien zu Grunde.
Man betrachtet zun\"{a}chst ein Teilchen, dass bestimmte zus\"{a}tzliche Eigenschaften aufweist. Diese Klasse von Eigenschaften 
beschreibt man dann formal durch Zuordnung einer neuen Quantenzahl, welche die Beschreibung durch einen neuen inneren 
Raum zur Folge hat, der nat\"{u}rlich gem\"{a}\ss\ den Postulaten der Quantentheorie eine Hilbertraumstruktur aufweisen 
muss. Da aber nur innere Produkte zwischen Zust\"{a}nden und nicht Zust\"{a}nde an sich physikalisch relevant sind, muss 
die Theorie invariant unter der bez\"{u}glich dieses Raumes fundamentalen unit\"{a}ren Transformationsgruppe sein. 
Mit einer Symmetrietransformation bez\"{u}glich des inneren Raumes, die einer solchen Transformationsgruppe angeh\"{o}rt,
meint man zun\"{a}chst eine Transformation, die keine Raumzeitabh\"{a}ngigkeit aufweist und die man daher als global 
bezeichnet. Ein im Hinblick auf Eichtheorien relevantes Beispiel eines zus\"{a}tzlichen inneren Freiheitsgrades liefert 
die Farbladung der Quarks. 

In Rahmen von Eichfeldtheorien fordert man nun die Invarianz nicht nur unter globalen Symmetrien, sondern unter
inneren Symmetrietransformationen, die vom speziellen Raumzeitpunkt des Feldes abh\"{a}ngen. Die freien Materiefeldgleichungen 
erf\"{u}llen diese Forderung zun\"{a}chst nicht. Durch Einf\"{u}hrung einer kovarianten Ableitung, welche einen Kopplungsterm 
in der Lagrangedichte zur Folge hat, kann man jedoch Invarianz auch unter lokalen Symmetrien gew\"{a}hrleisten.
Die Zust\"{a}nde der Felder leben mathematisch in einem Raum, der dem Tensorprodukt des Hilbertraumes der quadratintegrablen
Funktionen mit dem Spinraum und dem entsprechenden Raum der zus\"{a}tzlichen inneren Symmetrien entspricht. Wenn man also 
eine Symmetrietransformation betrachtet, die an jedem Raumzeitpunkt anders wirkt, betrachtet man in gewisser Weise an jedem 
Raumzeitpunkt einen separaten inneren Raum gleicher Struktur. Man hat es also nicht mehr mit einem globalen Tensorprodukt 
zu tun, sondern es muss eine mathematische Struktur geben, welche diese einzelnen R\"{a}ume miteinander verbindet. Es handelt
sich hierbei um die gleiche formale Beschreibungsweise wie im Falle der Allgemeinen Relativit\"{a}tstheorie. Die in der kovarianten 
Ableitung auftauchenden Zusammenhangkoeffizienten bestimmen, wie zwei Elemente der entsprechenden R\"{a}ume verglichen 
werden m\"{u}ssen. Der Unterschied zur Allgemeinen Relativit\"{a}tstheorie besteht darin, dass es dort die 
Tangentialr\"{a}ume an die einzelnen Raumzeitpunkte, also Verschiebungsvektoren auf der Raumzeit sind, die miteinander 
verglichen werden. Im Falle von Eichtheorien sind es die Elemente der inneren R\"{a}ume. Im folgenden soll es nun um die
konkrete Beschreibungsweise von Eichtheorien gehen, welche von Yang und Mills in die Teilchenphysik eingef\"{u}hrt wurden und beispielsweise in \cite{Ramond}, \cite{WeinbergQTF2} und \cite{Pokorski} dargestellt ist.

\section{Das Prinzip lokaler Eichinvarianz}

\subsection{Innere Symmetrien einer Lagrangedichte}

$\Psi$ sei ein Materiefeld, dass einer freien Materiefeldgleichung gehorchen m\"{o}ge. Da alle fundamentalen Materiefelder au\ss er
dem Higgsfeld, auf das sp\"{a}ter noch zu sprechen zu kommen sein wird, Fermionen sind, ist dies die Diracgleichung. Desweiteren enthalte $\Psi$ einen zus\"{a}tzlichen inneren Freiheitsgrad und soll damit also n Komponenten beschreiben, die ihrerseits jeweils wieder einen Diracspinor darstellen. Die entsprechende Lagrangedichte lautet damit

\begin{equation}
\mathcal{L}=\bar \Psi ^n (i\gamma^\mu \partial_\mu-m) \Psi _n,
\end{equation}
wobei n die Komponenten des inneren Freiheitsgrades beschreiben soll. Wenn man $\Psi$ nun einer Transformation der unit\"{a}ren Symmetriegruppe des inneren Raumes aussetzt, also der SU(N)

\begin{equation}
\Psi(x)\quad\rightarrow\quad U \Psi(x)\quad,\quad \Psi^{\dagger}(x) \quad
\rightarrow \quad \Psi^{\dagger}(x) U^{\dagger},
\end{equation}
so heben sich die beiden Operatoren U und $U^{\dagger}$ gegenseitig auf, denn der Operator U wirkt auf einen anderen Raum als die Gammamatrizen und der Ableitungsoperator und kommutiert daher mit diesen. Damit ist die Lagrangedichte invariant unter dieser Transformation. Der tiefere Grund hierf\"{u}r ist, wie bereits im letzten Abschnitt erw\"{a}hnt, die Tatsache, dass nur innere Produkte zwischen Zust\"{a}nden physikalisch von Bedeutung sind, und diese sind unter unit\"{a}ren Transformationen invariant.
Die unit\"{a}ren Operatoren haben folgende Gestalt

\begin{equation}
U=e^{i\omega_a T^{a}},
\end{equation}
wobei die $T^{a}$ die Generatoren der Gruppe darstellen, welche hermitesch sind und eine Liealgebra bilden. Diese wiederum ist durch die Vertauschungsrelationen gegeben

\begin{equation}
[T^a,T^b]=if_{c}^{ab} T^c.
\end{equation}
Hierbei beschreiben die $f_c^{ab}$ die sogenannten Strukturkonstanten der Gruppe.

\subsection{\"{U}bergang zu lokalen Symmetrien und Wechselwirkung}

Wenn der Transformationsparameter $\omega$ aber jetzt nicht mehr konstant ist, sondern vom Raumzeitpunkt abh\"{a}ngt, was zu einer
Transformation der folgenden Gestalt f\"{u}hrt 

\begin{equation}
\Psi \quad \rightarrow \quad U(x) \Psi,
\end{equation}
so gilt f\"{u}r die Transformation der in der Lagrangedichte auftauchenden Ableitung des Feldes $\partial_\mu \Psi$

\begin{equation}
\partial_\mu \Psi(x) \rightarrow \partial_\mu U(x) \Psi(x)=[\partial_\mu U(x)] \Psi(x)+U(x)\partial_\mu \Psi(x) \neq
U(x) \partial_\mu \Psi(x).  
\end{equation}
Dies hat zur Folge, dass die Lagrangedichte nicht mehr invariant unter der nun lokalen unit\"{a}ren Transformation ist. Um die Invarianz auch unter lokalen Transformationen $U(x)$ zu gew\"{a}hrleisten, muss die Ableitung $\partial_\mu$ durch eine kovariante Ableitung $D_\mu$ ersetzt werden, die folgende Transformationseigenschaft erf\"{u}llt

\begin{equation}
D_\mu \quad \rightarrow \quad U(x) D_\mu U^{\dagger}(x). 
\label{TranskovAbleitung}
\end{equation}
Wenn diese Transformationseigenschaft erf\"{u}llt ist, gilt n\"{a}mlich

\begin{equation}
D_\mu \Psi(x) \quad \rightarrow \quad U(x) D_\mu U^{\dagger}(x) U(x) \Psi(x)=U(x) D_\mu \Psi(x).
\end{equation}
Damit ist die Lagrangedichte

\begin{equation}
\mathcal{L}=\bar \Psi^n(i\gamma^\mu D_\mu-m)\Psi_n 
\label{LagrangekovAbl}
\end{equation}
invariant unter den entsprechenden lokalen Symmetrietransformationen. Es muss also eine entsprechende kovariante Ableitung gebildet werden, welche die Bedingung ($\ref{TranskovAbleitung}$) erf\"{u}llt. Hierzu sei folgender Ansatz gew\"{a}lt

\begin{equation}
D_\mu=\partial_\mu+iA_\mu(x),
\label{kovAbleitung}
\end{equation}
wobei $A_\mu(x)$ ein Vektorfeld ist, dass sp\"{a}ter mit dem Wechselwirkungsfeld identifiziert wird und Werte in der Liealgebra der Symmetriegruppe annimmt. Es kann daher als Linearkombination der Generatoren der Gruppe ausgedr\"{u}ckt werden

\begin{equation}
A_\mu(x)=A_\mu^a(x) T^a.
\end{equation}
Einsetzen des Ausdruckes f\"{u}r die kovariante Ableitung ($\ref{kovAbleitung}$) in (\ref{TranskovAbleitung}) ergibt

\begin{equation}
\partial_\mu+iA_\mu \quad\rightarrow\quad U(x)(\partial_\mu+iA_\mu)U^{\dagger}(x).
\end{equation}
Dies kann umgeformt werden zu

\begin{equation}
A_\mu \quad \rightarrow \quad -iU(x)[\partial_\mu U(x)^{\dagger}]
+U(x) A_\mu U(x)^{\dagger}.
\label{TransVektorfeld}
\end{equation}
Es ist also eine Transformationsbedingung f\"{u}r $A_\mu$
gefunden. Um dies weiter umzuformen betrachten wir nun infinitesimale Transformationen. Unter dieser Voraussetzung  k\"{o}nnen wir die Entwicklung des Transformationsoperators 
$U(x)$ in erster Ordnung verwenden

\begin{equation}
U(x)=1+i\omega^a T^a+\mathcal{O}(\omega^2).
\end{equation}
Einsetzen in die Transformationsbedingung f\"{u}r das Vektorfeld $A_\mu$ ergibt

\begin{equation}
A_\mu \quad \rightarrow \quad A_\mu-\partial_\mu \omega^a T^a+i[\omega^a T^a, A_\mu]+\mathcal{O}(\omega^2).
\end{equation}
Die Lagrangedichte ($\ref{LagrangekovAbl}$) mit der kovarianten
Ableitung ($\ref{kovAbleitung}$) wird also unter einer infinitesimalen Transformation der Form

\begin{eqnarray}
\delta \Psi &=& i \omega^a T^a \Psi\nonumber,\\
\delta A_\mu &=& -\partial_\mu \omega^a T^a+i[\omega^a T^a, A_\mu]
\end{eqnarray}
invariant gelassen. Der erste Teil der Transformation des Vektorfeldes entspricht also der Form nach einer Eichtransformation des elektromagnetischen Feldes. Hinzu kommt jedoch, da das Vektorfeld liealgebrawertig ist und die Generatoren nicht kommutieren, der Kommutator des Eichparameters mit dem Vektorfeld.
Es ist nun aufgrund der Transformationseigenschaften sinnvoll, das Vektorfeld in der kovarianten Ableitung mit einem neuen Wechselwirkungsfeld zu identifizieren, dessen Existenz so gewisserma\ss en aus der Forderung nach lokaler Eichinvarianz unter
einer gewissen Symmetriegruppe hergeleitet wurde. Die Nichtkommutativit\"{a}t der Generatoren der Eichgruppe und damit der entsprechenden Felder entspricht hierbei ihrer Selbstwechselwirkung. Nat\"{u}rlich ist dies eigentlich so nicht ganz richtig, denn bisher wurde eigentlich nichts anderes getan, als an jedem Raumzeitpunkt ein anderes Koordinatensystem bez\"{u}glich des inneren Raumes eingef\"{u}hrt. 
Die kovariante Ableitung enth\"{a}lt die Information, wie die Komponenten des inneren Raumes an verschiedenen Raumzeitpunkten miteinander verglichen werden m\"{u}ssen. Dies hat aber zun\"{a}chst eine rein mathematische Bedeutung. Um der kovarianten Ableitung $D_\mu$ und dem mit ihr eingef\"{u}hrten Feld $A_\mu$  physikalische Signifikanz zu verleihen, bedarf es einer zus\"{a}tzlichen Annahme, die analog dem \"{A}quivalenzprinzip in der Allgemeinen Relativit\"{a}tstheorie ist. Auf die mit dieser Frage verbundenen tiefgr\"{u}ndigen philosophischen Probleme, denen eine herausragende Bedeutung in der zeitgen\"{o}ssischen Naturphilosophie zukommt, kann an dieser Stelle nicht n\"{a}her eingegangen werden. Hierzu sei auf \cite{Lyre} verwiesen.

\subsection{Die Dynamik des Wechselwirkungsfeldes}

Was bisher beschrieben wurde, ist die Art und Weise wie ein zun\"{a}chst freies Materiefeld an das jeweilige Wechselwirkungsfeld koppelt. In der neuen Lagrangedichte erscheint aber bisher noch kein Term, der die innere Dynamik des Wechselwirkungsfeldes selbst beschreibt. Ein solcher Term muss nat\"{u}rlich auch die Forderung nach lokaler Eichinvarianz erf\"{u}llen.
Zun\"{a}chst soll ein Feldst\"{a}rketensor $F_{\mu\nu}$ wie folgt definiert werden

\begin{equation}
F_{\mu\nu}=-i[D_\mu, D_\nu].
\label{Feldstaerketensor}
\end{equation}
Ein solcher Ausdruck muss eichinvariant sein, da die in ihm enthaltenen kovarianten Ableitungen eichinvariant sind. Durch Einsetzen des Ausdruckes ($\ref{kovAbleitung}$) in ($\ref{Feldstaerketensor}$) erh\"{a}lt man

\begin{equation} 
F_{\mu\nu}=\partial_\mu A_\nu-\partial_\nu A_\mu+i[A_\mu, A_\nu].
\end{equation}
Es handelt sich also um eine Verallgemeinerung des elektromagnetischen Feldst\"{a}rketensors. Im Spezialfall kommutierender Felder $A_\mu$ geht er in diesen \"{u}ber.
Da der Feldst\"{a}rketensor ebenso wie das Vektorfeld $A_\mu$
liealgebrawertig ist, kann er ebenso nach den Generatoren der Eichgruppe entwickelt werden

\begin{equation}
F_{\mu\nu}=F^a_{\mu\nu}T^a, 
\end{equation}
wobei f\"{u}r die einzelnen Komponenten $F^a_{\mu\nu}$ gilt

\begin{equation}
F^a_{\mu\nu}=\partial_\mu A^a_\nu-\partial_\nu A^a_\mu-f^{abc} A^b_\mu A^c_\nu.
\end{equation}
Um nun den entsprechenden Term in der Lagrangedichte zu erhalten, muss ein skalarer Ausdruck aus dem Feldst\"{a}rketensor konstruiert werden. Nur eine Lagrangedichte der folgenden Form ist mit Lorentzinvarianz und Parit\"{a}tserhaltung vertr\"{a}glich 

\begin{equation}
\mathcal{L}=-\frac{1}{4} g_{ab} F^a_{\mu\nu}F^{b \mu\nu},
\end{equation}
wobei durch Definition einer geeigneten Skala grunds\"{a}tzlich 
erreicht werden kann, dass $g_{ab}=\delta_{ab}$.
Damit nimmt die Gesamtlagrangedichte die folgende Gestalt an

\begin{equation}
\mathcal{L}=\bar \Psi^n(i\gamma^\mu D_\mu-m)\Psi_n-\frac{1}{4} F^a_{\mu\nu} F^{a \mu\nu}.
\end{equation}
Es ist nun wichtig, dass Eichinvarianz nur dann gew\"{a}hrleistet
werden kann, wenn die entsprechenden Vektorbosonen des Wechselwirkungsfeldes $A_\mu$ keine Masse besitzen. Ein zus\"{a}tzlicher Term etwa der Form $m^2 A_\mu A^\mu$ in der Lagrangedichte w\"{u}rde die Eichinvarianz aufheben. 

\section[Higgsmechanismus]{Elektroschwache Vereinheitlichung und Higgsmechanismus}

Alle bekannten in der Natur vorkommenden Wechselwirkungen k\"{o}nnen durch Eichtheorien beschrieben werden. Im Falle des 
Elektromagnetismus ist es die $U(1)$-Gruppe, bei der schwachen Wechselwirkung die Symmetriegruppe des schwachen Isospins $SU(2)$, bei der starken Wechselwirkung die $SU(3)$ colour und bei der Gravitation schlie\ss lich die Lorentzgruppe (siehe Kapitel 3), welche als Eichgruppen zu Grunde gelegt werden. Nun ist es aber eine empirische Tatsache, dass die Austauschteilchen der schwachen Wechselwirkung eine Masse aufweisen. Dies steht jedoch im Widerspruch zu der erw\"{a}hnten Eigenschaft, dass Massenterme die Eichinvarianz brechen. Eine L\"{o}sung dieses Problems liefert die Generierung der Massen durch eine sogenannte spontane Symmetriebrechung, die auch als Higgsmechanismus bezeichnet wird, benannt nach dem schottischen Physiker Peter Higgs \cite{Higgs:1964pj}.

\subsection{Die Elektroschwache Theorie}

Im folgenden soll nun der Higgsmechanismus im Rahmen der Eichtheorie, welche die elektromagnetische und die schwache Wechselwirkung in einem einheitlichen Schema beschreibt, betrachtet werden. Sie wurde von Glashow \cite{Glashow:1961tr}, Weinberg \cite{Weinberg:1967tq} und Salam \cite{Salam:1968rm} entwickelt und ist beispielsweise in \cite{WeinbergQTF2},\cite{Greiner} und \cite{Quigg} zu finden.
Als Eichgruppe wird hier die Gruppe $SU(2)$ $\times$ $U(1)$ zu Grunde gelegt, wobei sich die $SU(2)$-Gruppe auf den schwachen Isospin, also
beispielsweise das Dublett bestehend aus Elektron und Elektronneutrino bezieht. Da Neutrinos nur linksh\"{a}ndig vorkommen und der rechtsh\"{a}ndige Teil des entsprechenden Diracspinors demnach bez\"{u}glich des schwachen Isospins ein Singlett darstellt, muss in diesem Falle die $SU(2)$ genau genommen auf die linksh\"{a}ndige Komponente des Spinors eingeschr\"{a}nkt werden. Die gesamte Gruppe besitzt insgesamt vier Transformationsfreiheitsgrade, die den drei Generatoren der $SU(2)_L$ und der Phasentransformation der $U(1)$ entsprechen.
Eine allgemeine Symmetrietransformation angewandt auf einen  Zustand, welcher ein Dublett aus Elektron und Elektronneutrino beschreibt, hat also folgende Gestalt

\begin{equation}
\Psi\quad\rightarrow\quad exp\left(i\left[\omega^a T^a+\omega^y Y \right]\right) \Psi,
\end{equation}
wobei die drei Generatoren $T^a$ der $SU(2)_L$ und der Generator $Y$ der $U(1)$, auch als schwache Hyperladung bezeichnet, wie folgt definiert sind 

\begin{eqnarray}
T^a&=&g\left[\frac{1}{2}\left(\frac{1+\gamma_5}{2}\right)
\sigma^a\right] \\
Y&=&g^{'}\left[\frac{1}{2}\left(\frac{1+\gamma_5}{2}\right)
{\bf 1}+\left(\frac{1-\gamma_5}{2}\right)\right].
\end{eqnarray}
Hierbei ist ${\bf 1}$ die Einheitsmatrix in zwei Dimensionen und die $\sigma^a$ bezeichnen die Paulischen Spinmatrizen 

\begin{equation}
{\bf 1}=\left(\begin{array}{cc} 1&0\\0&1\end{array}\right),
\sigma^1=\left(\begin{array}{cc} 0&1\\1&0\end{array}\right),
\sigma^2=\left(\begin{array}{cc} 0&-i\\i&0\end{array}\right),
\sigma^3=\left(\begin{array}{cc} 1&0\\0&-1\end{array}\right).
\end{equation}
Die Operatoren $\frac{1+\gamma_5}{2}$ und $\frac{1-\gamma_5}{2}$
stellen die Projektion auf die links- bzw. rechtsh\"{a}ndige Komponente des Diracspinors dar.
Der elektrische Ladungsoperator ist eine Linearkombination des Transformationsoperators der $U(1)$, n\"{a}mlich Y, und der dritten Komponente der Transformation im schwachen Isospinraum $T^3$  

\begin{equation}
Q=\frac{e}{g}T^3-\frac{e}{g^{'}}Y.
\label{Ladung-Hyperladung}
\end{equation}
Damit ist ein reiner Elektronzustand im schwachen Isospinraum Eigenzustand zur elektrischen Ladung, worin sich widerspiegelt, dass nur die Elektronen, nicht aber die Neutrinos eine Ladung tragen. 
Nun werden die neuen Felder $W^\mu, W^{\mu\dagger}, Z^\mu$ und $A^\mu$  eingef\"{u}hrt, die wie folgt definiert sind

\begin{eqnarray}
W^\mu&=&\frac{1}{\sqrt{2}}(\mathcal{A}_1^\mu+i\mathcal{A}_2^\mu)
\label{DefinitionW+Feld}\\
W^{\mu\dagger}&=&\frac{1}{\sqrt{2}}(\mathcal{A}_1^\mu-i\mathcal{A}_2^\mu),
\label{DefinitionW-Feld}
\end{eqnarray}
die dem $W^+$-und dem $W^-$-Teilchen der schwachen Wechselwirkung entsprechen, sowie

\begin{eqnarray}
Z^\mu&=&cos(\theta_W)\mathcal{A}_3^\mu+sin(\theta_W)\mathcal{B}^\mu
\label{DefinitionZ-Feld}\\
A^\mu&=&-sin(\theta_W)\mathcal{A}_3^\mu+cos(\theta_W)\mathcal{B}^\mu.
\end{eqnarray}
Hier beschreibt $\theta$ den sogenannten Weinbergwinkel, den Mischungswinkel zwischen dem elektromagnetischen Feld $A^\mu$, das dem Photon und dem Feld, das dem Z-Teilchen der schwachen Wechselwirkung entspricht.
Der Weinbergwinkel ist auf folgende Weise mit den Parametern $g$ und $g^{'}$ verkn\"{u}pft, welche das Verh\"{a}ltnis der $SU(2)_L$- zu den $U(1)$ Transformationen bestimmen

\begin{equation}
g=-\frac{e}{sin{\theta_W}}\quad,\quad g^{'}=-\frac{e}{cos{\theta_W}}.
\end{equation}

\subsection{Generierung der Massen der Eichbosonen}

In der bisherigen Beschreibung sind die verschiedenen Wechselwirkungsfelder zwangsl\"{a}ufig noch masselos geblieben, da wie bereits erw\"{a}hnt die Forderung der Eichinvarianz keine Massenterme zul\"{a}sst.  
Die Austauschteilchen der schwachen Wechselwirkung m\"{u}ssen ihre Massen also gewisserma\ss en auf indirektem Wege erhalten.
Aus diesem Grunde postuliert man ein skalares Hintergrundfeld, das Higgsfeld, dessen Selbstwechselwirkung durch einen zus\"atzlichen Potentialterm beschrieben wird. 
Das Higgsfeld stellt ebenfalls ein Dublett bez\"{u}glich des schwachen Isospins dar, wobei vorausgesetzt wird, dass der obere Zustand Eigenzustand zum Ladungsoperator mit positiver Ladung ist
\begin{equation}
\Phi=\left(\begin{array}{c}\Phi^{+}\\ \Phi^0 \end{array}\right).
\end{equation}
Der Grund hat damit zu tun, dass auch das Elektron seine Masse durch das Higgsfeld durch eine Yukawakopplung erhalten soll. Da im Falle des Higgsfeldes beide Komponenten sowohl rechts- alsauch linksh\"{a}ndig vorkommen sollen, tauchen in den Generatoren der Transformationen keine Projektionsoperatoren auf die rechts- bzw. linksh\"{a}ndige Komponente auf. Unter Ber\"{u}cksichtigung des Zusammenhangs zwischen Ladung und Hyperladung ($\ref{Ladung-Hyperladung}$) und der Gestalt des Ladungsoperators des Higgsfeldes 
\begin{equation}
q=-\frac{g^{'}}{2}\left(\begin{array}{cc}1&0\\0&1\end{array}\right)
\end{equation}
ergeben sich damit folgende Generatoren f\"{u}r die Transformation des Higgsfeldes

\begin{eqnarray}
\bar T^a=\frac{g}{2}\sigma^a,\\
\bar Y=-\frac{g^{'}}{2}{\bf 1}.
\label{GeneratorenHiggsFeld}
\end{eqnarray}
Die entsprechende eichinvariante Lagrangedichte lautet

\begin{equation}
\mathcal{L}_{Higgs}=\frac{1}{2}(D_\mu \Phi)(D^\mu \Phi)^\dagger+
\frac{\mu^2}{2}\Phi^\dagger\Phi-\frac{\lambda}{4}(\Phi^\dagger \Phi)^2,
\end{equation}
wobei f\"{u}r die kovariante Ableitung $D_\mu$ gilt

\begin{equation}
D_\mu=\partial_{\mu}+i\mathcal{A}_\mu^a T^a+i\mathcal{B}_\mu Y.
\end{equation}
Nun besitzt das skalare Feld bei dieser Lagrangedichte allerdings einen von 0 verschiedenen Vakuumerwartungswert $\nu$, der dem Minimum des Potentialterms in der Higgslagrangedichte $\frac{\mu^2}{2}\Phi^\dagger\Phi-\frac{\lambda}{4}(\Phi^\dagger \Phi)^2$ entspricht. F\"{u}r das Betragsquadrat gilt

\begin{equation}
\nu^2= \langle \Phi \rangle ^{\dagger} \langle \Phi \rangle = \frac{\mu^2}{\lambda}.
\end{equation}
Es ist grunds\"{a}tzlich m\"{o}glich, eine entsprechende $SU(2) \times U(1)$ Eichtransformation durchzuf\"{u}hren, welche die obere Komponente des Higgsfeldes 
zum verschwinden bringt und die untere reell macht. Eine solche Eichung wird als unit\"{a}re Eichung bezeichnet. Damit gilt f\"{u}r die Vakuumerwartungswerte

\begin{equation}
\langle \Phi^+ \rangle=0\quad,\quad \langle \Phi^0 \rangle=\nu.
\end{equation}
Wenn man nun das Feld $\Phi$ um den Vakuumerwartungswert entwickelt

\begin{equation}
\Phi=\left(\begin{array}{c}0\\ \nu+\phi \end{array}\right),
\end{equation}
so taucht im Kopplungsterm der Wechselwirkungsfelder an das Higgsfeld der Ausdruck

\begin{equation}
\left(i\left[ (\mathcal{A}_\mu^a \bar T^a+\mathcal{B}_\mu \bar Y)\left(\begin{array}{c}0\\v \end{array}\right)\right]\right)
\left(i\left[ (\mathcal{A}_\mu^a \bar T^a+\mathcal{B}_\mu \bar Y)\left(\begin{array}{c}0\\v \end{array}\right)\right]\right)^{\dagger}
\end{equation}
auf.
Dieser kann unter Verwendung von ($\ref{GeneratorenHiggsFeld}$) wie folgt umgeschrieben werden

\begin{equation}
\left|\frac{g}{2}A_\mu^1\left(\begin{array}{c}\nu\\0\end{array}\right) 
-i\frac{g}{2}A_\mu^2\left(\begin{array}{c}\nu\\0\end{array}\right)    
-\frac{g}{2}A_\mu^3\left(\begin{array}{c}0\\\nu\end{array}\right) 
-\frac{g^{'}}{2}YB_\mu^1\left(\begin{array}{c}0\\\nu\end{array}\right)\right|^2.
\label{KopplungVakuum}
\end{equation}
Wenn man nun ($\ref{DefinitionW+Feld}$) und 
($\ref{DefinitionW-Feld}$)
in Erinnerung ruft, sowie ($\ref{DefinitionZ-Feld}$) nach $\mathcal{B}$ umstellt und in obige Gleichung ($\ref{KopplungVakuum}$) einsetzt, so heben sich die $\mathcal{A}_\mu^3$-Terme auf und man erh\"{a}lt 

\begin{equation}
\left|\frac{g}{\sqrt{2}} W_\mu^\dagger
\left(\begin{array}{c}\nu\\0\end{array}\right) 
+\frac{g g^{'}}{e} Z_\mu
\left(\begin{array}{c}0\\\nu\end{array}\right)\right|^2
=\frac{\nu^2 g^2}{4}W_\mu^\dagger W^\mu
+\frac{\nu^2 (g^2+g^{'2})}{8} Z_\mu Z^\mu.
\label{KopplungVakuumEndausdruck}
\end{equation}
Diese haben die Gestalt von Massentermen f\"{u}r die W-Felder und das Z-Feld, welche so also eine Masse erhalten haben, wobei man unter Ber\"{u}cksichtigung der Tatsache, dass die Massenterme von Skalar- und Vektorfeldern immer das Massenquadrat enthalten,  f\"{u}r die Massen der W- und Z-Teilchen aus obigem Ausdruck 
folgende Werte herausliest 

\begin{equation}
m_W=\frac{\nu g}{2}\quad,\quad m_Z=\frac{\nu\sqrt{g^2+g{'}^2}}{2}.
\end{equation}
Da kein Kopplungsterm f\"{u}r das Photonenfeld $A_\mu$ an den Vakuumerwartungswert des Higgsfeldes auftaucht, bleiben diese  masselos.
Die Eigenschaft der Masse stellt sich also als Wechselwirkung mit einem skalaren Hintergrundfeld dar. Dies bedeutet in gewisser Weise die R\"{u}ckf\"{u}hrung des Begriffes der Masse auf den der Wechselwirkung.

\chapter[Das Grundprinzip der n.k. Geometrie]{Das Grundprinzip der nichtkommutativen Geometrie}

\section{Die Einf\"{u}hrung einer neuen Algebra f\"{u}r Ort und Impuls}

Beim \"{U}bergang von einer klassischen Theorie zu der entsprechenden Quantentheorie ersetzt 
man die klassischen Gr\"{o}\ss en durch hermitesche Operatoren, die man bestimmten Vertauschungsrelationen 
unterwirft und deren Eigenwerte die m\"{o}glichen Messwerte der entsprechenden Gr\"{o}\ss en darstellen.
Im speziellen Fall des \"{U}bergangs von der klassischen Punktteilchenmechanik zur Quantenmechanik, dem historisch ersten 
Beispiel einer Quantisierung, werden an die Operatoren, die den Ort x und den Impuls p des Teilchens beschreiben sollen,
die folgenden Vertauschungsrelationen gefordert

\begin{equation}
[\hat x^{i},\hat x_{j}]_-=0\quad,\quad [\hat p^{i},\hat p_{j}]_-=0\quad,\quad [\hat x^{i},\hat p_{j}]_-=\delta^{i}_{j}.
\label{Vertauschungsrelationen}
\end{equation} 

\subsection{Vertauschungsrelationen zwischen den Koordinaten}

Die Idee der nichtkommutativen Geometrie geht nun davon aus, die Vertauschungsrelation ($\ref{Vertauschungsrelationen}$) 
dahingehend abzu\"{a}ndern, dass der Kommutator der Raumzeitkoordinaten nicht mehr verschwindet. Man fordert also eine neue Vertauschungsrelation zwischen den Ortskoordianten

\begin{equation} 
[\hat x^{i},\hat x^{j}]_-\neq 0.
\end{equation}
Eine solche Vertauschungsrelationen l\"{a}sst jedoch die Vertauschungsrelationen zwischen den Ortskoordinaten und den Ableitungen nach den Ortskoordinaten und damit auch den Impulsoperatoren unber\"{u}hrt. Die Auswirkung auf die Vertauschungsrelation der Ableitungsoperatoren untereinander
ist zun\"{a}chst variabel. Die Idee, einen nichtverschwindenden Kommutator der Ortskoordinaten einzuf\"{u}hren, geht auf \cite{Snyder:1946qz} zur\"{u}ck. In neuerer Zeit wurde durch Seiberg und Witten gezeigt, dass sich eine Nichtkommutativit\"{a}t als Konsequenz aus Stringtheorien ergibt \cite{Seiberg:1999vs}. Die Anwesenheit eines magnetischen Hintergrundfeldes kann daf\"{u}r sorgen, dass sich normale Feldtheorien effektiv so verhalten, als l\"{a}ge eine nichtkommutative Geometrie vor \cite{Gorbar:2004ck}.  
Es gibt verschiedene Formen der Nichtkommutativit\"{a}t der Ortskoordinaten.
Eine M\"{o}glichkeit ist, dass sie eine Liealgebra bilden und die Vertauschungsrelation folgende Gestalt hat

\begin{equation}
[\hat x^{i},\hat x^{j}]_-=i\lambda^{ij}_k \hat x^k,
\end{equation}
wobei die $\lambda^{ij}_k$ die Strukturkonstanten der Gruppe darstellen. Dem Kommutator zweier Ortskoordinaten wird also selbst wieder eine Ortskoordinate zugeordnet. Eine solche Struktur ist analog den Liealgebren der Generatoren der Eichgruppen.
Eine weitere M\"{o}glichkeit ist die Zuordnung eines bez\"{u}glich der Koordinaten quadratischen Ausdrucks der folgenden Form

\begin{equation}
[\hat x^{i},\hat x^{j}]_-=\left(\frac{1}{q}\hat R^{ij}_{kl}
-\delta^i_l \delta^j_k\right)\hat x^k \hat x^l.
\end{equation}
Im kanonischen Fall wird der Kommutator gleich einem antisymmetrischen Tensor $i\theta^{ij}$ gesetzt, sodass sich die 
folgende Vertauschungsrelation ergibt

\begin{equation} 
[\hat x^{i},\hat x^{j}]_-=i\theta^{ij}.
\label{Koordinatenvertauschungsrelation}
\end{equation}
In den folgenden Betrachtungen soll grunds\"{a}tzlich eine kanonische Vertauschungsrelation zu Grunde gelegt werden. Desweiteren soll davon ausgegangen werden, dass das in diesem Falle auftretende $\theta^{ij}$ konstant ist, also nicht 
von den Koordinaten abh\"{a}ngt.

\subsection{Implikation f\"{u}r die Ableitungsoperatoren}

Es stellt sich nun noch die Frage nach den Vertauschungsrelationen der Ableitungsoperatoren. 
Die Gr\"{o}\ss e $\hat x^i-i\theta^{ij}\hat \partial_j$ kommutiert mit den Ortskoordinaten, denn

\begin{eqnarray}
[\hat x^i-i\theta^{ij}\hat \partial_j,\hat x^{k}]_-
=[\hat x^i,\hat x^{k}]_--[i\theta^{ij}\hat \partial_j,\hat x^{k}]_-\nonumber\\
=i\theta^{ik}-i\theta^{ij}\delta^k_j
=i\theta^{ik}-i\theta^{ik}=0.
\end{eqnarray}
Es ist also sinnvoll, diese Gr\"{o}\ss e gleich einer Konstanten zu setzen. Wenn man nun 0 als Konstante w\"{a}hlt, ergibt sich

\begin{equation}
\hat \partial_j=-i\theta_{ij}^{-1} \hat x^i.
\label{DefnkAbleitung}
\end{equation}
Dies hat folgende Vertauschungsrelation f\"{u}r die Ableitungsoperatoren zur Folge

\begin{eqnarray}
[\hat \partial_i,\hat \partial_j]_-&
=&[-i\theta_{ik}^{-1} \hat x^k,-i\theta_{jl}^{-1} \hat x^l]_-
=-\theta_{ik}^{-1}\theta_{jl}^{-1}[x^k,x^l]_-\nonumber\\
&=&-\theta_{ik}^{-1}\theta_{jl}^{-1}i\theta^{kl}
=-i\theta_{ik}^{-1}\delta_j^k
=-i\theta_{ij}^{-1}.
\end{eqnarray}
Bei der folgenden Definition 

\begin{equation}
\hat \partial_i \hat f=-i\theta_{ij}^{-1}[\hat x^j,\hat f]_-,
\end{equation}
erhielte man kommutierende Ableitungsoperatoren

\begin{equation}
[\hat \partial_i,\hat \partial_j]_-=0.
\end{equation}
Wie dem auch sei. Falls eine kanonische Vertauschungsrelation 
($\ref{Koordinatenvertauschungsrelation}$) sowie Zusammenhang ($\ref{DefnkAbleitung}$) als Definition des Ableitungsoperators zu Grunde gelegt werden, erh\"{a}lt man insgesamt folgenden Satz von Vertauschungsrelationen zwischen Orts- und Impulsoperatoren

\begin{equation}
[\hat x^i,\hat x_j]_-=i\theta^i_j\quad,\quad [x^i,p_j]_-=i\delta^i_j \quad,\quad [p^i,p_j]=i(\theta^i_j)^{-1}.
\end{equation}
Weitere allgemeine Aspekte der nichtkommutativen Geometrie werden in \cite{Wohlgenannt:2003de} ausgef\"{u}hrt.

\section{Das Sternprodukt}

Wenn man nun eine Quantenfeldtheorie auf einer Raumzeit, welche die Struktur einer nichtkommutativen Geometrie tr\"{a}gt, 
formulieren m\"{o}chte, so stellt sich die Frage, wie Produkte von Feldern zu formulieren sind, die von nicht-kommutativen 
Koordinaten abh\"{a}ngen und in den Lagrangedichten auftauchen. Die Antwort liefert die Einf\"{u}hrung des Sternproduktes, 
wie es beispielsweise in \cite{Wohlgenannt:2003de} beschrieben wird. Hierzu wird die Methode der sogenannten Weylquantisierung verwendet. 
Mit Weylquantisierung bezeichnet man die Beschreibung der Felder durch eine \"{U}berlagerung ebener Wellen, die nun von 
Operatoren abh\"{a}ngen

\begin{equation}
\psi(\hat x)=\frac{1}{2\pi}\int d^n k e^{ik_{j} \hat x^{j}} \bar \psi(k),
\label{Weyl}
\end{equation}
wobei $\bar \psi(k)$ die Gewichtungsfunktion der ebenen Wellen darstellt. Man erh\"{a}lt sie durch Fouriertransformation 
des von den \"{u}blichen Koordinaten abh\"{a}ngigen Feldes $\psi(x)$

\begin{equation} 
\psi(k)=\frac{1}{2\pi}\int d^n x e^{-ik_j x^j} \psi(x).
\label{Fouriertransformation}
\end{equation}
Multipliziert man nun zwei Felder auf der nichtkommutativen Raumzeit, so ergibt sich unter Ber\"{u}cksichtigung von ($\ref{Weyl}$) 

\begin{equation}
\psi(\hat x)\cdot \phi(\hat x)=\frac{1}{(2\pi)^4}\int d^n k d^n p e^{ik_{j}\hat x^{j}} e^{ip_{j}\hat x^{j}} \bar \psi(k) \bar \phi(p).
\label{ProduktnkF}
\end{equation}
Unter Verwendung der Campbell-Baker-Hausdorff-Formel 

\begin{equation}
e^{A}\cdot e^{B}=e^{A+B+\frac{1}{2}[A,B]_+\frac{1}{12}[[A,B],B]-\frac{1}{12}[[A,B],A]+...},
\label{C-B-H_F}
\end{equation}
welche das Produkt zweier Exponentialfunktionen bestimmt, in deren Exponenten Matrizen stehen, erh\"{a}lt man

\begin{equation}
exp(ik_j \hat x^{j})exp(ip_j \hat x^{j})=exp ( i(k_j+p_j)\hat x^{j}-\frac{i}{2} k_{i} \theta^{ij} p_{j} ).
\end{equation}
Hierbei wurde die kanonische Vertauschungsrelation zwischen den Koordinaten ($\ref{Koordinatenvertauschungsrelation}$) zu Grunde gelegt und die Tatsache ausgenutzt, dass alle Kommutatoren in der Baker-Campbell-Hausdorff-Formel, die selbst einen Kommutator enthalten, in diesem Falle verschwinden. Dies liegt daran, dass gem\"{a}\ss\ der oben getroffenen Annahme der Kommutator zweier Koordinaten ($\ref{Koordinatenvertauschungsrelation}$) konstant sein soll, also selbst nicht von den Koordinaten abh\"{a}ngen soll. 
Wenn man diesen Ausdruck in ($\ref{ProduktnkF}$) verwendet, so ergibt sich

\begin{equation}
\psi(\hat x)\cdot \phi(\hat x)=\frac{1}{(2\pi)^4}\int d^n k d^n p e^{i(k_j+p_j)\hat x^{j}-\frac{i}{2}k_{i}\theta^{ij}p_{j}} \bar \psi(k) \bar \phi(p).
\end{equation}
Unter Einsetzung von ($\ref{Fouriertransformation}$) und Verwendung der Ortsdarstellung f\"{u}r die Impulsoperatoren ergibt sich 

\begin{equation}
\psi(\hat x)\cdot \phi(\hat x)=exp\left(\frac{i}{2}\frac{\partial}{\partial x^{i}}\theta^{ij}\frac{\partial}{\partial y^{j}} \right)
\psi(x)\phi(y)\mid_{y \rightarrow x}.
\end{equation}
Das Produkt von Feldern auf einer Raumzeit mit nichtkommutativer Geometrie ist also auf einen Ausdruck zur\"{u}ckgef\"{u}hrt 
worden, der die von kommutativen Koordinaten abh\"{a}ngigen Felder enth\"{a}lt. Es ist deshalb sinnvoll, das sogenannte 
Sternprodukt zwischen Feldern zu definieren

\begin{equation}
\psi \star \phi (x)=exp\left(\frac{i}{2}\frac{\partial}{\partial x^{i}}\theta^{ij}\frac{\partial}{\partial y^{j}}\right)
\psi(x)\phi(y)\mid_{y \rightarrow x}.
\end{equation}
Multipliziert man also zwei normale Felder auf einer kommutativen Raumzeit mit dem Sternprodukt entspricht dies der 
Multiplikation der entsprechenden Felder, die von nichtkommutativen Koordinaten abh\"{a}ngen.

\chapter{Nichtkommutative Eichtheorien}

Nachdem das vorige Kapitel die Grundidee der nichtkommutativen Geometrie im Allgemeinen zum Thema hatte, sollen hier nun im Speziellen die Konsequenzen f\"{u}r die Formulierung von Eichtheorien er\"{o}rtert werden. Hierbei werden die Darstellungen in \cite{Wohlgenannt:2003de}, \cite{Calmet:2003jv}, \cite{Calmet:2004yj},\cite{Madore:2000en},\cite{Jurco:2000ja} und \cite{Jurco:2001rq} zu Grunde gelegt. Dies macht schlie\ss lich die Einf\"{u}hrung von sogenannten Seiberg-Witten-Abbildungen notwendig. Im ersten Kapitel wurde das Prinzip lokaler Eichtheorien beschrieben. Dort werden Symmetrietransformationen beschrieben, deren (liealgebrawertiger) Parameter von Raumzeitpunkt zu Raumzeitpunkt verschieden ist und damit eine Funktion auf der Raumzeit darstellt. Wenn man nun eine infinitesimale Eichtransformation eines Materiefeldes $\Psi$ zun\"{a}chst im kommutativen Fall betrachtet

\begin{equation}
\delta \Psi(x)=i\omega^a(x) T^a \Psi(x)=i\alpha(x)\Psi(x)\quad,\quad \alpha(x)=\omega^a (x) T^a,
\end{equation}
so werden beim \"{U}bergang zu einer nichtkommutativen Raumzeit sowohl der Eichparameter $\alpha$ alsauch das Materiefeld $\Psi$ zu Feldern, welche von nichtkommutativen Koordinaten abh\"{a}ngen und daher muss bei einer Multiplikation das Sternprodukt zu Grunde gelegt werden. Als infinitesimale Transformation f\"{u}r das Materiefeld ergibt sich damit

\begin{equation}
\hat \delta \Psi=i\alpha(\hat x)\Psi(\hat x)=i \alpha\star\Psi,
\end{equation}
wobei $\hat \delta$ eine Eichtransformation auf einer nichtkommutativen Raumzeit bezeichnet.
Damit hat die Nichtkommutativit\"{a}t der Raumzeit also auch eine Auswirkung auf eine lokale Eichtransformation. Im folgenden soll nun den sich daraus ergebenden Konsequenzen Rechnung getragen werden. 

\section{Kovariante Koordinaten}

Zun\"{a}chst muss darauf hingewiesen werden, dass sich die Koordinaten der nichtkommutativen Raumzeit $\hat x$ unter einer Eichtransformation nicht transformieren, also $\hat \delta \hat x=0$. Wenn man sich nun die Wirkung einer infinitesimalen Eichtransformation auf das Sternprodukt einer Koordinate mit einem Materiefeld ansieht, so erkennt man, dass dieses sich nicht wie das Materiefeld selbst transformiert, denn

\begin{equation}
\hat \delta(x\star\Psi)=\hat\delta x\star\Psi+x\star\hat\delta \Psi=ix\star\alpha(x)\star\Psi \neq i\alpha(x)\star x\star\Psi.
\label{Transformationprodukt}
\end{equation}
Dies verh\"{a}lt sich in Analogie zur Transformation der Ableitung des Materiefeldes innerhalb gew\"{o}hnlicher Eichtheorien, die im vorletzten Kapitel thematisiert wurden und wo die einfache Ableitung $\partial_\mu$ durch eine kovariante Ableitung $D_\mu=\partial_\mu+iA_\mu$ ersetzt werden musste. Bei nichtkommutativen Eichtheorien m\"{u}ssen daher die Koordinaten $\hat x^\mu$ durch kovariante Koordinaten $\hat X^\mu$ der folgenden Form ersetzt werden

\begin{equation}
\hat x^\mu \rightarrow \hat X^\mu=\hat x^\mu+B^\mu,  
\end{equation}
wobei an $B_\mu$ eine bestimmte Transformationsforderungen zu stellen ist. Wenn man diese Ersetzung in ($\ref{Transformationprodukt}$) vornimmt, so ergibt sich

\begin{eqnarray}
\hat \delta(X^\mu \star \Psi)=\hat \delta((x^\mu+B^\mu) \star \Psi)=(\hat \delta B^\mu)\star\Psi+(x^\mu+B^\mu)\star\delta \Psi\nonumber\\
=\hat \delta B^\mu \star\Psi+i(x^\mu+B^\mu)\star\alpha\star\Psi.
\end{eqnarray}
Um zu gew\"{a}hrleisten, dass sich das Produkt $X^\mu*\Psi$ gem\"{a}\ss\ 

\begin{equation}
\hat \delta (X^\mu \star \Psi)=i\alpha\star X^\mu \star \Psi
\end{equation}
transformiert, muss sich $B^\mu$ also wie folgt transformieren

\begin{equation}
\delta B^\mu=i[\alpha\ _{,}^{*}\ x^\mu]+i[\alpha\ _{,}^{*}\ B^\mu]=i[\alpha\ _{,}^{*}\ X^\mu].
\end{equation}
Das Symbol $[\ _{,}^{*}\ ]$ bezeichnet hierbei den Kommutator bez\"{u}glich des Sternproduktes.
Es ist insofern nicht \"{u}berraschend, dass die Koordinaten bei nichtkommutativen Eichtheorien durch kovariante Koordinaten ersetzt werden m\"{u}ssen, als sie ja wie im letzten Kapitel gezeigt als proportional zu den entsprechenden Ableitungen definiert werden k\"{o}nnen, um die entsprechenden Vertauschungsrelationen zu gew\"{a}hrleisten. Unter Verwendung dieser Definition und Betrachtung der kovarianten Ableitung
$\partial_\mu+iA_\mu$ ergibt sich folgender Zusammenhang zwischen $B^\mu$ und dem Eichpotential $A^\mu$ 

\begin{equation}
B^\mu=-\theta^{\mu\nu} A_\nu.
\end{equation}
Bei der Transformation des Vektorpotentials und der Definition des Feldst\"{a}rketensors tauchen naturgem\"{a}\ss\ ebenfalls die Kommutatoren bez\"{u}glich des Sternprodukts auf. F\"{u}r die Transformation des Vektorpotentials $A_\mu$ gilt damit

\begin{equation}
\hat \delta A_\mu=\partial_\mu \alpha+i[\alpha\ _{,}^{*}\ A_\mu].
\end{equation}
Der Feldst\"{a}rketensor ist nat\"{u}rlich wie \"{u}blich \"{u}ber den Kommutator der kovarianten Ableitungen definiert. Auf einer nichtkommutativen
Raumzeit enth\"{a}lt er aber nun das Sternprodukt

\begin{equation}
F_{\mu\nu}=i[D_\mu\ _{,}^{*}\ D_\nu],
\end{equation}
was auf folgenden Ausdruck f\"{u}r den Feldst\"{a}rketensor f\"{u}hrt

\begin{equation}
F_{\mu\nu}=\partial_\mu A_\nu-\partial_\nu A_\mu+i[A_\mu\ _{,}^{*}\ A_\nu].
\end{equation}
Dies bedeutet insbesondere, dass auch im Falle einer Abelschen Eichtheorie aufgrund der Nichtkommutativit\"{a}t des Sternproduktes Kommutatorterme auftauchen. 

\section{Seiberg-Witten-Abbildungen}

\subsection{Die Bedingung der Geschlossenheit}

Nat\"{u}rlich ist die Voraussetzung einer jeden Symmetriegruppe die Geschlossenheit, was bedeutet, dass die Anwendung zweier Symmetrietransformationen
wieder zu einer Symmetrietransformation f\"{u}hrt, die Element der Gruppe ist. Im Falle kommutativer Eichtheorien gilt f\"{u}r den Kommutator zweier
Eichtransformationen bez\"{u}glich der Eichparameter $\alpha$ und $\beta$

\begin{equation}
(\delta_\alpha \delta_\beta-\delta_\beta \delta_\alpha)\Psi=\delta_{i[\alpha,\beta]}\Psi. 
\label{TransformationKommutator}
\end{equation}
Der Kommutator zweier zu den Parametern $\alpha$ und $\beta$ geh\"{o}riger Eichtransformationen ist also gleich der Eichtransformation des Kommutators von $\alpha$ und $\beta$ multipliziert mit i. Da die Generatoren $T^a$ der Eichgruppe aber eine Liealgebra bilden $[T^a,T^b]=if^{abc} T^c$, sieht der Kommutator zweier Eichtransformationen wie folgt aus
   
\begin{equation}
i[\alpha,\beta]=i\omega^a \omega^b [T^a,T^b]=\omega^a \omega^b f^{bac} T^c.  
\end{equation}
Es ergibt sich also erneut eine Eichtransformation innerhalb der Gruppe. 
Nun stellt sich die Frage, wie dies bei nichtkommutativen Eichtheorien aussieht. Prinzipiell ist die Situation hier nat\"{u}rlich analog. Allerdings enth\"{a}lt der Kommutator jetzt nat\"{u}rlich das Sternprodukt. F\"{u}r nichtkommutative Eichtransformationen ergibt sich damit als Kommutator
der Eichparameter

\begin{equation} 
[\alpha\ _{,}^{*}\ \beta]=[\omega^a T^a\ _{,}^{*}\ \omega^b T^b]=\frac{1}{2}\{\alpha_a\ _,^{*}\ \beta_b\}[T^a\ ,\ T^b]+\frac{1}{2}[\alpha_a\ _,^{*}\  \beta_b]\{T^a\ ,\ T^b\}.
\end{equation}
Die Kommutatoren und Antikommutatoren der Vorfaktoren sind bez\"{u}glich der Frage, ob sich die Gruppe geschlossen ist, nicht von Bedeutung. Der Kommutator der Generatoren im ersten Term ist nat\"{u}rlich wie bei gew\"{o}hnlichen Eichtheorien auch hier umproblematisch. Probleme bereitet jedoch der Antikommutator im zweiten Term.
Dieser ist nur liealgebrawertig im Falle der Generatoren der U(N), nicht jedoch bei der SU(N).

\subsection{Erweiterung zur einh\"{u}llenden Algebra}

Es gibt allerdings die M\"{o}glichkeit, eine geschlossene Algebra zu erreichen, indem man die Gruppe der Eichparameter zu der Einh\"{u}llenden der Liealgebra erweitert. Diese beinhaltet im Gegensatz zu einer einfachen Liealgebra nicht nur beliebige Linearkombinationen eines Satzes von Generatoren, sondern enth\"{a}lt auch beliebige Produkte der Generatoren. Ein Eichparameter $\hat \alpha$ in der Einh\"{u}llenden der Liealgebra mit Generatoren $T^a$ hat dann folgende allgemeine Gestalt

\begin{equation}    
\hat \alpha=\hat \alpha_0^a T^a+\hat \alpha_1^{ab} \{T^a,T^b\}+\hat \alpha_2^{abc} \{T^a,T^b,T^c\}+...,
\end{equation}
wobei die geschwungenen Klammern eine Summierung \"{u}ber alle Reihenfolgen der Generatoren darstellt und insofern eine Verallgemeinerung des Antikommutators f\"{u}r mehr als zwei Elemente beschreibt. Gem\"{a}\ss\ den Eichparametern sind nat\"{u}rlich auch die nichtkommutativen Eichpotentiale und damit die Feldst\"{a}rketensoren in der einh\"{u}llenden Algebra definiert.
Ebenso wie das im letzten Kapitel behandelte Produkt zwischen Feldern auf einer nichtkommutativen Raumzeit, k\"{o}nnen auch die Eichparameter und die nichtkommutativen Felder selbst nach dem Parameter $\theta^{\mu\nu}$ entwickelt werden, welcher die Nichtkommutativit\"{a}t der Koordinaten angibt. Damit werden die kommutativen Gr\"{o}\ss en auf nichtkommutative Gr\"{o}\ss en abgebildet. Diese Abbildungen wurden erstmals von Seiberg und Witten gefunden \cite{Seiberg:1999vs} und werden daher nach ihnen als Seiberg-Witten-Abbildungen bezeichnet. 
Es soll nun beschrieben werden, wie die Seiberg-Witten-Abbildungen zu bestimmen sind. Aus der Relation ($\ref{TransformationKommutator}$) f\"{u}r die Eichparameter wird auf einer nichtkommutativen Raumzeit die folgende Bedingung, welche nun das Sternprodukt enth\"{a}lt 

\begin{eqnarray}
(i\hat \delta_\alpha \hat \delta_\beta-\hat \delta_\beta \hat \delta_\alpha)\star\Psi=\hat \delta_{i[\alpha,\beta]}\hat \Psi \nonumber\\
\Leftrightarrow i\hat \delta_\alpha \hat \beta-i\hat \delta_\beta \hat \alpha+[\hat \alpha\ _{,}^*\ \hat \beta]*\Psi=\hat {[\alpha,\beta]}\star\hat \Psi.
\end{eqnarray}
Wenn man nun den Eichparameter $\hat \alpha$ in der einh\"{u}llenden Algebra in einer Reihe nach der Gr\"{o}\ss e $\theta^{\mu\nu}$ entwickelt, welche die Nichtkommutativit\"{a}t der Raumzeit angibt und damit im Sternprodukt auftaucht

\begin{equation}
\hat \alpha=\alpha_0+\alpha_1(\theta)+...,
\end{equation}
so ergibt sich bei einer Entwicklung in erster Ordnung in $\theta^{\mu\nu}$ f\"{u}r den Term $\alpha_1$

\begin{equation}
\alpha_1=\frac{1}{4}\theta^{\mu\nu}\{\partial_\mu,A_\nu \}.
\end{equation}
Die Entwicklung von $\hat \alpha$ in erster Ordnung hat also folgende Form

\begin{equation}
\hat \alpha=\alpha+\frac{1}{4}\theta^{\mu\nu}\{\partial_\mu,A_\nu \}+\mathcal{O}(\theta^2).
\end{equation}

\subsection{Abbildungen der Felder}

Die Seiberg-Witten-Abbildungen f\"{u}r die Felder sind nun dadurch bestimmt, dass eine gew\"{o}hnliche Eichtransformation der nichtkommutativen Felder, welche also auf die kommutativen Felder wirkt, von denen sie abh\"{a}ngen, gleich einer nichtkommutativen Eichtransformation der nichtkommutativen Felder ist. Dies bedeutet f\"{u}r eine infinitesimale Eichtransformation eines Feldes A 

\begin{equation} 
\hat A[A+\delta_\alpha A]=\hat A[A]+\hat \delta_\alpha \hat A[A].
\end{equation}
F\"{u}r ein Materiefeld $\hat \Psi$ ergibt sich damit konkret die Bedingung

\begin{equation}
\delta_\alpha \hat \Psi=\hat \delta_\alpha \hat \Psi=i\hat \alpha * \hat \Psi.
\end{equation}
Eine Entwicklung in erster Ordnung 

\begin{equation}
\hat \Psi=\Psi^0+\Psi^1+\mathcal{O}(\theta^2)
\end{equation}
f\"{u}hrt hierbei auf folgende Bedingung 

\begin{equation}
\delta_\alpha \Psi_0+\delta_\alpha \Psi_1=i\alpha \Psi_0-\frac{1}{2}\theta^{\mu\nu}\partial_\mu \alpha \partial_\nu \Psi_0+i\alpha_1 \Psi_0+i\alpha \Psi_1+\mathcal{O}(\theta^2).
\end{equation}
Da im kommutativen Grenzfall, welcher $\theta=0$ impliziert, sich nat\"{u}rlich wieder $\Psi$ ergeben muss, gilt $\Psi_0$=$\Psi$. Insgesamt findet man f\"{u}r die Seiberg-Witten-Abbildung in erster Ordnung

\begin{equation}
\hat \Psi=\Psi-\frac{1}{2}\theta^{\mu\nu}A_\nu \partial_\mu \Psi+\frac{i}{8}\theta^{\mu\nu}[A_\mu,A_\nu] \Psi+\mathcal{O}(\theta^2).
\end{equation}
Hierbei ist zu beachten, dass bei einer kommutativen Eichtransformation des nichtkommutativen Materiefeldes $\hat \Psi$, welches nicht nur eine Funktion des kommutativen Materiefeldes $\Psi$, sondern auch des Eichpotentiales $A_\mu$ ist, letzteres bei einer Eichtransformation mittransformiert werden muss.  
Wenn man die analogen Bedingungen f\"{u}r das Vektorpotential und den Feldst\"{a}rketensor aufstellt

\begin{equation}
\delta_\alpha  \hat A_\mu=\partial_\mu \hat \alpha+i[\hat \alpha\ _,^*\ \hat A_\mu]
\end{equation}
und

\begin{equation}
\hat \delta \hat F_{\mu\nu}=i[\hat \alpha\ _{,}^*\ \hat F_{\mu\nu}],
\end{equation}
so erh\"{a}lt man die folgenden Seiberg-Witten-Abbildungen in erster Ordnung in $\theta^{\mu\nu}$

\begin{eqnarray}
\hat A_\mu=A_\mu-\frac{1}{4}\theta^{\rho\nu}\{A_\rho,\partial_\nu A_\mu+F_{\nu\mu}\}+\mathcal{O}(\theta^2)\nonumber\\
\hat F_{\mu\nu}=F_{\mu\nu}+\frac{1}{2}\theta^{\rho\sigma}\{F_{\mu\rho},F_{\nu\sigma}-\frac{1}{4}\theta^{\rho\sigma}\{A_\rho,(\partial_\sigma+D_\sigma)F_{\mu\nu}\}+\mathcal{O}(\theta^2).
\end{eqnarray}
Diese machen nun die Formulierung des Standardmodells auf einer nichtkommutativen Raumzeit m\"{o}glich, wie sie in \cite{Calmet:2001na}, \cite{Melic:2005fm} und \cite{Melic:2005am} gegeben wird.

\chapter[Higgsmechanismus und n.k. Geometrie]{Higgsmechanismus und nichtkommutative Geometrie}

Bei der Formulierung nichtkommutativer Eichtheorien spielen zwangsl\"{a}ufig die neu eingef\"{u}hrten nichtkommutativen Felder eine Rolle, welche die Eichinvarianz auch auf einer nichtkommutativen Raumzeit gew\"{a}hrleisten. Da nun aber dem Higgsmechanismus eine prinzipielle Bedeutung zukommt, weil die Austauschteilchen der schwachen Wechselwirkung nur so eine Masse bekommen k\"{o}nnen, ohne dass die Eichinvarianz unter der Symmetriegruppe des schwachen Isospins verletzt w\"{u}rde, muss dieser nat\"{u}rlich ebenfalls in die nichtkommutative Eichtheorie integriert werden. In \cite{Calmet:2001na},\cite{Melic:2005fm} und \cite{Petriello:2001mp} wurde auf den Higgsmechanismus im Rahmen nichtkommutativer Eichtheorien eingegangen. Wenn man zun\"{a}chst einmal die Lagrangedichte des Higgsfeldes auf einer nichtkommutativen Raumzeit formuliert, indem man das Produkt zwischen Feldern durch das Sternprodukt
und die Felder durch die nichtkommutativen Felder ersetzt, erh\"{a}lt man

\begin{equation}
\mathcal{L}_{Higgs}=(D_\mu \hat \Phi)^{\dagger}\star (D^\mu \hat \Phi)-\mu^2 (\hat \Phi)^\dagger \star \hat \Phi-\lambda(\hat \Phi ^\dagger \star \hat \Phi)\star (\hat \Phi ^\dagger \star \hat \Phi). 
\end{equation}  

Es sind nun zwei M\"{o}glichkeiten denkbar, wie man die entsprechende Lagrangedichte mit den entsprechenden Massentermen in Abh\"{a}ngigkeit der u\"{b}lichen Felder erh\"{a}lt. Man kann entweder zun\"{a}chst die Seiberg-Witten-Abbildungen ausf\"{u}hren und dann die spontane Symmetriebrechung betrachten. Dies ist die \"{u}bliche Reihenfolge. Es ist aber auch denkbar, die spontane Symmetriebrechung schon bei den nichtkommutativen Feldern auszuf\"{u}hren und anschlie\ss end die Seiberg-Witten-Abbildungen zu verwenden. Allerdings stellt sich die Frage, ob man in beiden F\"{a}llen das gleiche Resultat erh\"{a}lt. Es wird hier gezeigt werden, dass diese beiden Prozeduren tats\"{a}chlich \"{a}quivalent zueinander sind. Zun\"{a}chst soll der Abelsche Fall betrachtet werden. Hier f\"{a}llt der konstante Vakuumerwartungswert bei allen Termen mit $\theta$ heraus, sodass sich sie Situation als v\"{o}llig unproblematisch erweist. Bei nicht-Abelschen Eichtheorien muss die Bedingung an die nichtkommutativen Felder 

\begin{equation}
\delta \hat \Phi=\hat \delta \hat \Phi=i\hat \alpha \star \hat \Phi
\label{SWcondition}
\end{equation}
betrachtet werden, welche auf die Seiberg-Witten-Abbildung 

\begin{equation}
\hat \Phi=\Phi-\frac{1}{2}\theta^{\mu\nu}A_\mu \partial_\nu \Phi
+\frac{i}{8}\theta^{\mu\nu}[A_\mu,A_\nu]_-\Psi+\mathcal{O}(\theta^2)
\label{SeibergWittenPhi}
\end{equation}
f\"{u}hrt. 
Die Seiberg-Witten-Abbildung des Eichparameters ist durch

\begin{equation}
\hat \alpha=\alpha+\frac{1}{4}\theta^{\mu\nu}\{\partial_\mu \alpha,A_\nu\}+\mathcal{O}(\theta^2)
\end{equation}
gegeben.

\section{Abelscher Fall}
 
Im Abelschen Fall reduziert sich ($\ref{SeibergWittenPhi}$) auf folgende Seiberg-Witten-Abbildung f\"{u}r das Materiefeld $\Phi$ 

\begin{equation}
\hat \Phi=\Phi-\frac{1}{2}\theta^{\mu\nu}A_\mu \partial_\nu \Phi
+\mathcal{O}(\theta^2).
\label{AbelianSeiberg-Witten}
\end{equation}
Die Abbildung f\"{u}r den Eichparameter wird zu

\begin{equation}
\hat \alpha=\alpha-\frac{1}{2}\theta^{\mu\nu} A_\mu \partial_\nu \alpha+\mathcal{O}(\theta^2).
\end{equation}
Die Seiberg-Witten-Abbildungen sowohl f\"{u}r das Eichpotential $A_\mu$ alsauch f\"{u}r den Feldst\"{a}rketensor $F_{\mu\nu}$ m\"{u}ssen nicht betrachtet werden, da die spontane Symmetriebrechung keine direkte Auswirkung auf sie hat.

\subsection[S.-W.-Abbildung und dann Symmetriebrechung]{Zun\"{a}chst Seiberg-Witten-Abbildung und dann Symmetriebrechung}

Zun\"{a}chst soll die spontane Symmetriebrechung auf Seite der kommutativen Felder erfolgen. Dies f\"{u}hrt auf folgende Relation  

\begin{equation}
\Phi=\nu+h+i\sigma,
\label{symmetry-breaking}
\end{equation}
wobei $\nu$ der konstante Vakuumerwartungswert des Feldes $\Phi$ ist. Die spontane Symmetriebrechung hat das Erscheinen von Goldstonebosonen zur Folge. Dies spielt hier jedoch keine Rolle. Indem man definiert

\begin{equation}
\phi=h+i\sigma,
\end{equation}
wird aus ($\ref{symmetry-breaking}$)

\begin{equation}
\Phi=\nu+\phi.
\label{symmetry-breaking-ugauge}
\end{equation}
Einsetzen von ($\ref{symmetry-breaking-ugauge}$) in ($\ref{AbelianSeiberg-Witten}$) f\"{u}hrt auf

\begin{equation}
\hat \Phi=\nu+\phi-\frac{1}{2}\theta^{\mu\nu}A_\mu \partial_\nu (\nu+\phi)+\mathcal{O}(\theta^2).
\label{Abelianresult1}
\end{equation}
Da $\nu$ eine Konstante ist, bleibt dieser Ausdruck \"{u}brig

\begin{equation}
\hat \Phi=\nu+\phi-\frac{1}{2}\theta^{\mu\nu}A_\mu \partial_\nu h+\mathcal{O}(\theta^2).
\label{result1}
\end{equation}

\subsection[Symmetriebrechung und dann S.-W.-Abbildung]{Zun\"{a}chst Symmetriebrechung and dann Seiberg-Witten-Abbildung}

Die andere M\"{o}glichkeit besteht darin, eine Symmetriebrechung der nichtkommutativen Felder zu betrachten. 

\begin{equation}
\hat \Phi=\nu+\hat \phi.
\label{nonc-symmetry-breaking-ug}
\end{equation}
Nun muss man jedoch die Seiberg-Witten-Abbildung f\"{u}r $\hat \phi$ festlegen. Deshalb ist es notwendig, die Bedingung f\"{u}r die Seiberg-Witten-Abbildung von $\Phi$ zu betrachten.
Einsetzen von ($\ref{nonc-symmetry-breaking-ug}$) in ($\ref{SWcondition}$) liefert

\begin{equation}
\delta (\nu +\hat \phi)=\hat \delta (\nu+\hat \phi).
\end{equation}
Da $\nu$ eine Konstante ist, transformiert es sich nicht. Aus diesem Grund bleibt folgende Relation  

\begin{equation}
\delta \hat \phi=\hat \delta \hat \phi.
\label{SWcondition-h}
\end{equation}
Diese Bedingung ist von der gleichen Gestalt wie die Bedingung an $\Phi$, aber hierbei ist zu ber\"{u}cksichtigen, dass $\phi$ sich unter einer Eichtransformation anders als $\Phi$ transformiert. Dies kann man wie folgt sehen

\begin{equation}
\delta \Phi=i \alpha \Phi=i \alpha (\nu+\phi)\quad,\quad\delta \Phi=\delta(\nu+\phi)=\delta \phi.
\end{equation}
Gleichsetzen der beiden Ausdr\"{u}cke zeigt, dass sich $\phi$ wie folgt transformiert

\begin{equation}
\delta \phi=i\alpha(\nu+\phi).
\label{gauge-transformation-h}
\end{equation}
Indem man die analoge Argumentation im Falle des nichtkommutativen Feldes $\hat \phi$ unter einer nichtkommutative Eichtransformation anwendet, sieht man, dass sich $\hat \phi$ analog transformiert 

\begin{equation}
\hat \delta \hat \phi=i\hat \alpha \star(\nu+\hat \phi).
\label{gauge-transformation-hath}
\end{equation}
Wie dem auch sei, im Abelschen Falle sieht die Seiberg-Witten-Abbildung f\"{u}r $h$ wie die Seiberg-Witten-Abbildung f\"{u}r $\Phi$ aus 

\begin{equation}
\hat \phi=\phi-\frac{1}{2}\theta^{\mu\nu}A_\mu \partial_\nu \phi
+\mathcal{O}(\theta^2).
\end{equation}
\label{SWmaph}
Dies kann man sehen, indem man \"{u}berpr\"{u}ft, ob sie die Bedingung an die Seiberg-Witten-Abbildung ($\ref{SWcondition-h}$) erf\"{u}llt. Einsetzen von ($\ref{SWmaph}$) in ($\ref{SWcondition-h}$) f\"{u}hrt auf die folgende Gleichung in erster Ordnung in $\theta$

\begin{equation}
\delta(\phi-\frac{1}{2}\theta^{\mu\nu}A_\mu \partial_\nu \phi)
=i \hat \alpha \star (\phi-\frac{1}{2}\theta^{\mu\nu}A_\mu \partial_\nu \phi).
\end{equation}
Benutzung von ($\ref{gauge-transformation-h}$), ($\ref{gauge-transformation-hath}$) und der Transformationseigenschaft des Vektorpotentials $A_\mu$

\begin{equation}
\delta A_\mu=\partial_\mu \alpha
\end{equation}
liefert

\begin{eqnarray}
i\alpha(\nu+\phi)-\frac{1}{2}\theta^{\mu\nu}A_\mu \partial_\nu i\alpha(\nu+\phi)-\frac{1}{2}\theta^{\mu\nu}\partial_\mu \alpha \partial_\nu \phi =\nonumber\\
i\hat \alpha \star (\nu+\phi-\frac{1}{2}\theta^{\mu\nu}A_\mu \partial_\nu \phi).
\end{eqnarray}
Durch Ausf\"{u}hren der Ableitung auf der linken und des Sternproduktes auf der rechten Seite erh\"{a}lt man

\begin{eqnarray}
i\alpha(\nu+\phi)-i\frac{1}{2}\theta^{\mu\nu}A_\mu [(\partial_\nu \alpha)(\nu+\phi)+\alpha (\partial_\nu \phi)]-\frac{1}{2}\theta^{\mu\nu}\partial_\mu \alpha \partial_\nu \phi \nonumber\\
=i\alpha(\nu+\phi)-\frac{1}{2}\theta^{\mu\nu}\partial_\mu \alpha \partial_\nu \phi-i\alpha \frac{1}{2}\theta^{\mu\nu}A_\mu \partial_\nu \phi-i\frac{1}{2}\theta^{\mu\nu}A_\mu \partial_\nu \alpha(\nu+\phi).
\end{eqnarray}
Nun kann man sehen, dass die beiden Seiten gleich sind und ($\ref{SWmaph}$) in der Tat die richtige Seiberg-Witten-Abbildung f\"{u}r $\hat \phi$ ist.
Indem wir ($\ref{SWmaph}$) in ($\ref{nonc-symmetry-breaking-ug}$) einsetzen erhalten wir 
($\ref{result1}$)

\begin{equation}
\hat \Phi=\nu+\phi-\frac{1}{2}\theta^{\mu\nu}A_\mu \partial_\nu \phi+\mathcal{O}(\theta^2).
\end{equation}
Somit ist gezeigt, dass spontane Symmetriebrechung und Abbilden auf kommutative Felder im Abelschen Fall umkehrbar ist.

\section{Nicht-Abelscher Fall}

Im Falle Nicht-Abelscher Eichtheorien sieht die Situation etwas anders aus. 

\subsection{S.-W.-Abbildung und dann Symmetriebrechung}

Die folgende Betracht gilt f\"{u}r beliebige nichtabelsche Eichgruppen. Dazu geh\"{o}rt also auch die Eichgruppe, bei welcher der Higgsmechanismus im Standardmodell relevant ist. Es handelt sich also um die in Kapitel 5 thematisierte $SU(2) \times U(1)$ Eichgruppe. In einer allgemeinen Darstellung ohne Wahl einer speziellen Eichung f\"{u}hrt die Symmetriebrechung auf folgende Relation

\begin{equation}
\Phi=\nu_n+\phi_n,
\label{NAsymmetrybreaking}
\end{equation}
wobei der Index n andeutet, dass es sich bei dem Vakuumerwartungswert $\nu$ und dem Feld $\phi$ um Multipletts handelt.
Durch Einsetzen von ($\ref{NAsymmetrybreaking}$)
in ($\ref{SeibergWittenPhi}$), erh\"{a}lt man

\begin{equation}
\hat \Phi=\nu_n+\phi_n
-\frac{1}{2}\theta^{\mu\nu}A_\mu \partial_\nu (\nu_n+\phi_n)
+\frac{i}{8}\theta^{\mu\nu}[A_\mu,A_\nu]
(\nu_n+\phi_n)
+\mathcal{O}(\theta^2).
\end{equation}
Der $\nu$-Term im zweiten Term verschwindet und so ergibt sich folgender Ausdruck

\begin{equation}
\hat \Phi=\nu_n+\phi_n
-\frac{1}{2}\theta^{\mu\nu}A_\mu \partial_\nu (\nu_n+\phi_n)
+\frac{i}{8}\theta^{\mu\nu}[A_\mu,A_\nu]
(\nu_n+\phi_n)
+\mathcal{O}(\theta^2).
\label{NAresult1}
\end{equation}

\subsection{Symmetriebrechung und dann S.-W.-Abbildung}

Wenn man Symmetriebrechung auf der nichtkommutativen Seite betrachtet

\begin{equation}
\hat \Phi=\nu_n+\hat \phi_n,
\label{NAsymmetrybreakingnc}
\end{equation}
sieht die Bedingung an die Seiberg-Witten-Abbildung des Materiefeldes ($\ref{SWcondition}$) wie folgt aus

\begin{equation}
\delta (\nu_n+\hat \phi_n)
=i \hat \alpha \star (\nu_n+\hat \phi_n).
\label{SWconditionnA}
\end{equation}
Die Bedingung ($\ref{SWconditionnA}$) kann in folgender Weise umformuliert werden

\begin{equation}
\delta \hat \phi_n=\hat \delta \hat \phi_n.
\label{SWconditionh-d}
\end{equation}
$\phi_n$ und $\hat \phi_n$ transformieren sich in Analogie zum Abelschen Fall gem\"{a}\ss

\begin{equation}
\delta \phi_n=i\alpha(\nu_n+\phi_n)\quad,\quad \hat \delta \hat \phi_n=i\hat \alpha \star(\nu_n+\hat \phi_n).
\label{gauge-transformation-h-d}
\end{equation}
Es wird die Annahme gemacht, dass $\phi_n$ auf die folgende Art und Weise abgebildet wird

\begin{eqnarray}
\hat \phi_n&=&\phi_n-\frac{1}{2}\theta^{\mu\nu}A_\mu \partial_\nu \phi_n
+\frac{i}{8}\theta^{\mu\nu}[A_\mu,A_\nu]\phi_n\nonumber\\
&&+\frac{i}{8}\theta^{\mu\nu}[A_\mu,A_\nu]\nu_n
+\mathcal{O}(\theta^2).
\label{SeibergWittenhd}
\end{eqnarray}
Einsetzen von ($\ref{SeibergWittenhd}$) in ($\ref{SWconditionh-d}$) f\"{u}hrt zu der folgenden Gleichung in erster Ordnung in $\theta^{\mu\nu}$

\begin{eqnarray}
\delta 
(\phi_n-\frac{1}{2}\theta^{\mu\nu}A_\mu \partial_\nu \phi_n
+\frac{i}{8}\theta^{\mu\nu}[A_\mu,A_\nu]\phi_n
+\frac{i}{8}\theta^{\mu\nu}[A_\mu,A_\nu]\phi_n)\nonumber\\
=i \hat \alpha \star
(\phi_n-\frac{1}{2}\theta^{\mu\nu}A_\mu \partial_\nu \phi_n
+\frac{i}{8}\theta^{\mu\nu}[A_\mu,A_\nu]\phi_n
+\frac{i}{8}\theta^{\mu\nu}[A_\mu,A_\nu]\nu_n).
\end{eqnarray}
Durch Benutzung von ($\ref{gauge-transformation-h-d}$) und der Eichtransformation eines liealgebrawertigen Vektorfeldes

\begin{equation}
\delta A_\mu=\partial_\mu \alpha+i[\alpha,A_\mu]
\end{equation}
kann dies geschrieben werden als

\begin{eqnarray}
i\alpha(\nu_n+\phi_n)-\frac{1}{2}\theta^{\mu\nu} A_\mu \partial_\nu (i\alpha(\nu_n+\phi_n))-\frac{i}{8}\theta^{\mu\nu}[A_\mu,A_\nu]i\alpha(\nu_n+\phi_n)&&\nonumber\\
-\frac{1}{2}\theta^{\mu\nu}\partial_\mu \alpha \partial_\nu \phi_n-\frac{1}{2}\theta^{\mu\nu}i[\alpha,A_\mu]\partial_\nu \phi_n&&\nonumber\\
+\frac{i}{8}\theta^{\mu\nu}[\partial_\mu \alpha, A_\nu](\nu_n+\phi_n)+\frac{i}{8}\theta^{\mu\nu}[A_\mu,\partial_\nu \alpha](\nu_n+\phi_n)&&\nonumber\\
+\frac{i}{8}\theta^{\mu\nu}[i[\alpha,A_\mu],\partial_\nu \alpha](\nu_n+\phi_n)+\frac{i}{8}\theta^{\mu\nu}[A_\mu,i[\alpha,A_\nu]](\nu_n+\phi_n)&&\nonumber\\
=i\alpha(\nu_n+\phi_n)-\frac{1}{2}\theta^{\mu\nu}\partial_\mu \alpha \partial_\nu \phi_n-i\alpha\frac{1}{2}\theta^{\mu\nu}A_\mu \partial_\nu \phi_n\nonumber\\
-i\alpha\frac{i}{8}\theta^{\mu\nu}[A_\mu,A_\nu](\nu_n+\phi_n)+\frac{i}{4}\theta^{\mu\nu}\{\partial_\mu \alpha,A_\nu\}(\nu_n+\phi_n).
\end{eqnarray}
Die Ausdr\"{u}cke $\ i\alpha(\nu_n+\phi_n)\ $ und $\ \frac{1}{2}\theta^{\mu\nu}\partial_\mu \alpha \partial_\nu \phi_n\ $ erscheinen auf beiden Seiten und heben sich gegenseitig auf. Deshalb kann man schreiben

\begin{eqnarray}
-\frac{1}{2}\theta^{\mu\nu}A_\mu i(\partial_\nu \alpha)(\nu_n+\phi_n)-\frac{1}{2}\theta^{\mu\nu}A_\mu i\alpha \partial_\nu \phi_n
+\frac{i}{8}\theta^{\mu\nu}[A_\mu,A_\nu]i\alpha(\nu_n+\phi_n)&&\nonumber\\
-\frac{1}{2}\theta^{\mu\nu}i[\alpha,A_\mu]\partial_\nu \phi_n
+\frac{i}{8}\theta^{\mu\nu}[\partial_\mu \alpha,A_\nu](\nu_n+\phi_n)+\frac{i}{8}\theta^{\mu\nu}[A_\mu,\partial_\nu \alpha](\nu_n+\phi_n)&&\nonumber\\
+\frac{i}{8}\theta^{\mu\nu}[i[\alpha,A_\mu],A_\nu](\nu_n+\phi_n)+\frac{i}{8}\theta^{\mu\nu}[A_\mu,i[\alpha,A_\nu]](\nu_n+\phi_n)&&\nonumber\\
=-i\alpha\frac{1}{2}\theta^{\mu\nu}A_\mu\partial_\nu \phi_n+i\alpha\frac{i}{8}[A_\mu,A_\nu](\nu_n+\phi_n)+\frac{i}{4}\theta^{\mu\nu}\{\partial_\mu \alpha, A_\nu\}(\nu_n+\phi_n).&&\nonumber\\
\end{eqnarray}
Wenn man beachtet, dass die folgende Relation g\"{u}ltig ist

\begin{equation}
-\frac{1}{2}\theta^{\mu\nu}A_\mu i\alpha \partial_\nu \phi_n-\frac{1}{2}\theta^{\mu\nu}i[\alpha,A_\mu]\partial_\nu \phi_n=-i\alpha\frac{1}{2}\theta^{\mu\nu}A_\mu\partial_\nu \phi_n,
\end{equation}
heben sich die entsprechenden Terme auf und man wird auf folgende Gleichung gef\"{u}hrt

\begin{eqnarray}
-\frac{1}{2}\theta^{\mu\nu}A_\mu i(\partial_\nu \alpha)(\nu_n+\phi_n)
+\frac{i}{8}\theta^{\mu\nu}[A_\mu,A_\nu] i\alpha(\nu_n+\phi_n)\nonumber\\
+\frac{i}{8}\theta^{\mu\nu}[\partial_\mu \alpha,A_\nu](\nu_n+\phi_n)
+\frac{i}{8}\theta^{\mu\nu}[A_\mu,\partial_\nu \alpha](\nu_n+\phi_n)\nonumber\\
+\frac{i}{8}\theta^{\mu\nu}[i[\alpha,A_\mu],A_\nu](\nu_n+\phi_n)
+\frac{i}{8}\theta^{\mu\nu}[A_\mu,i[\alpha,A_\nu]](\nu_n+\phi_n)\nonumber\\
=i\alpha\frac{i}{8}[A_\mu,A_\nu](\nu_n+\phi_n)+\frac{i}{4}\theta^{\mu\nu}\{\partial_\mu \alpha, A_\nu\}(\nu_n+\phi_n).
\end{eqnarray}
Aufgrund der Antisymmetrie von $\theta^{\mu\nu}$ gilt

\begin{equation}
\frac{i}{8}\theta^{\mu\nu}[\partial_\mu \alpha, A_\nu]_-=\frac{i}{8}\theta^{\mu\nu}[A_\mu,\partial_\nu \alpha]
\end{equation}
und somit

\begin{eqnarray}
-\frac{1}{2}\theta^{\mu\nu}A_\mu i(\partial_\nu \alpha)(\nu_n+\phi_n)+\frac{i}{8}\theta^{\mu\nu}[\partial_\mu \alpha,A_\nu](\nu_n+\phi_n)\nonumber\\
+\frac{i}{8}\theta^{\mu\nu}[A_\mu,\partial_\nu \alpha](\nu_n+\phi_n)\nonumber\\
=-\frac{1}{2}\theta^{\mu\nu}A_\mu i(\partial_\nu \alpha)(\nu_n+\phi_n)+\frac{i}{4}\theta^{\mu\nu}[\partial_\mu \alpha, A_\nu](\nu_n+\phi_n)\nonumber\\  
=-\frac{1}{2}\theta^{\mu\nu}A_\mu i(\partial_\nu \alpha)(\nu_n+\phi_n)+\frac{i}{4}\theta^{\mu\nu}\partial_\mu \alpha A_\nu(\nu_n+\phi_n)\nonumber\\
-\frac{i}{4}\theta^{\mu\nu}A_\nu \partial_\mu \alpha(\nu_n+\phi_n)
=\frac{i}{4}\theta^{\mu\nu}\{\partial_\mu \alpha,A_\nu\}.
\end{eqnarray}
Es verbleibt die folgende Gleichung

\begin{eqnarray}
\frac{i}{8}\theta^{\mu\nu}[A_\mu,A_\nu](\nu_n+\phi_n)+\frac{i}{8}\theta^{\mu\nu}[i[\alpha,A_\mu],A_\nu](\nu_n+\phi_n)\nonumber\\
+\frac{i}{8}\theta^{\mu\nu}[A_\mu,i[\alpha,A_\nu]](\nu_n+\phi_n)
=i\alpha\frac{i}{8}[A_\mu,A_\nu]_-(\nu_n+\phi_n).
\end{eqnarray}
Indem man nun die verschachtelten Kommutatoren ausrechnet, ergibt sich die G\"{u}ltigkeit auch dieser Gleichung.
Dies bedeutet, dass ($\ref{SeibergWittenhd}$) in der Tat die richtige Seiberg-Witten-Abbildung liefert.
Indem man nun ($\ref{SeibergWittenhd}$) in ($\ref{NAsymmetrybreakingnc}$) einsetzt, erh\"{a}lt man

\begin{equation}
\hat \Phi=(\nu_n+\phi_n)
+\frac{1}{2}\theta^{\mu\nu}A_\mu \partial_\nu (\nu_n+\phi_n)
+\frac{i}{8}\theta^{\mu\nu}[A_\mu,A_\nu]
(\nu_n+\phi_n)
+\mathcal{O}(\theta^2).
\end{equation}
Die entspricht exakt dem Resultat aus (\ref{NAresult1}).
Damit wurde gezeigt, dass es keine Rolle spielt, ob man zun\"{a}chst die Seiberg-Witten-Abbildungen ausf\"{u}hrt und dann die Symmetrie bricht oder ob die Symmetriebrechung den Seiberg-Witten-Abbildungen voran geht.

\chapter*{Schlussbemerkungen}

\addcontentsline{toc}{chapter}{Schlussbemerkungen}

\pagestyle{plain}

Am Ende dieser Arbeit soll nun das hier Beschriebene in den gr\"{o}\ss eren Rahmen der zeitgen\"{o}ssischen Physik eingeordnet werden.
Die in dieser Diplomarbeit vorgestellten Erweiterungen der bisherigen Physik gehen letztlich aus dem Versuch hervor, das Geb\"{a}ude der theoretischen Physik in einer Weise zu erweitern, die einer einheitlicheren Beschreibung der Natur zustrebt. Dieses Bestreben besteht im Grunde aus zwei Teilen, die nat\"{u}rlich eng miteinander verwoben, aber nicht identisch sind. 
Die Quantentheorie stellt der heutigen Auffassung nach ein allgemeines Beschreibungsschema f\"{u}r beliebige dynamische Objekte in der Natur dar. In der Teilchenphysik ist es nun das Ziel, eine ebenso allgemeine Theorie zu finden, aus der hervorgeht, welche Arten von Objekten es gibt und welche Eigenschaften diese besitzen. Dies w\"{u}rde insbesondere die R\"{u}ckf\"{u}hrung der bekannten Wechselwirkungen auf ein einheitliches Prinzip bedeuten.
Durch diese Fragestellung wird man jedoch direkt auf das zweite Problem gef\"{u}hrt, dessen L\"{o}sung in gewisser Hinsicht die Voraussetzung f\"{u}r die Behandlung des ersten ist. Dieses besteht n\"{a}mlich in der Wesensfremdheit der Gravitation zu den anderen Wechselwirkungen. Denn einerseits wird die Gravitation im Rahmen der Allgemeinen Relativit\"{a}tstheorie als Eigenschaft der Raumzeit beschrieben, w\"{a}hrend die anderen fundamentalen Wechselwirkungen durch Felder auf der Raumzeit beschrieben werden. Andererseits ist die Allgemeine Relativit\"{a}tstheorie in ihrer bisher erreichten Formulierung eine klassische Theorie, w\"{a}hrend die anderen Wechselwirkungen in den Rahmen relativistischer Quantenfeldtheorien eingebettet sind.
Es muss also eine Quantentheorie der Gravitation, damit aber letztlich eine quantentheoretische Beschreibung der Raumzeitstruktur
selbst gefunden werden.
Die nichtkommutative Geometrie ist ein Versuch ein quantentheoretisches Element in die Beschreibung der Raumzeit hineinzubringen, indem man Vertauschungsrelationen zwischen den Koordinaten fordert, was in gewisser Hinsicht einem Prozess der Quantisierung \"{a}hnelt. Hierbei spielt die Gravitation aber im Grunde noch keine Rolle. Wie bereits erw\"{a}hnt ergibt sich im Rahmen von Stringtheorien eine nichtkommutative Geometrie \cite{Seiberg:1999vs}. In diesen wird die Gravitation grunds\"{a}tzlich garnicht als Struktur der Raumzeit, sondern als Konsequenz bestimmter Anregungszust\"{a}nde von Strings beschrieben. Aus ihnen geht nat\"{u}rlich auch die Motivation hervor, eine Raumzeit mit zus\"{a}tzlichen kompaktifizierten Dimensionen zu postulieren, auch wenn diese im Gegensatz zu der im ersten Teil dieser Arbeit untersuchten Annahme von einer Kompaktifizierung ausgehen, die nur Voraussagen jenseits des in absehbarer Zeit empirisch zug\"{a}nglichen zulassen \cite{Polchinsky}. Daneben steht das Hierarchieproblem, das durch die Annahme zus\"{a}tzlicher Dimensionen gel\"{o}st werden k\"{o}nnte. Ihm scheint aber keine so prinzipielle Bedeutung
zuzukommen wie den beiden ersten grunds\"{a}tzlichen Schwierigkeiten.
Bei beiden Annahmen, sowohl derer der nichtkommutativen Geometrie alsauch jener der zus\"{a}tzlichen kompaktifizierten Dimensionen, handelt es sich in jedem Falle um wichtige Ans\"{a}tze zu einer Erweiterung der Physik, denen weiterhin eine gro\ss e Bedeutung in der theoretischen Physik zukommen wird und die auch rein mathematisch von gro\ss em Interesse sind. Es muss aber auch kritisch erw\"{a}hnt werden, dass sie in Bezug zur L\"{o}sung der oben angesprochenen konzeptionellen Probleme grunds\"{a}tzlicher Natur nur teilweise einen Fortschritt darstellen. Diese bestehen eben gerade darin, die Dualit\"{a}t zwischen einer Beschreibung von Objekten auf einer vorgegebenen Raumzeit und der Beschreibung der Gravitation als Eigenschaft einer dynamischen Raumzeit aufzuheben und zu einer einheitlichen quantentheoretischen Beschreibung zu gelangen. Der Ansatz der kanonischen Quantisierung der Allgemeinen Relativit\"{a}tstheorie, wie er im Rahmen der Loop-Quantengravitation untersucht wird, scheint diesbez\"{u}glich noch \"{u}berzeugender. Dies liegt einerseits daran, dass er den wichtigsten Grundaussagen der Quantentheorie und der Allgemeinen Relativit\"{a}tstheorie gerecht wird, zu denen bez\"{u}glich letzterer Theorie vor allem die Hintergrundunabh\"{a}ngigkeit geh\"{o}rt. Andererseits kommt er ohne \"{u}ber die beiden Theorien hinausgehende Ad-hoc-Annahmen aus \cite{Rovelli}, was dem heuristischen Gebot Rechnung tr\"{a}gt, eine Theorie auf so wenige zus\"{a}tzliche Annahmen wie m\"{o}glich zu gr\"{u}nden. Dies gilt auch dann, wenn man hinzuf\"{u}gt, dass er zun\"{a}chst nichts \"{u}ber eine Vereinheitlichung mit den anderen Wechselwirkungen aussagt.

\chapter*{Anhang}
\addcontentsline{toc}{chapter}{Anhang}

\begin{appendix}
In diesem Anhang soll gezeigt werden, dass der dem Feynmangraphen ($\ref{Graph2}$) entsprechende Beitrag zur S-Matrix bzw. Feynmanamplitude f\"{u}r den Quarkprozess verschwindet. Unter Verwendung von ($\ref{GravitonpropagatorMasse}$),($\ref{VertexFFG}$)
und ($\ref{VertexVbVbG}$) ergibt sich folgender Ausdruck f\"{u}r den Beitrag zur S-Matrix

\begin{eqnarray}
S_{Graviton}(q(p_1,u_1)+\bar q(p_2,v_2) \rightarrow Z(k_1,Z_1)+Z(k_2,Z_2))\nonumber\\
=\frac{1}{(2\pi)^4}\int d^4 k \frac{u_1}{(2\pi)^{\frac{3}{2}}}\frac{\bar v_2}{(2\pi)^{\frac{3}{2}}}
\left(-\frac{i}{4\bar M_P}[(p_1+p_2)^\mu \gamma^\nu+(p_1+p_2)^\nu \gamma^\mu]\right)\nonumber\\
\cdot (2\pi)^4\delta^4(p_1+p_2-k)
\sum_n \frac{iP_{\mu\nu\rho\sigma}}{k^2-m^2}
(2\pi)^4\delta^4(k-k_1-k_2)\nonumber\\
\cdot \left(-\frac{i}{\bar M_P}\delta^{cd}\left[W^{\rho\sigma\gamma\delta}+W^{\sigma\rho\gamma\delta}\right] \right)
\frac{Z_{1\gamma c}}{(2\pi)^\frac{3}{2}\sqrt{k_{10}}} \frac{Z_{2\delta d}}{(2\pi)^\frac{3}{2}\sqrt{k_{20}}},\nonumber
\end{eqnarray}
wobei $W^{\rho\sigma\gamma\delta}$ und $P_{\mu\nu\rho\sigma}$ wieder gem\"{a}\ss\ ($\ref{VertexVbVbG}$) und ($\ref{PolarisationstensorGraviton}$) definiert sind. Dies kann vereinfacht werden zu

\begin{eqnarray}
S_G&=&\frac{-i}{(2\pi)^2 \sqrt{2k_{10}} \sqrt{2k_{20}}} \frac{1}{\bar M_P^2} \sum_n \frac{1}{p^2-m^2} \nonumber\\
&&\cdot u_1 \bar v_2 [(p_1+p_2)^\mu \gamma^\nu+(p_1+p_2)^\nu \gamma^\mu] P_{\mu\nu\rho\sigma} \left[W^{\rho\sigma\gamma\delta}+W^{\sigma\rho\gamma\delta}\right] Z_{1\gamma} Z_{2\delta} \nonumber\\
&&\cdot \delta^4(p-k_1-k_2).\nonumber
\end{eqnarray}
Mit $p=p_1+p_2$, $A=\frac{1}{(2\pi)^3 \sqrt{2k_{10}}\sqrt{2k_{20}}}\frac{1}{\bar M_P^2} \sum_n \frac{1}{p^2-m^2}$ und ($\ref{SMatrixFeynman}$) erh\"{a}lt man folgenden Ausdruck f\"{u}r den Beitrag zur Feynmanamplitude

\begin{equation}
M_G=A u_1 \bar v_2 [p^\mu \gamma^\nu+p^\nu \gamma^\mu]P_{\mu\nu\rho\sigma}
\left[W^{\rho\sigma\gamma\delta}+W^{\sigma\rho\gamma\delta}\right]Z_{1\gamma} Z_{2\delta}\nonumber
\end{equation}
und durch Ausnutzung der Symmetrieeigenschaften des Polarisationstensors des Gravitons $P_{\mu\nu\rho\sigma}$

\begin{equation}
M_G=4A\cdot u^1\bar v^2 \cdot p^\mu \gamma^\nu \cdot P_{\mu\nu\rho\sigma} \cdot W^{\rho\sigma\gamma\delta} \cdot Z_\gamma Z_\delta.\nonumber
\end{equation}
Im Schwerpunktsystem gilt 

\begin{equation}
k_1^\mu Z_1^\mu=0\quad,\quad k_2^\mu Z_2^\mu=0\nonumber
\end{equation}
und

\begin{equation}
\vec p_1=-\vec p_2\quad,\quad \vec p=0\quad,\quad \vec k_1=-\vec k_2.\nonumber 
\end{equation}
(Die fettgedruckten Gr\"{o}\ss en bezeichnen Dreiervektoren.)
Durch Einsetzen der Ausdr\"{u}cke f\"{u}r $W^{\rho\sigma\gamma\delta}$ 
und $P_{\mu\nu\rho\sigma}$ und Ausnutzen der Relationen f\"{u}r die Polarisation des Z-Teilchens ergibt sich

\begin{eqnarray}
M_G&=&4A u_1 \bar v_2 p^\mu \gamma^\nu
\cdot[\frac{1}{2}(\eta_{\mu\rho}\eta_{\nu\sigma}+\eta_{\mu\sigma}\eta_{\nu\rho})-\frac{1}{3}\eta_{\mu\nu}\eta_{\rho\sigma}\nonumber\\
&&-\frac{1}{2m^2}(\eta_{\mu\rho}p_{\nu}p_{\sigma}+\eta_{\nu\sigma}p_{\mu}p_{\rho}+\eta_{\mu\sigma}p_{\nu}p_{\rho}+\eta_{\nu\rho}p_{\mu}p_{\sigma})\nonumber\\
&&+\frac{1}{3m^2}\eta_{\rho\sigma}p_\mu p_\nu+\frac{1}{3m^2}\eta_{\mu\nu}p_\rho p_\sigma+\frac{2}{3m^4} p_\mu p_\nu p_\rho p_\sigma]\nonumber\\
&&\cdot\left[-\frac{1}{2}\eta^{\rho\sigma}(k_1 \cdot k_2)(Z_1 \cdot Z_2)+(Z_1 \cdot Z_2)(k_1^\rho k_2^\sigma)+(k_1 \cdot k_2) Z_1^\rho Z_2^\sigma \right].\nonumber
\end{eqnarray}
Da $\vec p=0$ gilt, liefern bei der Multiplikation des letzten Termes der unteren eckigen Klammer mit der oberen eckigen Klammer nur die Terme,
die $\eta_{\rho\sigma}$ enthalten (wobei $\rho$ und $\sigma$ hier die Summationsindizes bez\"{u}glich $Z^\rho$ und $Z^\sigma$ bezeichnen), einen Beitrag. Damit tauchen aber keine Terme auf, welche den Faktor $\gamma^\mu Z_\mu$ enthalten. Das bedeutet, dass mn der Form nach einen Ausdruck der folgenden Gestalt erh\"{a}lt

\begin{equation}
M_G=4A\cdot u_1 \bar v_2 \cdot [a \gamma^\mu p_\mu+b \gamma^\mu k_{1\mu}+c \gamma^\mu k_{2\mu}](Z_1 \cdot Z_2),\nonumber
\end{equation}
wobei a, b und c skalare Gr\"{o}\ss en sind, deren genaue Gestalt f\"{u}r die weitere Argumentation keine Rolle spielt.  
Da $M_G$ aber symmetrisch in $k_1$ und $k_2$ ist, muss $b=c$ gelten. Es gilt aber auch $k_1+k_2=p$ und damit $\gamma^\mu k^{1}_\mu+\gamma^\mu k^2_\mu=\gamma^\mu p_\mu$. Das f\"{u}hrt auf folgenden Ausdruck

\begin{equation}
M_G=4A\cdot u_1 \bar v_2 \cdot d \gamma^\mu p_\mu (Z_1 \cdot Z_2),\nonumber
\end{equation}
mit $d=a+2b$. Wenn man nun das Quadrat der Feynmanamplitude bildet, wobei \"{u}ber die Spineinstellungen summiert wird, ergibt sich

\begin{equation}
\sum_\sigma |M|_G^2=16 A^2 d^2 \sum_\sigma (\bar v_2 \gamma^\mu p_\mu u_1)(\bar u_1 \gamma^\mu p_\mu v_2)(Z_1 \cdot Z_2).\nonumber
\end{equation}
Wenn man sich nun den Term $\sum_\sigma (\bar v_2 \gamma^\mu p_\mu u_1)(\bar u_1 \gamma^\nu p_\nu v_2)$ alleine betrachtet und 

\begin{equation}
\sum_\sigma(\bar v_2 \gamma^\mu u_1)(\bar u_1 \gamma^\nu v_2)=Tr\{\gamma^\mu(\gamma^\rho p_{2\rho}-m_f) \gamma^\nu(\gamma^\sigma p^1_\sigma+m_f)\}\nonumber
\end{equation}
ausnutzt (wobei $m_f$ die Fermionenmasse bezeichnet), erh\"{a}lt man

\begin{eqnarray}
\sum_\sigma (\bar v_2 \gamma^\mu p_\mu u_1)(\bar u_1 \gamma^\nu p_\nu v_2)\nonumber\\
=p_\mu p_\nu Tr\{\gamma^\mu(\gamma^\rho p_{2\rho}-m_f) \gamma^\nu(\gamma^\sigma p_{1\sigma}+m_f)\}.\nonumber
\end{eqnarray}
Die Spur einer ungeraden Anzahl von Gammamatrizen verschwindet. Desweiteren gelten die beiden Relationen

\begin{equation}
Tr\{\gamma^\mu \gamma^\nu\}=\eta^{\mu\nu}\quad,\quad Tr\{\gamma^\mu \gamma^\rho \gamma^\nu \gamma^\sigma\}=\eta^{\mu\rho}\eta^{\nu\sigma}-\eta^{\nu\mu}\eta^{\rho\sigma}+\eta^{\mu\sigma}\eta^{\rho\nu}.\nonumber
\end{equation} 
Damit erh\"{a}lt man

\begin{eqnarray}
\sum_\sigma (\bar v_2 \gamma^\mu p_\mu u_1)(\bar u_1 \gamma^\nu p_\nu v_2)\nonumber\\
=p_\mu p_\nu [p_{2\rho} p_{1\sigma} (\eta^{\mu\rho}\eta^{\nu\sigma}-\eta^{\nu\mu}\eta^{\rho\sigma}+\eta^{\mu\sigma}\eta^{\rho\nu})-m_f^2\eta^{\mu\nu}]\nonumber\\
=p_\mu p_\nu[p_2^\mu p_1^\nu-(p_1 \cdot p_2)\eta^{\mu\nu}+p_2^\mu p_1^\nu)]-m_f^2 p^2\nonumber\\
=(p\cdot p_1)(p\cdot p_2)-p^2(p_1\cdot p_2)+(p_1\cdot p)(p_2\cdot p)-m_f^2 p^2\nonumber\\
=4E^4-4 E^2(E^2-\vec p_1 \vec p_2)+4E^4-4E^2 m_f^2\nonumber\\
=8E^4-4 E^2(2 E^2-m_f^2)-4E^2 m_f^2=0.\nonumber
\end{eqnarray}
Im vorletzten Schritt wurde $\vec p_2=-\vec p_1$ und $E^2=(p_1)^2+m_f^2=(p_2)^2+m_f^2$ ausgenutzt.
Damit verschwindet aber $\sum_\sigma |M|_G^2$ und damit auch $M_G$ selbst.

\end{appendix}

\chapter*{Danksagung}

\addcontentsline{toc}{chapter}{Danksagung}

\pagestyle{plain}

Nat\"{u}rlich ist es unm\"{o}glich, all jenen zu danken, die letztlich indirekt zur Entstehung dieser Diplomarbeit beigetragen haben, indem sie mir beim Erwerb des Wissens halfen, das die unabdingbare Voraussetzung f\"{u}r die Behandlung der hier vorgestellten Themen
darstellt. Hier w\"{a}ren zahllose Tutoren, Professoren und Buchautoren und wenn man die Entwicklung nur weit genug zur\"{u}ckverfolgt, auch viele ehemalige Lehrer aufzuf\"{u}hren, denen mein innigster Dank ausgesprochen werden m\"{u}sste.
Vier Namen sind jedoch zu nennen, denen in besonderer Weise Dank geb\"{u}hrt. Es handelt sich um Benjamin Koch, Xavier Calmet, Marcus Bleicher und Horst St\"{o}cker. 

Auf Benjamin Koch geht die Idee zur\"{u}ck, die ZZ-Produktionsrate in der beschriebenen effektiven Quantenfeldtheorie der Gravitation unter Einbeziehung zus\"{a}tzlicher Dimensionen zu untersuchen. Ihm verdanke ich unz\"{a}hlige Ratschl\"{a}ge und hilfreiche Diskussionen. Xavier Calmet hat mir den Vorschlag gemacht, die oben durchgef\"{u}hrte Untersuchung bez\"{u}glich des Higgsmechanismus auf einer nichtkommutativen Raumzeit durchzuf\"{u}hren. W\"{a}hrend zweier sehr angenehmer einw\"{o}chiger Aufenthalte bei ihm in Br\"{u}ssel und \"{u}ber das Internet kam es zu einem fruchtbaren wissenschaftlichen Dialog. Marcus Bleicher danke ich daf\"{u}r, dass er mich zu Xavier Calmet vermittelte und mir die beiden Reisen nach Br\"{u}ssel erm\"{o}glichte. Dar\"{u}ber hinaus hat er mich in den letzten Monaten in vielerlei Hinsicht unterst\"{u}tzt. Hierbei ragen vor allem seine vielen wertvollen Hinweise in Zusamenhang mit der Ver\"{o}ffentlichung eines gemeinsamen Artikels hervor. Schlie\ss lich danke ich meinem offiziellen Betreuer Horst St\"{o}cker. Er hat mir durch die Aufnahme ans Institut f\"{u}r Theoretische Physik \"{u}berhaupt erst die M\"{o}glichkeit er\"{o}ffnet, in diesem Gebiet zu arbeiten. Der erw\"{a}hnte Artikel geht auf seine Idee zu einem interessanten Sommerprojekt zur\"{u}ck. Vor allem aber hat er mit einer hervorragenden sechs Semester w\"{a}hrenden Einf\"{u}hrungsvorlesung in die Theoretische Physik die Grundlage f\"{u}r alle weitere Besch\"{a}ftigung mit dieser faszinierenden Wissenschaft gelegt. 

Am meisten aber schulde ich meinen Eltern Dank. Sie sorgten daf\"{u}r, dass ich mich w\"{a}hrend meines Studiums ganz auf die Physik konzentrieren konnte. Noch bedeutender aber ist, dass ich es im Grunde ihnen zu verdanken habe, mit jener Literatur in Ber\"{u}hrung gekommen zu sein, die mir die Dimension so vieler Fragen in Naturwissenschaft und Philosophie er\"{o}ffnete und mich die f\"{u}r unser Weltbild konstitutive Rolle der Physik als basalster aller Naturwissenschaften erkennen lie\ss.  

\newpage
\noindent
Ich versichere hiermit, dass ich die vorliegende Arbeit selbst\"{a}ndig verfasst, keine anderen als die angegebenen Hilfsmittel verwendet und s\"{a}mtliche Stellen, die benutzten Werken im Wortlaut oder dem Sinne nach entnommen sind, mit Quellen- bzw. Herkunftsangaben kenntlich gemacht habe.\\
\\
Martin Kober

\end{document}